\documentstyle[twocolumn,epsf,aps,prb]{revtex} \begin{document} 
\title{Mode-Coupling Model of Mott Gap Collapse in the Cuprates:\\
Natural Phase Boundary for Quantum Critical Points}
\author{R.S. Markiewicz}
\address{Physics Department, Northeastern University, Boston MA 02115, USA}
\maketitle
\begin{abstract}
A simple antiferromagnetic approach to the Mott transition was recently
shown to provide a satisfactory explanation for the Mott gap collapse with
doping observed in photoemission experiments on electron-doped cuprates.
Here this approach is extended in a number of ways.  RPA, mode
coupling (via self-consistent renormalization), and (to a limited extent)
self-consistent Born approximation calculations are compared to assess the
roles of hot-spot fluctuations and interaction with spin waves.  When
fluctuations are included, the calculation satisfies the Mermin-Wagner
theorem (N\'eel transition at $T=0$ only -- unless interlayer coupling
effects are included), and the mean-field gap and transition temperature
are replaced by pseudogap and onset temperature.  The model is in
excellent agreement with experiments on the doping dependence of both 
photoemission dispersion and magnetic properties.
The magnetic phase terminates in a quantum critical point (QCP), with 
a {\it natural phase boundary} for this QCP arising from hot-spot physics.

Since the resulting T=0 antiferromagnetic transition is controlled by a
generalized Stoner factor, an {\it ansatz} is made of dividing the Stoner
factor up into a {\it material}-dependent part, the bare susceptibility
and a {\it correlation}-dependent part, the Hubbard U, which depends only
weakly on doping. From the material dependent part of the interaction, it
is possible to explain the striking differences between electron- and
hole-doping, despite an approximate symmetry in the doping of the QCP.
The slower divergence of the magnetic correlation length in hole doped
cuprates may be an indication of more Mott-like physics.

Discussions of interlayer coupling, doping dependence of $U$, extension to a 
three-band model, and polaronic effects are included.
\end{abstract}

\narrowtext

\section{Introduction}

Schrieffer, Wen, and Zhang\cite{SWZ} originally proposed that the magnetic
insulating phase in underdoped cuprates could be understood via a spin
density wave (SDW) approach to the Mott transition, and successfully described 
the spin wave spectrum of the undoped parent compound, which is an 
antiferromagnetic (AFM) insulator.  Kampf and Schrieffer\cite{KaSch} showed 
that precursors of the Mott transition could give rise to a {\it pseudogap} in 
the quasiparticle spectrum, between incipient upper and lower Hubbard bands 
(U/LHBs).  Attempts were quickly made to go beyond mean field theories by
incorporating fluctuation effects, but a number of problems soon arose.  While 
some calculations found evidence for pseudogaps\cite{VAT}, others did
not\cite{FLEX}.  

The rapid disappearence of Neel order with hole doping created more
problems: many calculations, even including strong fluctuations, predicted
magnetic order as $T\rightarrow 0$, coupled with diverging magnetic
correlation length $\xi$, whereas $\xi$ is found to remain finite even in
the presence of the pseudogap.  This has been used as evidence that the
band structure picture of the Mott transition breaks down, and must be
replaced by a local picture: `Mott physics' instead of `Slater physics'.
On the other hand, other calculations find evidence for instabilities --
either to incommensurate magnetism\cite{ShSi} or to phase
separation\cite{Schulz,RM3}, and the saturation of $\xi$ could be due to
nanoscale phase separation physics.  The situation is at a stalemate, with
some models neglecting both phase separation and magnetic effects, and
explaining the pseudogap in terms of purely superconducting precursor
effects, while others find a magnetic quantum critical point (QCP) in the
deeply underdoped regime, and yet others find a QCP above optimal doping.

Clearly, a simpler alternative is an important desideratum, and one has
recently been proposed.  While phase separation is a significant
complication for {\it hole doping}, this instability appears to be greatly
reduced or absent in {\it electron-doped} materials\cite{nparm,KLBM},
allowing a much simpler analysis.  Moreover, for electron doping, the band
picture involving short-range commensurate AFM order seems justified, in
that magnetic correlations remain commensurate, while the correlation
length diverges for all dopings up to the QCP. The
desirability of a reference system free of phase separation complications
coheres with Laughlin and Pines' observation\cite{LP}: ``This problem [of
identifying the correct quantum protectorate] is exacerbated when the
principles of self-organization ... compete. ... [H]igher organizing
principles are best identified in the limiting case in which the
competition is turned off, and the key breakthroughs are almost always
associated with the serendipitous discovery of such limits.''

While many models attempt to describe the properties of the cuprates over
a limited doping range, it has proven difficult to systematically
reproduce the changes over an extended doping range.  Remarkably, simple
mean field calculations\cite{KLBM} were able to reproduce the full doping
dependence of ARPES spectra in the electron-doped cuprates\cite{nparm} in
terms of a Mott gap collapse (QCP) near optimal doping.  Here, these
results are expanded upon in a number of ways.  First, a number of
models are applied to the electron-doped system, to see the effects of
various correlations.  A key issue is finite temperature effects: the RPA
predicts a Neel temperature $T_N\sim U$ -- much larger than found
experimentally.  Proper inclusion of thermal fluctuations, introduced via
a self-consistent renormalization (SCR) model\cite{Mor,ICTP}, drives the
N\'eel temperature $T_N$ to zero (Mermin-Wagner theorem), replacing the
RPA gap $\Delta_{mf}$ by a SCR pseudogap $\Delta^*$ and the mean-field
N\'eel temperature $T^{mf}_N$ with a crossover temperature $T^*$, with 
$\Delta_{mf}\simeq\Delta^*$, $T^{mf}_N\simeq T^*$.  Inclusion of spin wave
scattering, via the self-consistent Born approximation (SCBA) produces a
large incoherent background, but the coherent part of the spectrum is
recognizably the same as the spectrum found in the RPA and SCR approaches,
with only moderate band renormalizations.  The present conclusions
(including Ref.~\onlinecite{ICTP}) are consistent with more recent
findings\cite{KuR,SeMSTr}.

In summary, there is a QCP near optimal doping in the {\it electron}-doped
cuprates, associated with Mott (pseudo)gap collapse.  The transition is
characterized by three concurrent factors: termination of a
zero-temperature AFM transition (which can be associated with a finite-T
N\'eel transition due to weak interlayer coupling); collapse of a
pseudogap centered on $(\pi ,\pi )$; and crossover of the Fermi surface
from small pockets to large barrel.  Good agreement with experiment
requires a weak Kanamori-style\cite{Kana} renormalization of the Hubbard
$U$ with doping.  The same model can describe both ARPES and magnetization
results.  A similar QCP is predicted at a comparable {\it hole}
doping -- indeed a {\it natural phase boundary} for magnetism exists,
associated with hot spot physics.  There is however, a striking difference
in the hole doping case: saturation of a spin sum rule leads to much
smaller correlation lengths and absence of finite-T N\'eel order.  Stripe
physics appears to play a lesser role -- turning on at lower temperatures
-- possibly as a form of interaction of the doped polarons.
 
This paper is organized as follows.  Section II describes the SCR
formalism.  Related Appendices discuss the extension to a
three band model (Appendix A), the doping dependence of $U$ (Appendix
B), and a more accurate solution of the self-consistency equation
(Appendix C). Since the transition occurs when a Stoner factor
equals unity, it is controlled by the {\it real} part of the bare
susceptibility.  Hence Section III reviews the properties of $Re\chi$, 
showing that plateaus in $\chi$ as a function of doping, $\vec q$, or
$\omega$ are all controlled by the physics of hot spots.  In turn, these
plateaus provide {\it natural phase boundaries} for QCPs.  The 
resulting susceptibility has a form similar to that postulated for a
nearly antiferromagnetic Fermi liquid (NAFL), but a calculation of the
NAFL parameters (Appendix D) finds that there are {\it extra} (cutoff)
parameters, which cannot be neglected.  In Section IV, this renormalized
susceptibility is incorporated into the lowest-order correction to the
electronic self energy, allowing a calculation of the spectral function
associated with the {\it pseudogap} ($T_N=0$).  Excellent agreement is
found with the ARPES spectra of Nd$_{2-x}$Ce$_x $CuO$_{4\pm\delta}$
(NCCO).  A remaining problem lies in magnetic polaron effects which are
expected at very low dopings: these bear some resemblance to 
nanoscale phase separation, and can lead to anomalous localization effects.  
Section V offers a brief introduction to these effects, by analyzing the
self-consistent Born approximation at half filling.  It is found that only
minor quantitative changes to the earlier results are expected.  An
extension of the results to the hole-doped regime is considered in Section
VI.  The model also provides a good description of magnetic properties, as
discussed in Section VII.  Section VIII shows that inclusion of interlayer
hopping leads to a finite $T_N$ (Appendix E).  Results are discussed in
Section IX, and Conclusions in Section X.  Some of these results have
been reported previously in the discussion of the mean-field
results\cite{KLBM} and in a conference procedings\cite{ICTP}. 

\section{Mode-Coupling Calculation}

\subsection{Model Dispersion and Doping Dependence of $U$}

In the present paper the mean field results are extended by incorporating
fluctuations via mode-coupling theory\cite{MuDo}, following Moriya's
self-consistent renormalization (SCR)\cite{Mor,HasMo,MoreMor} procedure.
Mode coupling theories have been applied to charge density wave (CDW)
systems\cite{LRA,RM5}, and have led to a successful theory of weak
itinerant magnetic systems\cite{Mor,HasMo}.  They have also been used to
study glass transitions\cite{glass}, and recently extended to glasses in
cuprates\cite{Cast2}.  The mode coupling analysis is particularly
convenient, being the simplest model for which the Mermin-Wagner theorem
is satisfied.  The resulting pseudogaps compare well with recent
photoemission experiments in electron-doped cuprates. While the SCR
technique can be generalized to deal with competing phases\cite{OnI}, only
the antiferromagnetic fluctuations will be treated here. 

The cuprates are treated in a one-band model.  By comparison with a
3-band model (Appendix A), this can be shown to be an excellent 
approximation for the magnetic properties.  The bare electronic dispersion is 
\begin{equation}
\epsilon_k=-2t(c_x+c_y)-4t'c_xc_y, 
\label{eq:0}
\end{equation}
with $c_i=\cos{k_ia}$.  The dispersions for undoped Sr$_2$CuO$_2$Cl$_2$
(SCOC) and electron-doped NCCO can be fit by assuming $t=0.326eV$,
$t'/t=-0.276$, with $U$ taken as an effective doping dependent
parameter\cite{KLBM}, with $U=6t$ at half filling.  Similar parameters are
found\cite{PeAr} to describe the spin wave spectrum\cite{RoRi} in
La$_2$CuO$_4$: $t=0.34eV$, $t'/t=-0.25$, and $U/t=6.2$.  The former values
will be used here.

Many textbooks on strong correlation physics\cite{Ful,NNag} note that the 
Hubbard $U$ should be doping dependent, based on the original results of
Kanamori\cite{Kana}, but there are no satisfactory results for the doping
dependence in the cuprates.  A simple model calculation, which gives
semiquantitative agreement with experiment in NCCO\cite{nparm,KLBM}, is
described in Appendix B.

\subsection{Self-Consistent Equation}

The SCR scheme is introduced to incorporate strong fluctuations near the 
antiferromagnetic wave vector $\vec Q$.  The (path integral) formalism is 
standard\cite{NNag} and only the main results are given here.  The quartic
Hubbard contribution to the Hamiltonian is decoupled by a
Hubbard-Stratonovich transformation introducing spin wave fields $\phi$.
The Fermion fields are then integrated out, leaving an approximate
quartic effective action, which describes fluctuations about the mean
field solution due to mode coupling.  In the SCR model, the dynamical
susceptibility is found self-consistently as 
\begin{equation}
\chi (\vec q,i\omega_n)={\chi_0(\vec q,i\omega_n)\over 
1-U\chi_0(\vec q,i\omega_n)+\lambda},
\label{eq:B14}
\end{equation}
with the bare susceptibility
\begin{equation}
\chi_0(\vec q,\omega) =-\sum_{\vec k}{f(\epsilon_{\vec k})-f(\epsilon_{\vec k+
\vec q})\over\epsilon_{\vec k}-\epsilon_{\vec k+\vec q}+\omega+i\delta},
\label{eq:0a}
\end{equation}
where $\delta$ is a positive infinitesimal, 
and the RPA susceptibility given by Eq.~\ref{eq:B14} with $\lambda =0$.

The leading divergence corresponds to AFM at $\vec q=\vec Q$, so the
denominator of Eq.~\ref{eq:B14} -- the (inverse) Stoner factor -- is
expanded in terms of the small parameters $\omega$ and $\vec q'\equiv\vec
q-\vec Q$ (analytically continuing $i\omega_n\rightarrow \omega
+i\epsilon$): 
\begin{equation}
\delta_q(\omega )=1-U\chi_0(\vec q,\omega )+\lambda =
\delta +Aq'^2-B\omega^2-iC\omega,
\label{eq:B18}
\end{equation}
where 
\begin{equation}
\delta =1-U\chi_0(\vec Q,0)+\lambda ,
\label{eq:B17}
\end{equation}
and $\delta_0=\delta -\lambda$. The self-consistent equation for $\delta$ is 
\begin{equation}
\delta =\delta_0+{12u\over\beta V}\sum_{\vec q,i\omega_n}D_0(\vec q,i\omega_n),
\label{eq:B30}
\end{equation}
where $u$ is a measure of the quartic mode-mode coupling (Appendix D4) and 
\begin{equation}
D_0^{-1}(\vec q,i\omega_n)=\delta+Aq'^2+C|\omega_n|.
\label{eq:B25}
\end{equation}
The sum over Matsubara frequencies can be carried out using
\begin{eqnarray}
{1\over\beta}\sum_{i\omega_n}X(i\omega_n)=-{1\over\beta\pi}\sum_{i\omega_n}\int
_{-\infty}^{\infty}d\epsilon{ImX(\epsilon +i\delta )\over i\omega_n-\epsilon}
\nonumber \\
=-\int_0^{\infty}d{\epsilon\over \pi}coth{\epsilon\over 2T}ImX(
\epsilon +i\delta ).
\label{eq:B31}
\end{eqnarray}
Then
\begin{eqnarray}
{1\over\beta V}\sum_{\vec q,i\omega_n}D_0(\vec q,i\omega_n)
\nonumber \\
=\int{d^2\vec qa^2\over(2\pi )^2}\int_0^{\alpha_{\omega}/C}{d\epsilon\over\pi}
coth{\epsilon\over 2T}{C\epsilon\over (\delta +Aq^{'2})^2+(C\epsilon )^2}.
\label{eq:B32}
\end{eqnarray}
Note the sharp energy cutoff in Eq.~\ref{eq:B32}.  This comes about because the
linear-in-$\omega$ dissipation is a result of Landau damping of the spin
waves by electrons near the hot spots, and therefore the dissipation cuts off
when the spin wave spectrum gets out of the electron-hole continuum.  The
cutoff parameter $\alpha_{\omega}$ is defined in Appendix D2, above
Eq.~\ref{eq:C8b}.  Numerical calculations (Fig.~\ref{fig:11}) show that
the cutoff can be quite sharp, particularly near the VHS.  

\subsection{Approximate Solutions}

Equations~\ref{eq:B30},~\ref{eq:B32} can easily be solved in the limit $T=0$.  
In this case, there is a transition at
\begin{eqnarray}
\delta_0=-12u\int_0^{q_c^2}{dq'^2a^2\over 4\pi}\int_0^{\alpha_{\omega}/C}{d
\epsilon\over\pi}{C\epsilon\over (Aq'^2)^2+(C\epsilon )^2}
\nonumber \\
=-{3uq_c^2a^2\over \pi^2C}R_0\equiv 1-\eta ,
\label{eq:B35}
\end{eqnarray}
\begin{equation}
R_0={1\over 2}ln[1+a_q^{-2}]+{tan^{-1}(a_q)\over a_q},
\label{eq:B34a}
\end{equation}
with $a_q=Aq_c^2/\alpha_{\omega}$.  Since the right-hand side is finite
and negative, fluctuations reduce but in general do not eliminate the
order at $T=0$.  At the RPA level ($\lambda =0$), the AFM instability is
controlled by the Stoner criterion, $\delta_0\rightarrow 0$.  
The quantum corrected Stoner criterion is $U\chi_0=\eta$,
where representative values of $\eta$ are listed in Table I.

However, for finite $T$, there are corrections $\sim ln(\delta )$, so $\delta$
cannot be set to zero, and there is no finite temperature transition (the
Mermin-Wagner theorem is satisfied).  To see this, it is adequate to
approximate $coth(x)$ as $1/x$ for $x\le 1$ and 1 for $x>1$. In this case, 
Eq.~\ref{eq:B30} can be solved exactly, Appendix C.  However, this exact
solution is not very illuminating, and a simpler approximate solution will
be given here. Since only the term proportional to $T$ is singular, $T$
and $\delta$ can be set to zero in the remaining term.  Defining 
\begin{equation}
\bar\delta_0=\delta_0+\eta -1,
\label{eq:B34g}
\end{equation}
Eq.~\ref{eq:B30} becomes
\begin{eqnarray}
\delta-\bar\delta_0={6uTa^2\over\pi^2A}\int_{\delta}^{\delta+Aq_c^2}{dy\over
y}tan^{-1}({2TC\over y})
\nonumber \\
\simeq{3uTa^2\over\pi A}ln({2CT\over\delta}),
\label{eq:B34h}
\end{eqnarray}
where the second line uses Eq.~\ref{eq:B34bx}, below.
Hence, there is no finite temperature phase transition, and $\delta$ only
approaches zero asymptotically as $T\rightarrow 0$: approximately,
\begin{equation}
\delta =2CTe^{-\pi A|\bar\delta_0|/3uTa^2}.
\label{eq:B34f}
\end{equation}

\subsection{Susceptibility}                      

Given the (inverse) Stoner factor $\delta_q$, Eq.~\ref{eq:B18}, the 
renormalized susceptibility can be written in nearly-antiferromagnetic
Fermi liquid (NAFL)\cite{NAFL} form,
\begin{equation}
\chi (\vec q,\omega )={\chi_Q\over 1+\xi^2(\vec q-\vec Q)^2
-\omega^2/\Delta^2-i\omega /\omega_{sf}},
\label{eq:15}
\end{equation}
with coefficients 
\begin{equation}
\chi_{\vec Q}={\chi_0\over\delta}
\label{eq:B36}
\end{equation}
\begin{equation}
\xi^2={A\over\delta} 
\label{eq:B37}
\end{equation}
\begin{equation}
\Delta^2={\delta\over B} 
\label{eq:B38}
\end{equation}
\begin{equation}
\omega_{sf}={\delta\over C} 
\label{eq:B39}
\end{equation}
The similarity of Eq.~\ref{eq:15} to the corresponding result for
CDW's\cite{RM5} should be noted -- the SCR is a form of mode coupling theory.

In the renormalized classical regime, the vanishing of $\delta$ as $T\rightarrow
0$ is controlled by a correlation length, Eq.~\ref{eq:B37}, which can be
written as\cite{CHN}
\begin{equation}
\xi =\xi_0e^{2\pi\rho_s/k_BT}.
\label{eq:16}
\end{equation}
Numerically solving Eq.~\ref{eq:B34h} (or Eq.~\ref{eq:B34e}) for $\delta$,
then the spin stiffness $\rho_s$ is exactly given by 
\begin{equation}
\rho_s={k_BT\over 4\pi}\ln({A\over\xi_0^2\delta}),
\label{eq:B40a}
\end{equation}
with $\xi_0=\sqrt{eA\over 2TC}$.  Using Eq.~\ref{eq:B34f}, an approximate 
$\rho_s$ is:
\begin{equation}
\rho_s^a={A|\bar\delta_0|\over 12ua^2}.
\label{eq:B40}
\end{equation}
$\rho_s$ is plotted in Fig.~\ref{fig:11a1}b, with $u^{-1}=0.384eV$, chosen
to give a $\rho_s$ in agreement with experiment for $x=0$, $T=0$ (Section 
VII).  The T-dependence of the prefactor $\xi_0$ agrees with {\it
one-loop} $\sigma$-model results\cite{KoCha} rather than the more accurate 
two-loop results\cite{CHN,Tkhsh}.  This difference is presumably a deficiency of
the present model in not using fully self consistent parameters; it will
be discussed further in Section VII.  

\subsection{Parameter Evaluation}

The susceptibility Eq.~\ref{eq:15} is well-known in NAFL\cite{NAFL,MTU} and 
spin fermion\cite{Chu,Chu2} theories and in renormalization group (RG)
calculations of quantum phase transitions\cite{Her,Mil}.  In these
calculations, the parameters of Eq.~\ref{eq:15} (equivalently, $A$,
$B$, and $C$) are usually determined empirically from fits to experiments.
However, the good agreement between experiment and mean field theory for
electron doped cuprates encourages us to try to {\it calculate} these
parameters from first principles, following Ref.~\onlinecite{HasMo}, in
terms of a renormalized Hubbard parameter $U_{eff}$ and a mode coupling
parameter $u$.  This puts a special premium on the {\it bare susceptibility}
$\chi_0$, which is assumed to control the main material, doping, and
pressure dependence of the Mott transition, while the {\it interaction 
parameters} $U$ and $u$ are relatively constant.  In fact, it is found
that $U$ has a weak but important doping dependence, estimated in Appendix
B, which is consistent with experiment.  A single, doping-independent
value of $u$ is chosen to agree with $t-J$ results at half filling.

The motivation for this approach comes from experience with another strongly 
correlated system: electron-hole droplets in photoexcited semiconductors.
Here it was found\cite{MaKe,KiSi} that the correlation effects were
controlled by an isotropic density-dependent interaction potential,
whereas the material, anisotropy, and uniaxial pressure dependence were
controlled by the kinetic energy -- i.e., by the bare band structure.  A
similar approach was applied to CDW systems\cite{SaRS}.

\section{Hot Spot Plateaus and Generic QCPs}

While the properties of $\chi_0$ are now reasonably well understood, they
remarkably do not seem to have been used to derive the parameters of
SCR or NAFL theory.  Here this oversight is corrected.  In particular,
calculation of the curvature parameter $A$ is discussed below.  A new
cutoff parameter $q_c$ is introduced, which is essential in explaining the
differences between the QCPs for hole and electron dopings.  The
corresponding frequency parameter $C$, Eq~\ref{eq:C8}, and its associated
cutoff parameter $\alpha_{\omega}$ (below Eq~\ref{eq:C8b}) are discussed
in Appendix D2.


\subsection{Plateaus in Doping Dependence}

\subsubsection{Hot Spots}

\begin{figure}
\leavevmode
   \epsfxsize=0.38\textwidth\epsfbox{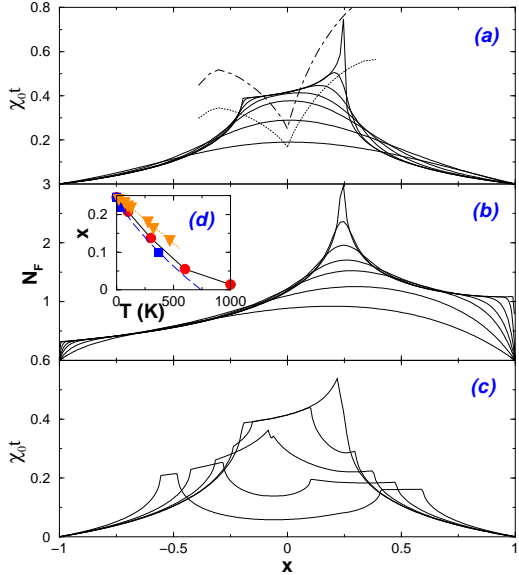}
\vskip1.5cm
\caption{(a) Susceptibility $\chi_0$ at $\vec Q$ as a function of doping for 
several temperatures.  From highest to lowest curves near $x=0.1$, the
temperatures are $T$ = 1, 100, 300, 600, 1000, 2000, and 4000 K. Dotted line =
$1/U_{eff}$, dot-dashed line = $1.5/U_{eff}$.  (b) Density of 
states $N_F$ for the same temperatures.  (c) Susceptibility $\chi_0$ at $\vec Q$
as a function of doping for several frequencies at $T=1K$: $\omega$ = 0.01, 0.1,
0.3, 0.6, 1.0 eV.  (d) Pseudo-VHS (peak of $\chi_0$) as a function of 
temperature $T_V$ (circles) or scaled frequency $T_c^-=\omega_c^-/\pi$ 
(squares); triangles = $T_{incomm}$.}
\label{fig:0a}
\end{figure}

In the self-consistent renormalization scheme, the $T=0$ AFM transition is
controlled by a Stoner factor, $URe(\chi_0)=\eta$, where $\eta>1$ includes
a quantum correction, Table I.  Hence, the relevant quantity on which the
study is based is the real part of the bare magnetic susceptibility,
Eq.~\ref{eq:0a}.  This susceptibility has been analyzed in a number of
papers.  Whereas usually only $Im(\chi )$ is explored in detail (e.g.,
Refs.~\onlinecite{BCT,SiZLL,LaSt}), $Re(\chi )$ was studied in
Ref.~\onlinecite{OPfeut}.  The extended discussion which follows is
intended to bring out salient features for the computation of the NAFL
parameters.  

The doping dependence of $\chi_0(\vec Q,\omega)$ is
illustrated in Fig.~\ref{fig:0a}a, where $\vec Q=(\pi ,\pi )$. 
At low $T$, the susceptibility has a plateau shape, which is not present
in the density of states, $N_F$, Fig.~\ref{fig:0a}b. Beyond the plateau
edges $\chi_0$ falls off sharply on both electron and hole doping sides of
half filling.  This sharp falloff explains the appearence of QCPs: the
Stoner criterion is satisfied on the plateau, but fails when $\chi_0$
drops.  

The plateau shape is characteristic of hot spot physics.  Hot spots are
those points where the Fermi surface (FS) intersects the replica FS
shifted by $\vec Q$.  They are located at $c_x=-c_y=c_{x0}$, with
\begin{equation}
c_{x0}=\cos{ak_{x0}}=\sqrt{\mu\over 4t'},
\label{eq:C1}
\end{equation}
and equivalent points.  The edges of the plateau are those points at which the 
overlap terminates (hot spots cease to exist).  For the present band
structure, hot spots exist only when the chemical potential $\mu$ is in
the range $4t'\le\mu\le 0$, or for doping $0.25>x>-0.19$ (electron dopings
are considered as negative).  
Since the two end points play an important role, it is convenient 
to label them, and they are here called `hot' hot spot and `cold' hot spot
(or H-point and C-point) for the hole and electron-doped termination
points, respectively.  It will be demonstrated below that at each
doping, the hot spots also lead to a susceptibility plateau in momentum
space, around $\vec Q$, collapsing to a logarithmic (square root)
divergence at the H- (C-)point.  The $H$-point is the VHS, and hence also
involves a conventional ETT. The physics is simpler near the $C$-point,
where the topology hardly changes but the FS and $\vec Q$-FS become
decoupled (it is therefore a form of Kohn anomaly\cite{OPfeut}).  

\subsubsection{Mean Field Mott Transition}

\begin{figure}
\leavevmode
   \epsfxsize=0.33\textwidth\epsfbox{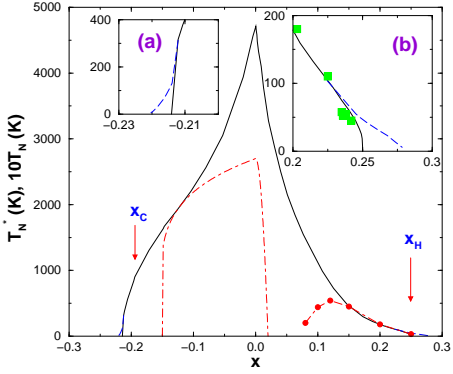}
\vskip0.5cm
\caption{Mean field magnetic transition temperatures determined from Stoner 
criterion using $U_{eff}$ of Fig.~\protect\ref{fig:0a}.  Solid line: 
commensurate (at $\vec Q$); long dashed line: incommensurate. Dot-dashed
line = 10$T_N$, where $T_N$ is the onset of long range AFM order, from 
[\protect\onlinecite{AIP}] and [\protect\onlinecite{Ich}] (with filled
circles). Insets = blowups near C- and H-points.  Squares in inset b =
pseudogap data of [\protect\onlinecite{Kras}].} 
\label{fig:0d}
\end{figure}

For the parameter values expected in the cuprates, these susceptibility
plateaus control the physics of the Mott gap collapse.  As a 
function of doping, the mean field Mott gap is found to close at a
doping just beyond the edge of the plateau, {\it for both electron and
hole doping}, Fig.~\ref{fig:0d}.  The solid and long dashed lines are the
commensurate and incommensurate mean field Mott transition temperatures
$T^*(x)$ calculated using the estimated $U_{eff}(x)$, dotted line in
Fig.~\ref{fig:0a}.  For electron doping, there is a double transition, 
first from commensurate to incommensurate antiferromagnetic order at the
plateau edge, then to the loss of any magnetic order at a slightly higher
doping (inset a).  For hole doping, the dominant antiferromagnetic order
is incommensurate for all dopings, but the difference in $T_N$ becomes
significant only near the H-point (inset b).  When fluctuations are
included (below), it is found that the N\'eel transition is shifted to
zero temperature, while a pseudogap first appears near the mean field
$T_N$.  Note that in the hole doped regime, there is good agreement
between the mean field transition and the pseudogap (squares in
Fig.~\ref{fig:0a}b = data of Krasnov\cite{Kras}, assuming $2\Delta =4.6T^*$).
For the real cuprates, the terminations of the Mott gaps are preempted by
superconducting transitions, close to the critical regime.  

\subsubsection{The Pseudo-VHS}

The susceptibility Fig.~\ref{fig:0a}a has a remarkable doping dependence,
with the large peak at
the Van Hove singularity (VHS) shifting\cite{OPfeut} to half filling with 
increasing temperature $T$. The peak position of this `pseudo-VHS' defines a 
temperature $T_V(x)$, Fig.~\ref{fig:0a}d (circles).  This behavior can readily 
be understood from the form of $\chi_0(\vec Q,0)$, Eq.~\ref{eq:0a}.  The
denominator $\epsilon_{\vec k}-\epsilon_{\vec k+\vec Q}=-4t(c_x+c_y)$, is {\it 
independent of $t'$}, and hence has a stronger divergence than the density
of states (dos).  Indeed, this divergence matches the strong VHS found for
$t'=0$ (perfect nesting), and like that VHS falls {\it at half filling},
$x=0$.  There is one crucial difference -- at low temperatures, this
divergence is {\it cut off} by the Fermi functions, which leave the
integrand non zero in a wedge which intercepts the zone diagonal (where
the denominator vanishes) only at isolated points: the hot spots.  Hence,
the residual divergence at low $T$ is still dominated by the conventional
VHS.  However, {\it at finite $T$}, excitations along the zone diagonal
become allowed, leading to a stronger divergence of $\chi_0(\vec Q,0)$
near $x=0$. 

The strong temperature dependence of the pseudo-VHS is in strong contrast to 
the density of states, $N_F$, Fig.~\ref{fig:0a}b, and also with the pairing
correlations\cite{OPfeut}.  The denominator of the pairing susceptibility 
involves the {\it sum} of the energies, $\epsilon_{\vec k}+\epsilon_{\vec k+\vec
Q}=-8t'c_xc_y$, rather than their difference (as in Eq.~\ref{eq:0a}), and hence 
always peaks at the ordinary VHS.

The difference between nesting and pairing susceptibilities has a
fundamental significance.  By mixing electron and hole-like excitations,
the superconducting gap is always pinned to the Fermi level, and can open
up a full gap at any doping.  On the other hand, a nesting gap need not be
centered on the Fermi surface, and is constrained to obey Luttinger's
theorem, conserving the net number of carriers in the resultant Fermi
surface.  Hence, the only way a nesting instability (such as
antiferromagnetism) can open a full gap at the Fermi level is for the
instability to migrate with increased coupling strength to integer filling
of a superlattice zone (e.g., half filling of the normal state).

Since the susceptibility has such a distinct {\it temperature} dependence from
the density of states, one might ask how the {\it frequency} dependence
compares.  This is illustrated in Fig.~\ref{fig:0a}c at low temperature (1K).
While the frequency introduces additional sharp features  and has
an overall very distinct appearence from the T-dependence, nevertheless the main
peak also shifts from the VHS toward lower doping with increasing $\omega$
-- in fact, the shift is almost the same when comparing $\hbar\omega$ and
$\pi k_BT$, Fig.~\ref{fig:0a}d.  The dashed line in Fig.~\ref{fig:0a}d is
$T_c^-=\hbar\omega_c^-/\pi k_B$, with\cite{BCT} 
\begin{equation}
\omega_c^-={4t(\hat\mu -\tau)\over 1-\tau},
\label{eq:00e}
\end{equation}
with $\tau=2t'/t$ and $\hat\mu=\mu /2t$. 
The proportionality of frequency and temperature dependences holds only
in the hole doped regime: temperature shifts the susceptibility peak only to
half filling, $x=0$, while frequency will shift the peak beyond half filling 
($x<0$).

The structure in the low temperature susceptibility, Fig.~\ref{fig:0a}, with its
largest peak at the H-point on the hole doped side, is in striking contrast to 
the calculated doping dependence of the N\'eel transition, Fig.~\ref{fig:0d},
which has a broad plateau on the electron-doped side, but falls off more 
quickly with hole doping, showing no sign of a peak near the VHS.  
This contrast can be accounted for by two effects.  First, the shift of 
spectral weight with temperature of the pseudo-VHS, noted in 
Fig.~\ref{fig:0a}, would tend to produce a symmetric falloff of $T_N$
with either electron or hole doping. But the dos peak at the VHS leads to
better screening of $U_{eff}$ for hole doping, thereby further depressing
$T_N$.  

\subsubsection{Neel Transition}

The mean field Neel transition is associated with {\it short-range}
magnetic order, and hence should be compared to the experimental pseudogap
transition $T^*$, while the experimental Neel transition involves {\it
long-range} magnetic order.  It is controlled by small parameters, such as
anisotropy and interlayer coupling (Section VIII) and need have no
connection to the mean field $T_N$.  Nevertheless, the mean   
field calculation provides an approximate envelope of the resulting data, 
but overestimates the transition temperatures by a factor of 10,
Fig.~\ref{fig:0d}.  The agreement is particularly good on the electron
doped side (except for overestimating the doping of the QCP), while for
hole doping the experimentally observed\cite{AIP} $T_N$ (dot-dashed line)
shows a stronger falloff, perhaps due to phase
separation.  Since stripes can frustrate magnetic order, the figure also
includes the magnetic ordering temperature of quasi-static stripe arrays,
from Nd-substituted La$_{2-x}$Sr$_x$CuO$_4$ (LSCO)\cite{Ich}, which is
taken as a lower bound for the N\'eel ordering transition in the absence
of stripes.  A possible explanation for the rough proportionality of the
mean field and long-range N\'eel transitions will be discussed in Section
VIII.  

\subsection{Plateaus in Momentum Space}

\subsubsection{Plateaus}

\begin{figure}
\leavevmode
   \epsfxsize=0.33\textwidth\epsfbox{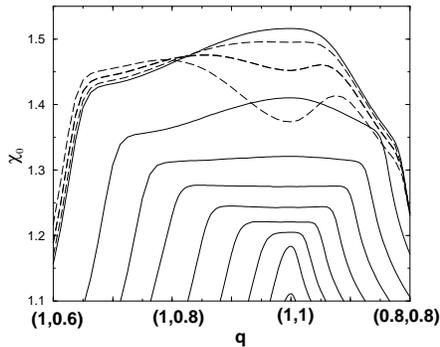}
\vskip0.5cm
\caption{Susceptibility $\chi_0$ near $\vec Q$ for a variety of dopings at 
$T=100K$.  From highest to lowest solid curves near $S\equiv\vec Q$, the 
chemical potentials are $\mu$ = -0.35, -0.30, -0.25, -0.20, -0.15, -0.10, 
-0.055, -0.02, and 0 eV.  For the dashed curves (top to bottom), $\mu$ = 
-0.352, -0.355, and -0.359eV.}
\label{fig:41}
\end{figure}
\begin{figure}
\leavevmode
   \epsfxsize=0.33\textwidth\epsfbox{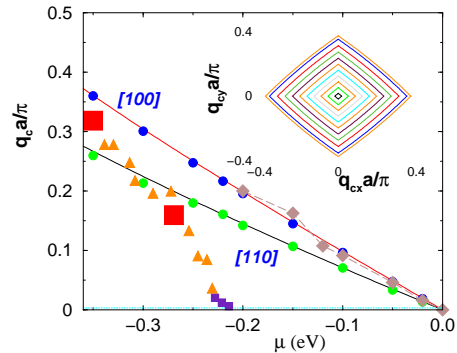}
\vskip0.5cm
\caption{Plateau width $q_c$, comparing Eq.\protect\ref{eq:0e} (solid lines)
and the measured widths (circles) from Fig.~\protect\ref{fig:41}.  Upper
curve along $[q_c,0]$ direction, lower along $[q_c,q_c]/\sqrt{2}$ direction.
Symbols = experimental inverse correlation lengths $\xi^{-1}$ from YBCO:
large squares = Ref.~\protect\onlinecite{BoRe}, triangles =
Ref.~\protect\onlinecite{BalBo}; LSCO: small squares =
Ref.~\protect\onlinecite{Keim}.  Diamonds = $T^*_A/5000K$. Dotted line:
$\xi =100a$.  Inset = plateau boundary for a series of chemical
potentials $\mu$ from 0 (smallest) to -0.359eV (largest).}
\label{fig:41a}
\end{figure}
\begin{figure}
\leavevmode
   \epsfxsize=0.33\textwidth\epsfbox{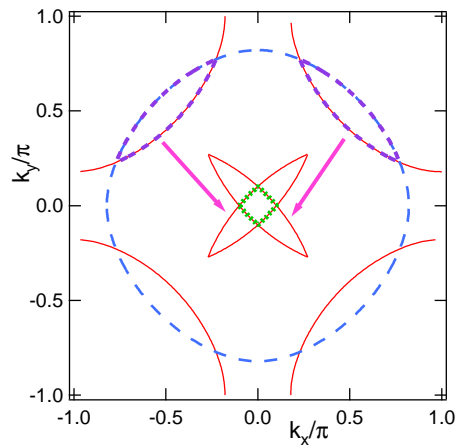}
\vskip0.5cm
\caption{Illustrating origin of plateaus (dotted line) from crossed
hole pockets (short dashed lines).}
\label{fig:41aa}
\end{figure}

In analyzing either thermal fluctuations or the quantum fluctuations associated 
with QCPs, it is necessary to understand the susceptibility near the AFM
vector $\vec Q$. At each doping, hot spot physics leads to a plateau in momentum
space, centered on $\vec Q$. Figure~\ref{fig:41} shows how $\chi_0$
varies near $\vec Q$ at a low temperature (100K) for a series of different
dopings.  Results near $T=0$ are presented in 
Ref.~\onlinecite{ICTP}.  For all dopings there is a plateau in $q$.  The
width of the plateau at $T=0$ can be readily determined: in any
direction, it is the minimum $q$ needed to shift the replica FS so that
the hot spots are eliminated.  This can be found from the dispersion,
Eq.~\ref{eq:0}, by substituting $\vec k\rightarrow (\vec Q+\vec q)/2$, or
\begin{equation}
-2t(\hat s_x+ \hat s_y)-4t'\hat s_x\hat s_y=\mu,
\label{eq:0e}
\end{equation}
with $\hat s_i=\sin{(q_ia/2)}$.  As shown in Fig.~\ref{fig:41a}, this formula
agrees with the (anisotropic) plateau width measured from
Fig.~\ref{fig:41} (circles). The inset shows the shape of the plateau as a
function of doping.  The diamond shape of the plateau, Eq.~\ref{eq:0e}, is
related to the profile of the hole pockets formed by the overlap of the
shifted and unshifted FSs.  Specifically, the plateau is the region of
overlap of the two hole pockets, shifted to have a common center, as
illustrated in Fig.~\ref{fig:41aa}. The remaining parts of the pockets
also show up, as ridges\cite{RMrec} in the susceptibility, radiating from
the corners of the diamond (similar to the peaks in the $\mu =0.05eV$ data
in Fig.~\ref{fig:42}, below).  As noted by B\'enard, et al.\cite{BCT}, the
susceptibility in two-dimensions acts as a FS caliper.  The plateau width
leads to a natural limit on the magnetic correlation length, $\xi_c\sim
1/q_c$, in agreement with experimental data from YBa$_2$Cu$_3$O$_{7
-\delta}$ (YBCO)\cite{BoRe,BalBo} (squares, triangles in
Fig.~\ref{fig:41a}), as noted previously\cite{LaSt,OPfeut}.  Related data
from LSCO\cite{Keim} are also shown.


\subsubsection{Cusps}

\begin{figure}
\leavevmode
   \epsfxsize=0.33\textwidth\epsfbox{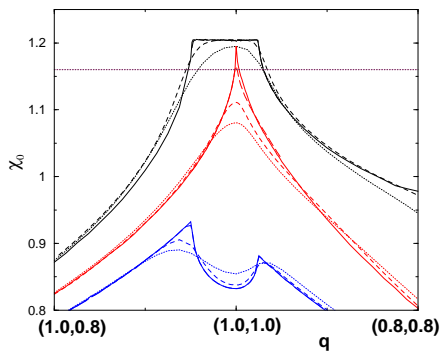}
\vskip0.5cm
\caption{Susceptibility $\chi_0$ near $\vec Q$ for several dopings near the
C-point.  Upper group at $\mu$ = -0.05eV, middle at $\mu$ = 0 (C-point), and
bottom at $\mu$ = +0.05eV.  Temperatures are $T=200K$ (dotted lines), 100K (short
dashed lines), 10K (long dashed lines), 1K (solid lines).  Horizontal line =
$U_{eff}(\mu =0)$.}
\label{fig:42}
\end{figure}

For electron doping, the plateaus in $q$ are particularly flat topped,
Fig.~\ref{fig:41}.  At low temperatures the edges sharpen up
Fig.~\ref{fig:42} and the falloff in $\chi_0$ acquires a square-root
singularity (Appendix D1).  The width of the plateau decreasing to zero
as $x\rightarrow x_C$, and for electron-doping beyond the C-point
($\mu>0$), the plateau ends and the susceptibility displays split peaks
away from $\vec Q$, Fig.~\ref{fig:42}, with a dip in between.  
Thus $x_C$ is a QCP\cite{OPfeut} where the magnetic order changes from
commensurate to incommensurate.   (There is a corresponding QCP at the
H-point\cite{OPfeut}.)  

However, the {\it magnitude} of $\chi_0$ also changes rapidly near $\mu=0$, 
so there should be an {\it independent} QCP from a magnetic to a non-magnetic 
phase near the same doping, as discussed in the previous subsection (note
the line depicting $1/U(\mu =0)$ in Fig.~\ref{fig:42}).

\begin{figure}
\leavevmode
   \epsfxsize=0.33\textwidth\epsfbox{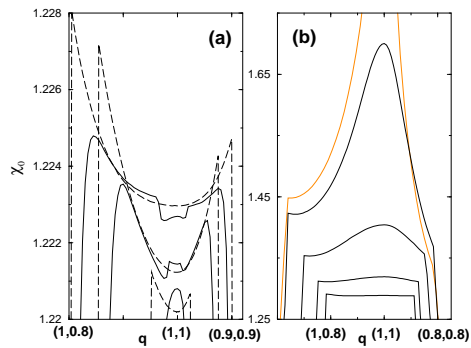}
\vskip0.5cm
\caption{(a): Expanded view of susceptibility $\chi_0$ on the plateaus near 
$\vec Q$ for a variety of dopings at $T=100K$ (solid curves) or 1K (dashed 
curves).  From highest to lowest curves near $\vec Q$, the chemical potentials 
are $\mu$ = -0.20, -0.15, and -0.05 eV (for both solid and dashed curves). 
All curves except $\mu=-0.20 eV$ have been shifted vertically to fit within the 
expanded frame. (b): Similar plateaus for the hole doped materials ($T=1K$),
with (from highest to lowest) $\mu$ = -0.359, -0.35, -0.3, -0.25, and -0.22 eV.}
\label{fig:44}
\end{figure}

Technically, similar cusps also arise at the plateau edges 
for electron doping, $0>\mu >-0.22eV$. The tops of the plateaus are not 
completely flat, Fig.~\ref{fig:44}a and the highest susceptibility is
shifted away from $\vec Q$ (Appendix D1).  However, these effects are
much weaker than those associated with $\mu >0$ 
($\Delta\chi /\chi\le 0.5\%$ -- compare the vertical scales of 
Figs.~\ref{fig:42},~\ref{fig:44}).  Thus near the mean-field transition any
structure on the plateaus is smeared out by thermal broadening.  Even at $T=0$, 
these features are likely to be negligible compared to dispersion in $U$ which
arises from renormalization effects\cite{AGGAH}.

\subsubsection{Curvature (A)}

\begin{figure}
\leavevmode
   \epsfxsize=0.33\textwidth\epsfbox{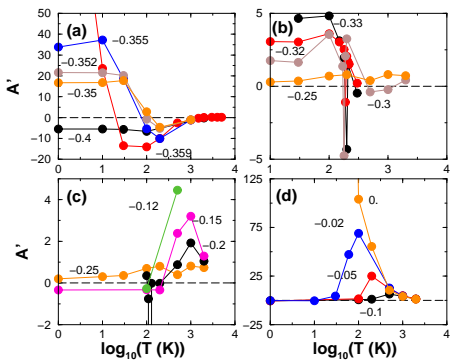}
\vskip0.5cm
\caption{Temperature dependence of $A'$ for several dopings. }
\label{fig:45}
\end{figure}

The plateau is a region of anomalously small local curvature $\hat A =A/U$ 
(Eq.~\ref{eq:0B18}) of the susceptibility, $\chi_0(\vec Q+\vec q)=\chi_Q-\hat 
Aq^2$, where $A$ is an important NAFL parameter. Clearly, at $T=100K$ the 
curvature $A$ has gone negative near the $H$-point, Fig.~\ref{fig:41}.  At
even lower temperatures, it reverts to positive values, Fig.~\ref{fig:44}b.  
The temperature dependence of the normalized parameter $A'=(\pi /a)^2(A/t)$ 
is illustrated in Fig.~\ref{fig:45} at several dopings.  The temperature 
dependence is dominated by divergences at both H- and C-points.  The divergence 
at the H-point Fig.~\ref{fig:45}a is the well-known logarithmic VHS.  However, 
at finite temperatures spectral weight is shifted away from the VHS and $A$ 
turns negative, only recovering a positive sign above $T\simeq 2000K$. The
temperature at which $A$ turns negative can be defined as $T_{incomm}$:
$A<0$ for $T>T_{incomm}$.  From Fig.~\ref{fig:0a}d, $T_{incomm}$ is
comparable to but larger than $T_V$ (for $x\le 0.06$ $A$ remains
positive).  This in fact explains the origin of $T_{incomm}$.
Figure~\ref{fig:45}a demonstrates that $A$ is negative at $T\rightarrow 0$
beyond the H-point ($\mu =-0.4eV$).  Thus, increasing $T$ above $T_V$
produces the same susceptibility crossover.  A similar crossover was
discussed by Sachdev, et al.\cite{SCS}, except that they assumed that in
the high temperature phase the AFM fluctuations remained centered on the
commensurate $\vec Q$, whereas here $A$ is negative. At sufficiently high
temperatures $A$ again becomes positive for all dopings -- i.e., the
leading singularity of $\chi_0$ is always at $\vec Q$.  

At the C-point, the collapse of the plateau width translates into a
divergence of the curvature at $\vec Q$ ($\hat A\rightarrow\infty$).  
This divergence of the high-temperature susceptibility is cut off at low $T$, 
Fig.~\ref{fig:45}d, when the thermal smearing becomes smaller than the
plateau width.  For smller $T$, $A$ is controlled by the curvature on the
plateau.  The temperature at which $A$ has a peak, defined as $T^*_A$, is
plotted as diamonds in Fig,~\ref{fig:41a} (the peak is only found for
$x\le 0$).  Rather surprisingly, $T^*_A$ scales with the plateau width $q_c
$, even though the dynamic exponent is $z=2$.  Further, the maximum slope scales
approximately as $A_{max}\sim T_A^{*-1.5}$, which follows from the fact that $A
\sim T^{-1.5}$ at the C-point.  

At intermediate doping, Fig.~\ref{fig:45}b,c, $A$ is generally a
scaled-down version of the behavior near the two end points, with a
crossover near $\mu =-0.25eV$, where the T-dependence is weak.  Also for
intermediate temperatures, there can be fine structure on the plateau
(e.g., solid lines in Fig.~\ref{fig:44}a) which can lead to wild swings in
$A(T)$.  However, at these dopings they are not relevant, since the
susceptibility peaks are away from $\vec Q$, and this fine structure 
is not generally reported in Fig.~\ref{fig:45}.

\subsection{Parameter Evaluation for Mode Coupling Theory}

\begin{figure}
\leavevmode
   \epsfxsize=0.33\textwidth\epsfbox{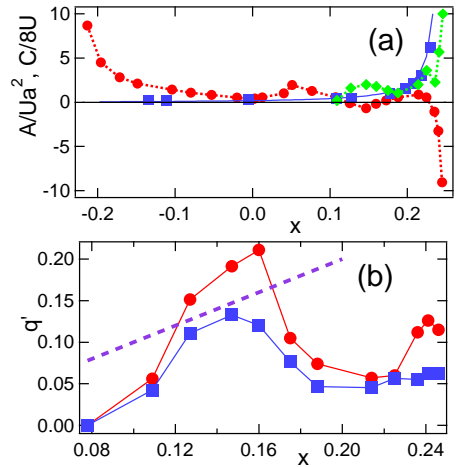}
\vskip0.5cm
\caption{(a) Calculated values of $A$ (circles for commensurate $\vec Q$, 
diamonds for incommensurate $\vec q=\vec Q+\vec q{\ }'$) and $C$
(squares).  Solid line = Eq.~\protect\ref{eq:C8}. (b)
Incommensurate wavevector $q'$ in two different directions: circles along 
$(1,0)$, squares along $(1,1)$; dashed line: $q'\propto x$.}
\label{fig:10} 
\end{figure}

\begin{figure}
\leavevmode
   \epsfxsize=0.33\textwidth\epsfbox{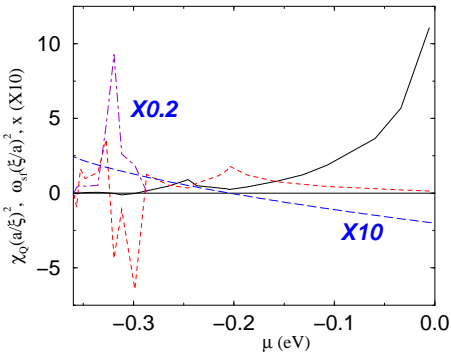}
\vskip0.5cm
\caption{Calculated values of $\chi_{\vec Q}/\xi^2$ (solid line) and $\omega_
{sf}\xi^2$ (short dashed line), assuming $U=6t$.  Long dashed line = doping 
$x(\mu )$ ($\times 10$); dot-dashed line = $\omega_{sf}\xi^2$ corrected
for incommensurate $\vec q\ne\vec Q$ ($\times 1/5$).}
\label{fig:9}
\end{figure}

The evaluation of the SCR parameters $A$ and $C$ was discussed above and
in Appendix D. The
collapse of the $\vec q$ and/or $\omega$ plateau widths near the H- and
C-points leads to the introduction of additional parameters $q_c$ and
$\alpha_{\omega}$.  The narrow width of the $\vec q$-plateau, particularly
for electron doping, leads to an additional complication not included in 
the conventional SCR analysis: the curvature of the bare susceptibility near $
\vec Q=(\pi ,\pi )$ (the $S$-point of the BZ) is strongly temperature dependent,
and for some dopings may even change sign.  
In principle, it is not difficult to incorporate an $A(T)$ into 
the analysis near the mean-field N\'eel temperature $T_N^*$ (pseudogap onset).  
But for the present 2D system, long range N\'eel order only sets in at $T_N=0$, 
and for $T<<T_N^*$, a self consistent value of $A$ should be found, by taking 
into account the effect of the pseudogap in modifying the electronic dispersion 
and hence $\chi$.  For the present, this complication is ignored, and in the 
following section $A$ is taken as $A=A(T_N^*)$, where $T_N^*$ is the
magnetic pseudogap onset, the temperature where $\chi_0(\vec Q)U_{eff}=1$, 
using the effective $U_{eff}$ found earlier\cite{KLBM} (Appendix B).  This 
should be the most important $A$ for controlling the pseudogap, and
moreover at lower temperatures the band renormalization should strongly
modify $A(T)$.  With this choice, the resulting $A(\mu )$ is plotted in
Fig.~\ref{fig:10}a, along with the $C$ parameter, evaluated at $T=0$. For
electron doping, this choice of $A$ is always positive and varies smoothly
with doping, diverging at the C-point.  By contrast, for hole doping $A$
is often negative, again illustrating the instability of the uniform AFM
phase.  Given $A$ and $C$, Fig.~\ref{fig:9} shows the calculated values of
$\chi_{\vec Q}$ and $\omega_{sf}$, normalized to $\xi^2$. 

For hole doping, the incommensurability creates difficulties in defining
the SCR model.  The mean field transition temperature at the
incommensurate vector $q$ is only marginally higher than that at $Q$,
Fig.~\ref{fig:0d}, suggesting that incommensurability should have only a
small efect on the phase diagram.  Thus, one might attempt to define a
positive $A$ by maasuring the curvature from an incommensurate nesting
vector.  However, this $A$ is highly anomalous, for a number of reasons.
First, the incommensurate $\vec q{\ }'$ ($\vec q=\vec Q+\vec q{\ }'$)
forms roughly a square around $\vec Q$, insert in Fig.~\ref{fig:41a}, with
the peak susceptibility generally along the $(\pi,0)$ axis (circles in
Fig.~\ref{fig:10}b).  In this case, by symmetry there are four peaks in
the susceptibility, at $(\pi\pm q',\pi )$ and at $(\pi ,\pi\pm q')$.
Moreover, the curvature measured from any incommensurate peak is highly
anisotropic, since the susceptibility is nearly constant along the ridge
of the square, with a shallow minimum at $(\pi ,\pi )$.  Thus parallel to
the ridge, $A_{\parallel}$ is nearly zero.  Moreover, perpendicular to the
ridge, $A_{\perp}$ takes on very different values on the sides of the
ridge  displaced toward or away from $(\pi ,\pi )$.  The curvature is
small, with significant deviations from quadratic (weaker curvature)
moving toward $(\pi ,\pi )$, while moving away from $(\pi ,\pi )$, the
curvature is larger, and deviating toward stronger curvature as the
susceptibility falls off the edge of the plateau.  For reference purposes,
the average value of the quadratic part of $A_{\perp}$ is plotted as
diamonds in Fig.~\ref{fig:10}a.  (In this case, $\omega_{sf}$ has a peak
near the VHS, Fig.~\ref{fig:9}.)  This definition of $A$ is almost
certainly an overestimate.  In the analysis of hole-doped cuprates in
Sections VI 
and VII, the commensurate SCR model will be applied, with $A$ as a free
parameter.  It will be found that agreement with measurement requires a 
somewhat smaller value for $A$ than the estimated value of $A_{\perp}$.
This small $A$ value, combined with the broad plateau, lead to a sum-rule
saturation for $\chi_0$ and a much slower divergence of $\xi (T)$ than
found for electron doped cuprates.

In the following section, the present results are applied to understanding
the ARPES
spectra of electron doped cuprates, concentrating on the four dopings analyzed
by Armitage, et al.\cite{nparm}.  For convenience, Table I summarizes the
parameters for these dopings.  From the mean-field analyses\cite{KLBM}, the 
effective Hubbard parameters were found to be $U_{eff}/t$ = 6 ($x=0$), 5 ($x=-0
.04$), 3 ($x=-0.10$), and 2.5 ($x=-0.15$).  [These numbers differ somewhat
from those of Ref.~\onlinecite{KLBM}, which included a second neighbor
hopping, $t''$, to give the best fit of the
Fermi surfaces.]  The Stoner factor has a quantum correction, $\eta$,
Eq.~\ref{eq:B35}, which tends to suppress the AFM transition; 
hence a smaller renormalization of $U$ is required.  This is reflected in Table 
1: for $x$ = -0.1, -0.15, there are two rows, the upper row using the 
mean-field $U$ parameters, the lower with the quantum correction.  Note
that the $U$'s are enhanced by essentially the quantum correction factor. 
These values will be used in the subsequent analysis.

The SCR analysis also involves a mode coupling parameter $u$.  An 
attempt to directly calculate $u$ (Appendix D4) failed, giving anomalously
small values (Table I) due to the flatness of the susceptibility plateau
($\partial\chi /\partial\omega\sim 0$), Fig.~\ref{fig:44b}d.  This problem
has been noted previously, although there is debate about whether $u$
diverges\cite{Chu} or vanishes\cite{Her,Mil}.  Here $\rho_s$ is estimated
from the measured correlation length for $x=0$, using Eq.~\ref{eq:16}, as
discussed in Section VII.  Since $\rho_s\propto u^{-1}$, Eq.~\ref{eq:B40}, 
this gives $u^{-1}=0.384eV$, which is assumed for all dopings.  The
calculated values of $\rho_s$ are illustrated in
Fig.~\ref{fig:11a1}b, based on Eqs.~\ref{eq:B34b},~\ref{eq:B40}.  
\vskip 0.1in
\begin{tabular}{||c|c|c|c|c|c|c|c|c||}        
\multicolumn{9}{c}{{\bf Table I: Electron Doped Cuprates}}\\
            \hline\hline
x& $U/t$&$A/a^2$&$\omega_1$(eV)&$\alpha_{\omega}$&$q_ca$&$\eta$&$T^*_A(K)$&$u^
{-1}$ (eV) \\
    \hline\hline
0&6&0.696&0.345&0.583&0.635&1.20&1020&760         \\     \hline
-0.04&5&1.16&0.540&0.455&0.518&1.17&850&3200    \\        \hline
-0.10&3&1.34&1.32&0.176&0.342&1.15&500&2700      \\     \hline
''&3.5&1.56&1.13&0.206&''&1.13&''&2300       \\     \hline
-0.15&2.5&1.75&2.16&0.054&0.172&1.09&56&4000      \\     \hline
''&2.9&2.03&1.86&0.062&''&1.05&''&3500\\     \hline
\end{tabular}
\vskip 0.1in
It is convenient to compare the present results with parameters estimated for
the SCR model\cite{MTU} from experimental data for (optimally) hole-doped
cuprates.  The parameters are defined as $T_0=Aq_B^2/2\pi C$, $T_A=Aq_B^2/2\chi_
0$, $y_0=\delta_0(T=0)/Aq_B^2$, and $y_1\simeq 12a^2u/\pi^3AC$.  The results
are listed in Table II, where the first line gives the hole-doped results
estimated in Ref.~\onlinecite{MTU}.  Moriya, et al.\cite{MTU} took $q_B^2=1/4\pi
a^2$ ($q_Ba=0.282$), while for Table II it is assumed that $q_B=q_c$.  A
key difference is that Moriya, et al.\cite{MTU} assume the system is in
the paramagnetic phase ($y_0>0$) at and above optimal (hole) doping, while
in the present work $y_0<0$, and the system is paramagnetic due to the
Mermin-Wagner theorem, with the Mott gap appearing as a pseudogap.  The
small magnitude of $y_0$ is suggestive of a system pinned close to a QCP.
Finally, the parameter $y_1$ is estimated using the value $u^{-1}
=0.384eV$ (above), and not the anomalous values of Table 1.
\vskip 0.1in
\begin{tabular}{||c|c|c|c|c||}        
\multicolumn{5}{c}{{\bf Table II: SCR Parameters}}\\
            \hline\hline
x& $T_0$ (K)&$T_A$ (K)&$y_0$&$y_1$ \\
    \hline\hline
$\sim0.2$&1600-4000&3000-10000&0.01-0.02&3         \\     \hline
0.0&180&1150&-5.27&0.75            \\      \hline
-0.04&310&1300&-3.31&0.7        \\        \hline
-0.10&380&670&-1.23&1.5         \\     \hline
-0.15&200&220&-0.31&1.85          \\     \hline
\end{tabular}

\section{ARPES Spectra}

\subsection {SCR Transition and Correlation Length}

\begin{figure}
\leavevmode
   \epsfxsize=0.33\textwidth\epsfbox{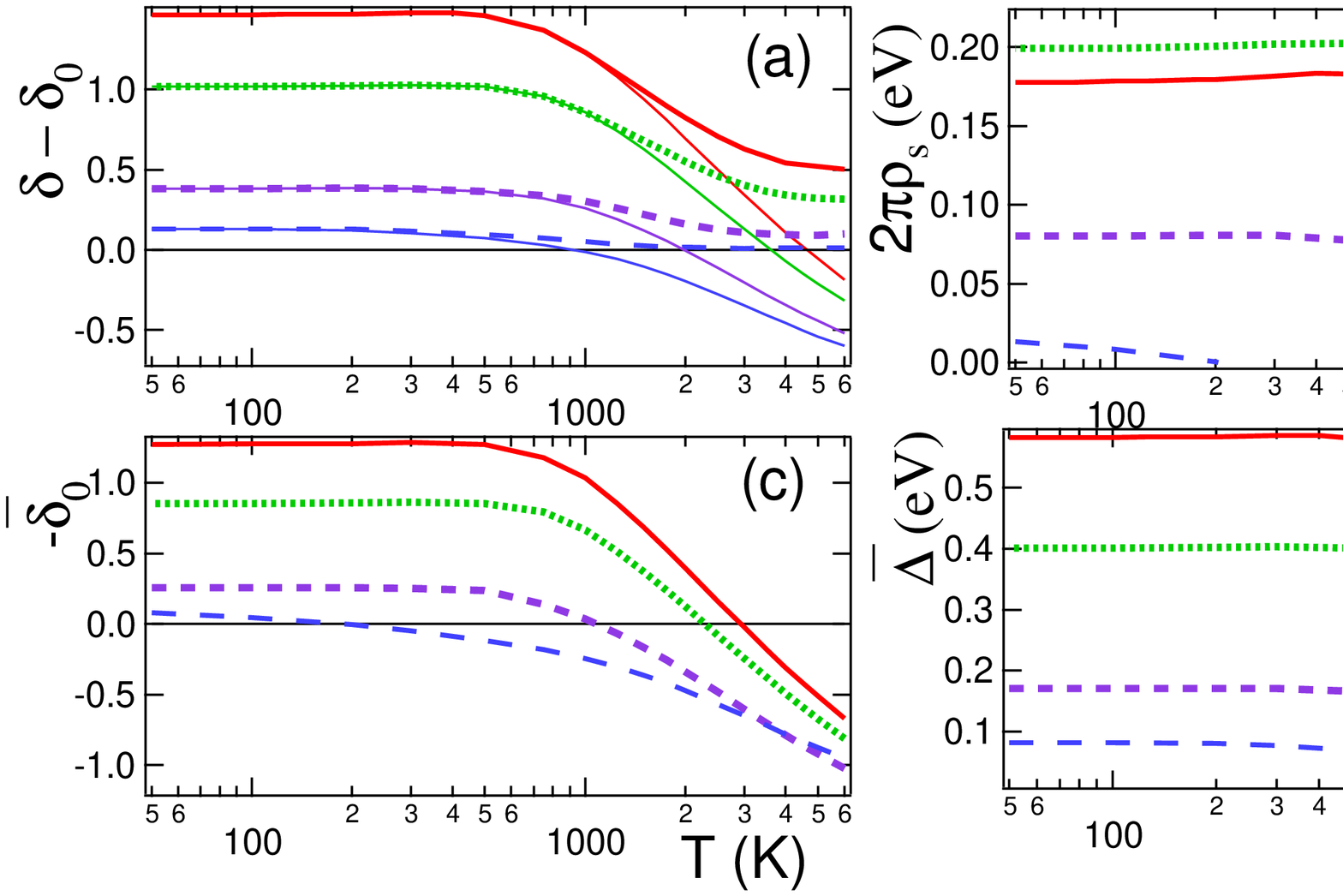}
\vskip0.5cm
\caption{(a) $\delta -\delta_0$ (thin solid lines = $-\delta_0$); (b) $\rho_s$ 
calculated from Eqs.~\protect\ref{eq:B40},~\protect\ref{eq:B34f}; (c) $-\bar
\delta_0$; (d) $\bar\Delta$, Eq.~\protect\ref{eq:20a}.
In all the plots, the solid curves correspond to $x=0.0$, dotted lines:
$x=-0.04$, short dashed lines: $x=-0.10$, long dashed lines: $x=-0.15$.}
\label{fig:11a1}
\end{figure}

\begin{figure}
\leavevmode
   \epsfxsize=0.33\textwidth\epsfbox{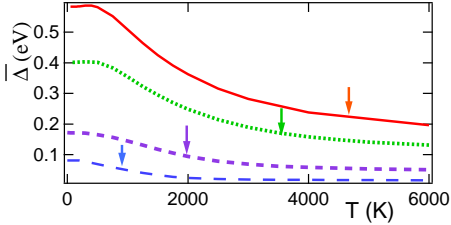}
\vskip0.5cm
\caption{Temperature dependence of gap $\bar\Delta$ for (from highest 
to lowest) $x=0$, -0.04, -0.10, and -0.15.  Arrows show mean field
transition temperature $T_N$.}
\label{fig:7da}
\end{figure}

Given the above parameters, the doping dependence of the MF and SCR
transitions is compared in Fig.~\ref{fig:11a1} for the four electron 
dopings studied in Refs.~\onlinecite{nparm},~\onlinecite{KLBM}.  The MF
transition occurs when the bare Stoner factor $\delta_0=1-\chi_{\vec Q0}U$
becomes negative, Fig.~\ref{fig:11a1}a.  However, in SCR the renormalized
Stoner factor $\delta$ stays positive, so there is no $T>0$ phase
transition (Mermin-Wagner theorem), although $\delta -\delta_0$ has a
strong increase near the temperature where $\delta_0$ changes sign.  There
is still a zero-T N\'eel transition, controlled by the quantum corrected
Stoner factor, $\bar\delta_0=\eta -\chi_{\vec Q0}U$. 
From Fig.~\ref{fig:11a1}c, it can be seen that at $x=-0.15$, the system is
close to a QCP, $\bar\delta_0(T=0)\rightarrow 0$.  This QCP is controlled
by the Stoner criterion of the zero-T antiferromagnet.  While there is no
long range order, there is still a Mott (pseudo)gap, controlled by {\it
short-range} order, Fig.~\ref{fig:11a1}d.  A direct comparison of the
transition temperatures is presented on a linear T scale in Fig.~\ref{fig:7da}.
The spin stiffness $\rho_s$ (Fig.~\ref{fig:11a1}b) is found to be nearly
$T$-independent below the pseudogap onset.

While the value of $U$ has been adjusted to fit the ARPES spectra, it is
important to note that good agreement has also now been found with
magnetic properties.  This is discussed in Section VII.

\subsection{General Results}

\begin{figure}
\leavevmode
   \epsfxsize=0.33\textwidth\epsfbox{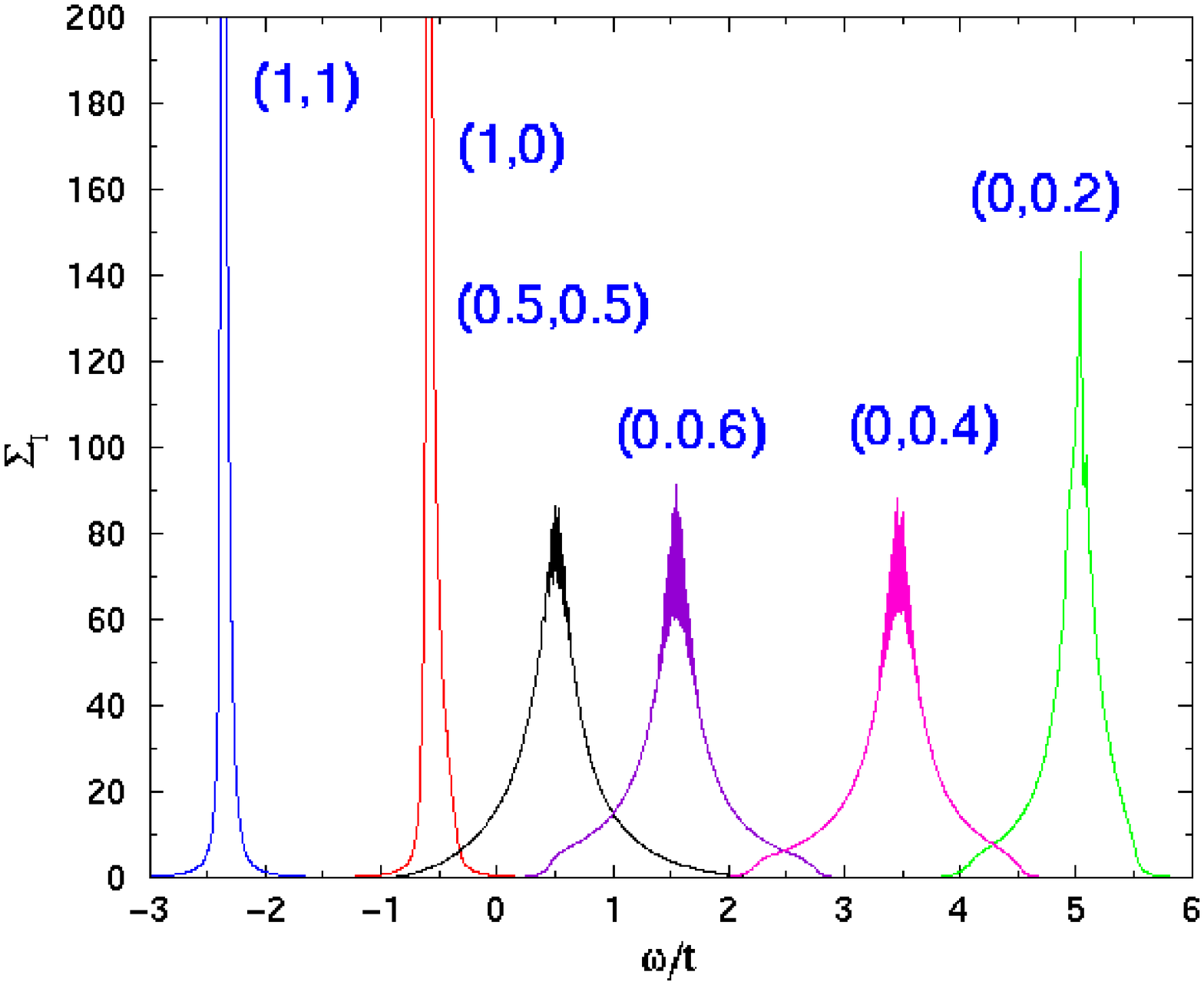}
\vskip0.5cm
\caption{Imaginary part of the self energy, Eq.~\protect\ref{eq:18}, assuming
$1/C=0.05t$, $\delta =0.002$, $\alpha_\omega =1$, $T=100K$.  The branches are 
labelled $(k_x,k_y)$, in units of $\pi$.}
\label{fig:6}
\end{figure}

Given the susceptibility, Eq.~\ref{eq:15}, the self energy can be calculated 
approximately as
\begin{eqnarray}
\Sigma (\vec k,i\omega_n)={g^2\chi_0\over\beta V}\sum_{\vec q, i\omega_m}G_0(
\vec k+\vec q,i\omega_n+i\omega_m)D_0(\vec q,i\omega_m)
\nonumber \\
={g^2\chi_0\over V}\sum_{\vec q}\int_{-\alpha_{\omega}/C}^{\alpha_{\omega}/C}
{d\epsilon\over\pi}{n(\epsilon )+f(\xi_{\vec k+\vec q})\over i\omega_n+
\epsilon -\xi_{\vec k+\vec q}}{C\epsilon\over (\delta +Aq^{'2})^2+(C\epsilon 
)^2},
\label{eq:17}
\end{eqnarray}
with bare Green's function $G_0(\vec k,i\omega_n)=1/(i\omega_n-\xi_{\vec k})$, 
$\xi_{\vec k}=\epsilon_{\vec k}-\mu$, and magnetic propagator $D_0$,
Eq.~\ref{eq:B25}; for the form of the integral, see the discussion near
Eq.~\ref{eq:B32}.  In addition, $\chi_0=\chi_0(\vec Q,0)$, $\vec q=\vec
Q+\vec q'$, $n$ is the Bose function, and
\begin{equation}
g^2\chi_0=U^2\chi_0(U\chi_0(\vec Q,i\omega_n)+{1\over 1+U\chi_0(\vec Q,i\omega_n
)})\simeq{3U\over 2}
\label{eq:17b}
\end{equation}
(Ref.~\onlinecite{BSW2}). The last form is an approximation based on the 
empirical substitution $\chi_0\rightarrow\simeq 1/U$ in the pseudogap 
regime.  [An improved approximation for $\Sigma$, ($G_0\rightarrow G$ in
Eq.~\ref{eq:17}) is discussed in Section V.]  After analytical 
continuation, the imaginary part of the retarded self energy is
\begin{eqnarray}
Im\Sigma^R(\vec k,\omega )
={-g^2\chi_0\over V}\sum_{\vec q}\int_{-\alpha_{\omega}/C}^{\alpha_{\omega}/C}
d\epsilon [n(\epsilon )+f(\xi_{\vec k+\vec q})]\times
\nonumber \\
\times\delta(\omega +\epsilon -\xi_{\vec k+\vec q}){C\epsilon 
\over (\delta +Aq^{'2})^2+(C\epsilon )^2}.
\label{eq:18}
\end{eqnarray}
The resulting self energy is plotted in Fig.~\ref{fig:6} for $T=100K$.  
(The weak oscillations seen in some branches of $\Sigma_I$ are an artifact due
to an insufficient density of points in the numerical integration.)  Note
that $Im\Sigma$ has the form of a broadened $\delta$-function peaked at $\omega
=\xi_{\vec k+\vec Q}$.  If it were a $\delta$-function, $Im\Sigma =-\pi\bar
\Delta^2\delta (\omega -\xi_{\vec k+\vec Q})$, then
\begin{equation}
Re\Sigma^R(\vec k,\omega )
={1\over\pi}\int_{-\infty}^{\infty}d\epsilon {Im\Sigma^R(\vec k,\epsilon )\over
\epsilon -\omega }={\bar\Delta^2\over\omega -\xi_{\vec k+\vec Q}},
\label{eq:19}
\end{equation}
so away from the $\delta$-function
\begin{equation}                           
G(\vec k,\omega )={1\over \omega -\xi_{\vec k}-Re\Sigma^R(\vec k,\omega )}=
{\omega -\xi_{\vec k+\vec Q}\over (\omega -\xi_{\vec k})(\omega -\xi_{\vec k+
\vec Q})-\bar\Delta^2}.
\label{eq:20}
\end{equation}
This is exactly the Green's function of the mean field 
calculation\cite{SWZ,MK3}, with the substitution
$\Delta\rightarrow\bar\Delta$, where $\bar\Delta$ can be evaluated by integrating 
\begin{eqnarray}
\bar\Delta^2=-{1\over\pi}\int_{-\infty}^{\infty}d\omega Im\Sigma^R(\vec k,\omega
 )
\nonumber \\
={U\over 8u}(\delta -\delta_0),
\label{eq:20a}
\end{eqnarray}
Fig.~\ref{fig:11a1}d.  This result is due to the Bose term $n(\epsilon )$
in the square bracket of Eq.~\ref{eq:18}, the Fermi function $f$ making no
contribution.  This leads to $\bar\Delta$ being independent of $\vec k$.

Equations~\ref{eq:20},\ref{eq:20a} constitute an important result, the 
connection between the Mott gap and short-range magnetic
order\cite{ChuM1,SPS}. Recalling
that $\Delta = U<M_i>$, or $\Delta^2=U^2<S_i>^2$, where $<M_i>=(-1)^i<S_i>$ 
is the staggered magnetization, then, in the spirit of an alloy analogy,
a {\it short-range order parameter} can be defined as
\begin{eqnarray}
\bar\Delta_{SR}^2(i\omega )={-g^2\over 4\beta}\int_0^{\beta}\sum_{<i,j>}<S_{i+}
(\tau )S_{j-}(0)>e^{i\omega\tau}d\tau
\nonumber \\
={-g^2\over 4\beta}\sum_k(c_x+c_y)\chi_{+-}(k,i\omega )
\simeq {g^2\over 2\beta}\sum_k\chi_{+-}(k,0)
\label{eq:20b}
\end{eqnarray}
which is equivalent to Eq.~\ref{eq:20a}.  (In the last equality in 
Eq.~\ref{eq:20b} the limit $i\omega\rightarrow 0$ is an adiabatic 
approximation\cite{Mor}, while the approximation is made that $\chi$ peaks near
$\vec Q$.)  Thus, {\it as long as there is short-range magnetic order ($\bar
\Delta$ or $\rho_s$ non-zero), there will be a Mott (pseudo)gap.}  

\subsection{Application to the Cuprates}

\begin{figure}
\leavevmode
   \epsfxsize=0.33\textwidth\epsfbox{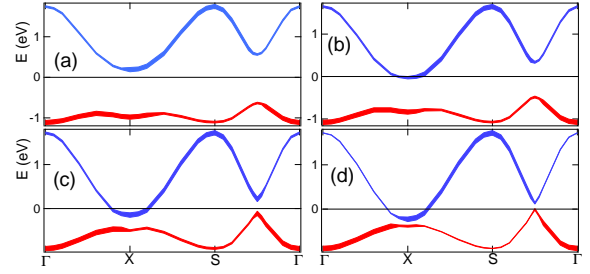}
\vskip0.5cm
\caption{SCR Dispersion relations for electron doped materials, calculated
at $T=100 K$: (a) $x=0$ ($U/t=6$), (b) $x=-0.04$ ($U/t=5$), (c) $x=-0.10$
($U/t=3.5$), and (d) $x=-0.15$ ($U/t=2.9$).  Linewidth indicates relative
intensity; for $x=-0.15$ all shadow features are extremely weak.}
\label{fig:7e}
\end{figure}

\begin{figure}
\leavevmode
   \epsfxsize=0.33\textwidth\epsfbox{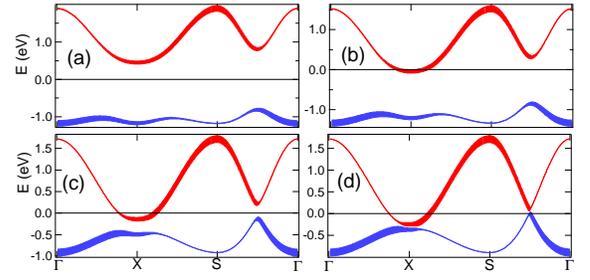}
\vskip0.5cm
\caption{Mean field dispersion relations for electron doped materials,
calculated at $T=1 K$: (a) $x=0$ ($U/t=6$), (b) $x=-0.04$ ($U/t=5$), (c)
$x=-0.10$ ($U/t=3$), and (d) $x=-0.15$ ($U/t=2.6$).  Linewidth indicates
relative intensity.}
\label{fig:7em}
\end{figure}

Using the correct $Im\Sigma^R$ from Eq.~\ref{eq:18}, and the calculated 
parameter values from Table I, ARPES spectra are calculated for electron-doped 
cuprates, at the four dopings for which detailed data are 
available\cite{nparm}.  The resulting dispersions are shown in 
Fig.~\ref{fig:7e}.  There is a well defined pseudogap, with two peaks in
the spectral function at a given $\vec k$.  It should be stressed that since
there is no interlayer coupling, long range antiferromagnetic order exists
only at $T=0K$.  
The agreement with the mean field results\cite{KLBM}, Fig.~\ref{fig:7em}, 
and experiment\cite{nparm} is quite good, except that the SCR gap is smaller
at half filling.  This is due to lack of self-consistency: in
calculating the self-energy, a susceptibility based on the bare Green's
function was used, neglecting the opening of a gap near the Fermi level. 
In Section V it will be shown that when this is accounted for (via the
self-consistent Born approximation) a larger gap is found.  For
completeness, Fig.~\ref{fig:7en} shows the mean field dispersion in the
{\it three-band} model, discussed in Appendix A.  The overall agreement in
all cases is quite striking.

\begin{figure}
\leavevmode
   \epsfxsize=0.33\textwidth\epsfbox{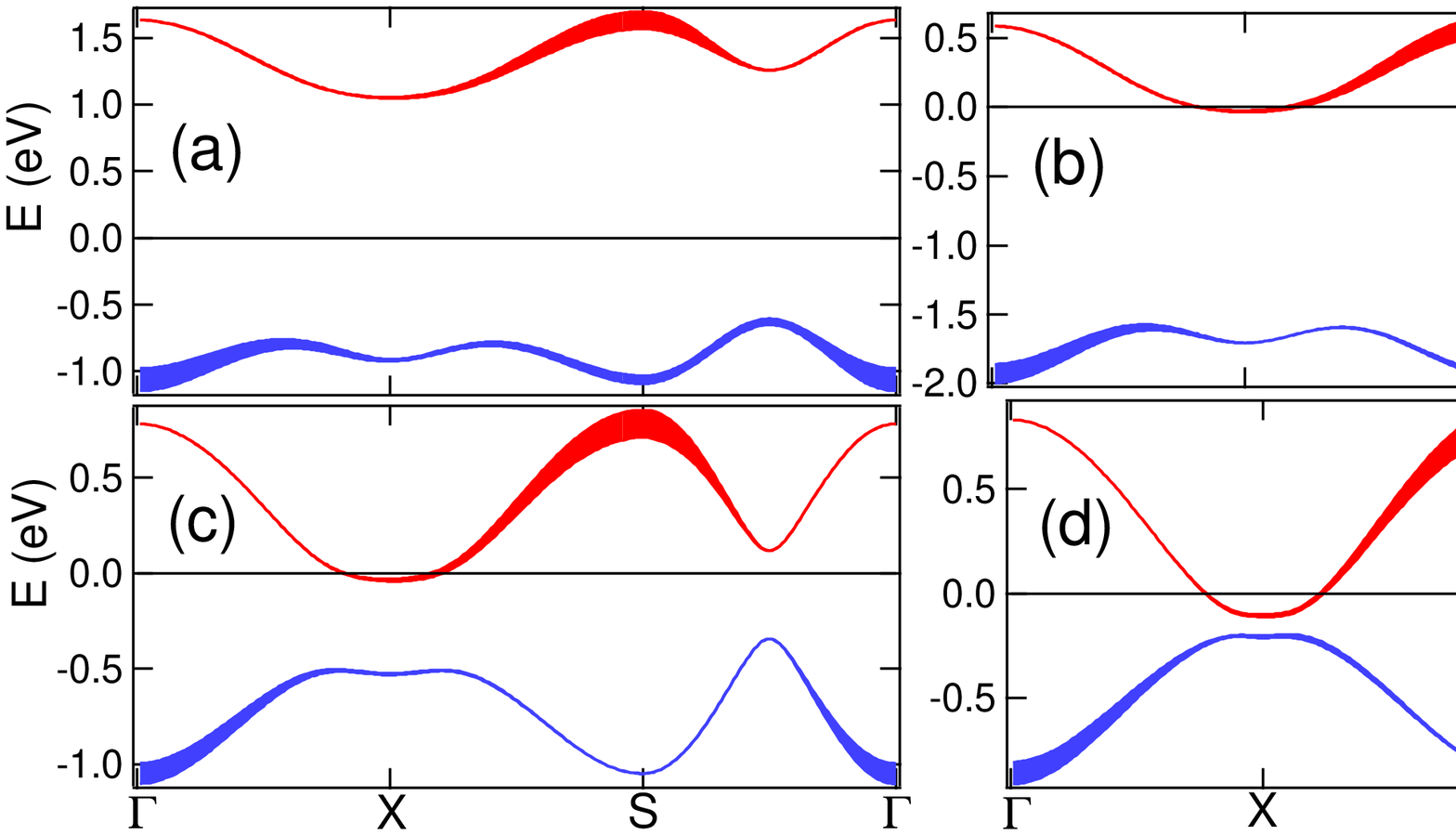}
\vskip0.5cm
\caption{Mean field dispersions in three-band model for electron doped
materials, showing the two antibonding bands, assuming 
$m_Q$ = 0.3 (a), 0.2 (b), 0.05 (c), and 0.01 (d). Other parameters are
discussed in Appendix A.}
\label{fig:7en}
\end{figure}

In an earlier calculation\cite{ICTP} a somewhat larger value of $u$ was
assumed, $u^{-1}=0.256eV$.  This leads to stronger quantum corrections:
the parameter $\eta -1$ (Table I) was about twice as large and the gaps in
Fig.~\ref{fig:7e} were smaller, particularly near half filling. 

\begin{figure}
\leavevmode
   \epsfxsize=0.33\textwidth\epsfbox{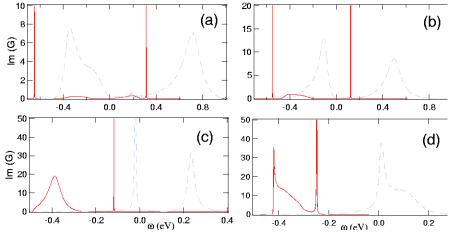}
\vskip0.5cm
\caption{Spectral functions for (a) $x=0$, (b) $x=-0.04$, (c) $x=-0.10$, and (d)
$x=-0.15$, at $T=100K$.  Solid lines at $(\pi ,0)$, 
and long dashed lines at $(\pi /2,\pi /2)$.}
\label{fig:7a1}
\end{figure}

Figure~\ref{fig:7a1} shows typical calculated spectra for several $\vec
k$-points in the a-b plane.  Broadened Hubbard bands are found, which
gradually smear out at high temperatures as $\delta$ increases ($\xi$
decreases). 

\begin{figure}
\leavevmode
   \epsfxsize=0.33\textwidth\epsfbox{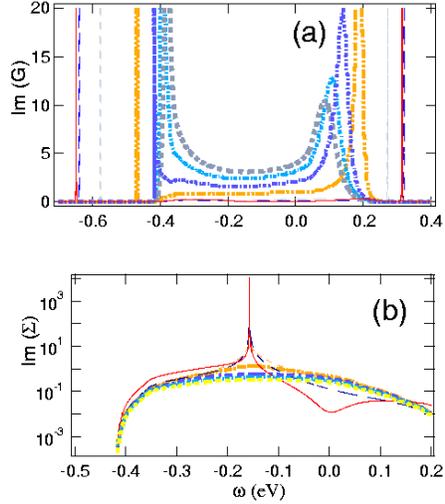}
\vskip0.5cm
\caption{Temperature dependence of (a) spectral function and (b) imaginary part
of self energy, for $x=0.0$ at $(\pi ,0)$.  Temperatures are 100, 500, 1000,
2000, 3000, 4000, and 5000K.}
\label{fig:7ee}
\end{figure}

\begin{figure}
\leavevmode
   \epsfxsize=0.33\textwidth\epsfbox{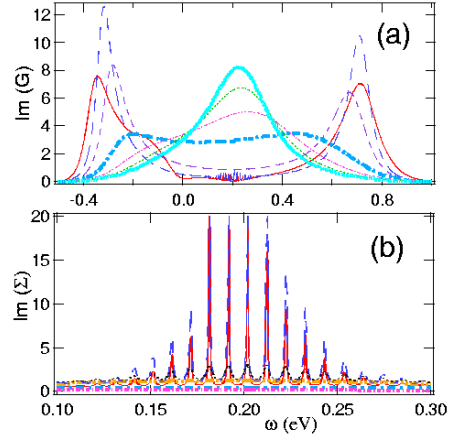}
\vskip0.5cm
\caption{Temperature dependence of (a) spectral function and (b) imaginary part
of self energy, for $x=0.0$ at $(\pi /2,\pi /2)$.  Temperatures are 100, 500, 
1000, 2000, 3000, 4000, and 5000K.}
\label{fig:7ef}
\end{figure}

\begin{figure}
\leavevmode
   \epsfxsize=0.33\textwidth\epsfbox{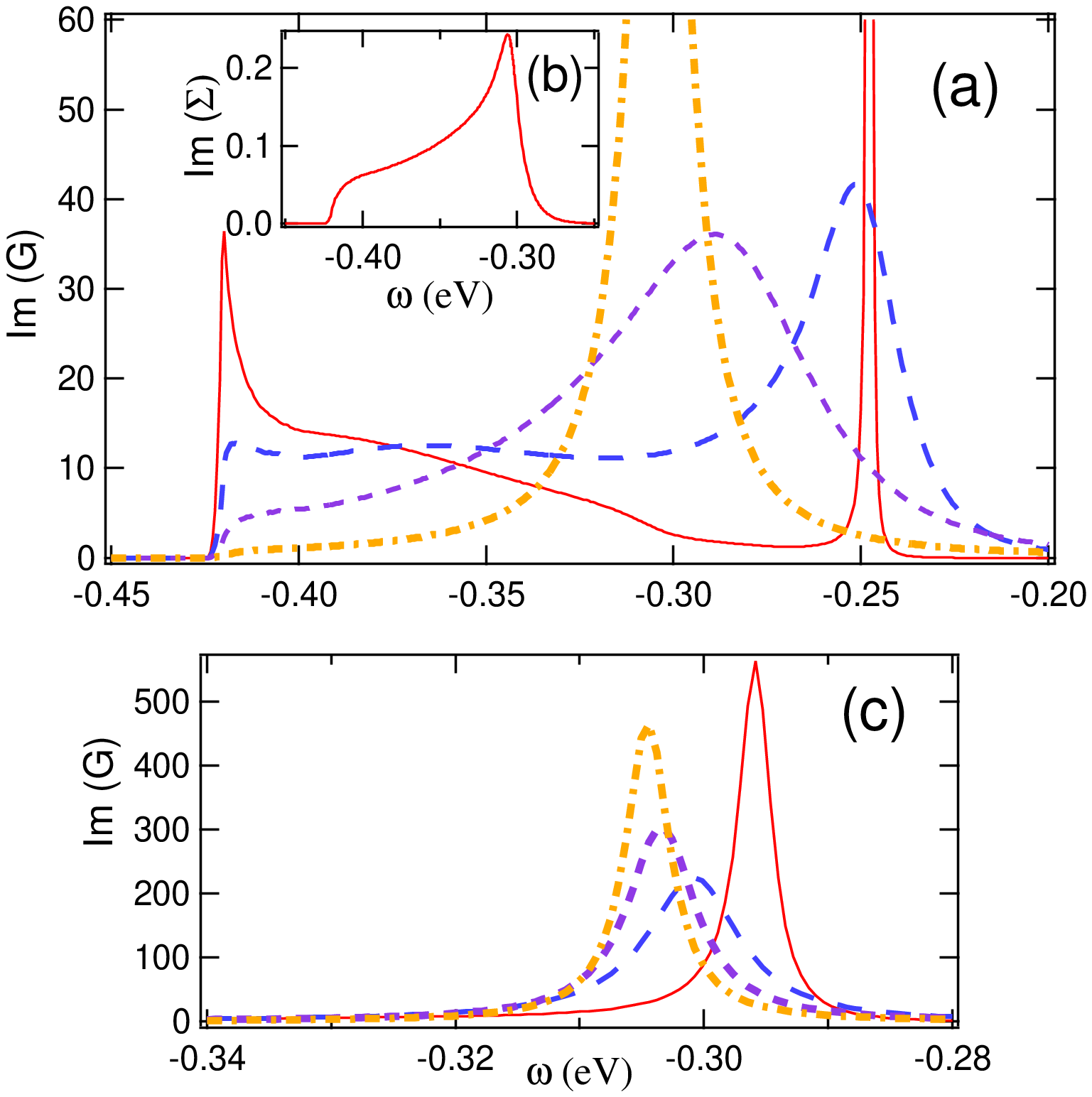}
\vskip0.5cm
\caption{Temperature dependence of spectral function 
for $x=-0.15$ at $(\pi ,0)$, for $U/t$ = 2.9 (a) and 2.5 (c).  Temperatures are 
100 (solid line), 500 (long-dashed line), 1000 (short-dashed line), and 2000K
dot-dashed line).  (b): imaginary part of self energy at $T$ = 100K,
$U/t$ = 2.9.}
\label{fig:7eg}
\end{figure}

Figures~\ref{fig:7ee}-~\ref{fig:7eg} illustrate the 
temperature dependence of $Im(G)$ and $Im(\Sigma)$ for two dopings, $x=0$ and 
-0.15.  The broadening of the peaks can be understood from 
Eq.~\ref{eq:18}: particle-hole excitations are present within a range
$\pm\alpha_{\omega}/C$ of $\xi_{\vec k+\vec q}$.  Away from this
particle-hole continuum the main peaks are sharp, while they broaden when
they enter the continuum.  

Note that the Mott gap collapse is anisotropic: for the undoped case, the nodal 
gap collapses between 2-3000K, while a gap persists near $(\pi ,0)$ above
5000K.  $Im(\Sigma )$ has striking oscillatory structure, particularly
near $(\pi /2,\pi /2)$, which produces a similar weak structure in $Im(G)$ at 
low T.  [Similar, weaker oscillations are present near $(\pi ,0)$, which can be
better seen in Fig. 4c of Ref.~\onlinecite{ICTP}.]  In addition, there is a very
intense, strongly T-dependent peak in $Im(\Sigma )$ exactly at $\xi_{\vec k+
\vec q}$ (Fig.~\ref{fig:7ee}b -- also present but not shown in 
Fig.~\ref{fig:7ef}b -- see Section IX.B).  It is the divergence of this
peak as $T\rightarrow 0$ which signals the AFM transition.  At low
temperatures, the peak positions in $Im(G)$ have a temperature dependence
consistent with the collapse of the Mott gap -- e.g., the LHB shifts to
higher energies (toward midgap) at higher temperatures.  Some experiments
on hole doped cuprates find the {\it opposite} dependence\cite{KiRo},
which can possibly be understood as a localization or phase separation
effect.

\begin{figure}
\leavevmode
   \epsfxsize=0.33\textwidth\epsfbox{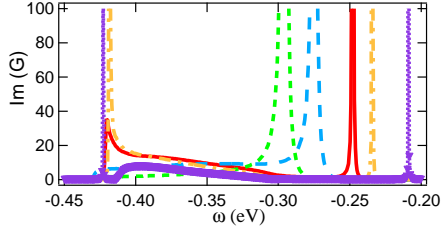}
\vskip0.5cm
\caption{$U$-dependence of spectral functions for $x=-0.15$ at $T=100K$ near the
$T=0$ QCP, for $U/t$ = 2.5 (short dashed line), 2.7 (long dashed line), 2.9
(solid line), 3.0 (dot-dashed line), and 3.2 (dotted line).}
\label{fig:7db}
\end{figure}

In contrast, for $x=-0.15$, Fig.~\ref{fig:7ef}, the splittings are absent near 
$(\pi /2,\pi /2)$, and vanish near $(\pi ,0)$ by $\sim$500K, and the lines 
actually {\it sharpen} on warming.  If the effective $U$ is reduced to $2.5t$, 
no splitting is found, but the peak position and broadening have an anomalous 
T dependence.  Clearly, the system is very close to a QCP.  Figure~\ref{fig:7db}
shows in more detail how the spectrum evolves with $U$ near this point.

\begin{figure}
\leavevmode
   \epsfxsize=0.33\textwidth\epsfbox{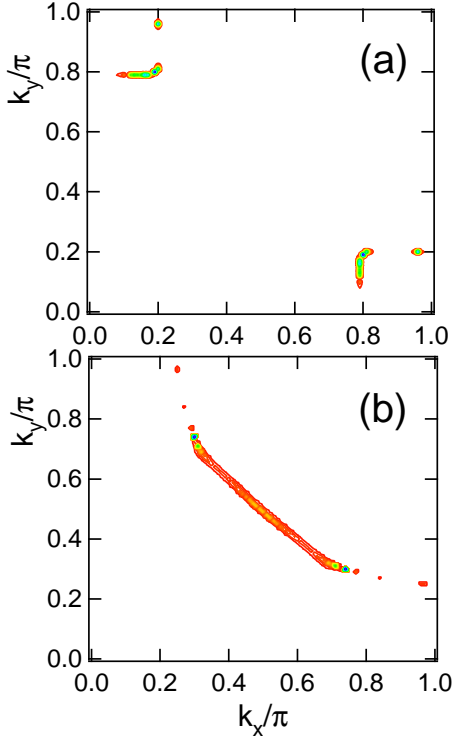}
\vskip0.5cm
\caption{Fermi surface map for $x=-0.10$ (a) and $-0.15$ (b).}
\label{fig:7g}
\end{figure}

Finally, Fig.~\ref{fig:7g} displays Fermi surface maps for $x=-0.10$ and 
$-0.15$, showing the crossover from small to large Fermi surface. Hot spot
effects are prominent at $x=-0.15$, pinning the Fermi surface to the zone
diagonal and broadening it at a pseudogap due to hot-spot 
scattering\cite{HlR}.  These should be compared with the mean-field\cite{KLBM} 
and experimental\cite{nparm} results.  It should be noted that in the mean
field calculation, it was necessary to include a $t''$ parameter to
reproduce the experimental hole pocket near the zone diagonal.  Such a
parameter would have shifted the Fermi surface across the zone diagonal,
leading to improved agreement with experiment here as well.

Thus, the SCR calculation agrees with the mean-field results\cite{KLBM}, if
the mean-field gaps and transition temperatures are interpreted as the opening
of a pseudogap at finite T, with the long-range AFM appearing only at T=0.  
Moreover, the overall dispersions, Fig.~\ref{fig:7e} are in quite good
agreement with the mean field results\cite{KLBM} and experiments\cite{nparm}.

\section{Self-Consistent Born Approximation}

A limitation of the above calculations is that the self energy Eq.~\ref{eq:17} 
is calculated using the bare susceptibility, whereas the full susceptibility
should be strongly modified by the opening of the Mott gap.  This can be
corrected for by including the full Green's function into the self energy
calculation.  This is conveniently done at $T=0$, since there is a
long-range ordered phase, and the sole difference between the SCR and
mean-field calculations is a weak renormalization of the Hubbard $U$.  In
this case, a renormalized Green's function can be found by summing all the
non-crossing diagrams; this is the self-consistent Born approximation
(SCBA).  By including the interaction of the quasiparticles with spin
waves, it also incorporates the physics of magnetic polarons.  Magnetic
polarons closely resemble lattice polarons, leading to both coherent and  
incoherent contributions to the spectral function, with considerable
bandwidth renormalization in the coherent spectrum.  In the $t-J$ model it
is known that the SCBA gives a good description of exact diagonalization
results on small lattices\cite{PRa}.  

Here, a simple calculation is presented to estimate the effect of the SCBA
corrections on the dispersion of the insulating phase.  Only the coherent band 
dispersion is included, and the SCBA is applied to the RPA 
solution\cite{VTwarn}, which should be similar to the ordered SCR phase at
$T=0$.  The calculation of Chubukov and Morr\cite{ChuM1} is extended to
include both lower and upper Hubbard bands.  
The RPA dispersions of the upper (c) and lower (v) Hubbard bands can be
written as follows.  If the bare dispersion is $\epsilon_k=-2t(cos(k_xa)
+cos(k_ya))-4t'cos(k_xa)cos(k_ya)-2t''(cos(2k_xa)+cos(2k_ya))$, then defining
$\epsilon_k^{(\pm )}=(\epsilon_k\pm\epsilon_{k+Q})/2$, $E^{(-)}_k=\sqrt{
\epsilon_k^{(-)2}+\Delta^2}$, $\Delta =U<S_z>$, then 
\begin{equation}
E^{c,v}_k=\epsilon_k^{(+)}\pm E^{(-)}_k.  
\label{eq:13z}
\end{equation}
Here $\Delta$ is the AFM gap, $\Delta\sim U/2$ at half filling.  For large
$\Delta$, this can be expanded as
\begin{eqnarray}
E^{c,v}_{\vec k}
=A_{00}+A_{01}\cos{k_xa}cos{k_ya}+
\nonumber\\
+A_{02}(\cos{2k_xa}+cos{2k_ya}),
\label{eq:6a}
\end{eqnarray}
with
\begin{equation}
A_{01}=J/2\pm t',
\label{eq:66a}
\end{equation}
\begin{equation}
A_{02}=J/2\pm 2t''.
\label{eq:66b}
\end{equation}
The  same dispersion is found in the $t-J$ model\cite{LiMan,MaHor},
suggesting that the SCBA will be an equally good approximation here.

The self-consistent equation (replacing Eq.~\ref{eq:17}) can be written
\begin{eqnarray}
G^{-1}(k,\omega )=\omega -(E^{c,v}_k-\mu )\nonumber\\
-\int{d^2q\over 4\pi^2}\Psi_{c,v} (k,q)G(k+q,\omega +\omega_q)
\label{eq:51}
\end{eqnarray}
where $\Psi_{c,v}$ is a vertex correction for the upper (c) or lower   
(v) Hubbard band and $\omega_q$ is the spin wave
dispersion.  As will be seen below (Eq.~\ref{eq:62}), $\Psi\propto t^2$,  
so Eq.~\ref{eq:51} is {\it independent of t}, depending only on ratios
$t'/t$, $t''/t$, and $J/t$.  However, the final dispersion also scales
with $t$, so any comparison with experiment requires all four parameters.
 For an arbitrary electronic dispersion, these quantities can
be evaluated as follows.  The transverse susceptibility in the RPA can be
written as\cite{SWZ,ChuF}
\begin{eqnarray}
\bar\chi^{+-}({\mathbf q},{\mathbf q},\omega )=\nonumber\\
{\chi^{+-}_0({\mathbf
q},\omega )[1-U\chi^{+-}_0({\mathbf q}+{\mathbf Q},\omega )]+U[\chi^{+-}_Q
({\mathbf q},\omega )]^2\over [1-U\chi^{+-}_0({\mathbf q},\omega )]
[1-U\chi^{+-}_0({\mathbf q}+{\mathbf Q},\omega ]-U^2[\chi^{+-}_Q
({\mathbf q},\omega )]^2},
\label{eq:52} 
\end{eqnarray}
\begin{eqnarray}
\bar\chi^{+-}({\mathbf q},{\mathbf q}+{\mathbf Q},\omega)=\nonumber\\     
{\chi^{+-}_Q
({\mathbf q},\omega )\over [1-U\chi^{+-}_0({\mathbf q},\omega )]
[1-U\chi^{+-}_0({\mathbf q}+{\mathbf Q},\omega ]-U^2[\chi^{+-}_Q
({\mathbf q},\omega )]^2},
\label{eq:53}
\end{eqnarray}
with
\begin{eqnarray}
\chi^{+-}_0({\mathbf q},\omega)=
{1\over 2N}\sum'_k\bigl[1-{\epsilon^{(-)}
_k\epsilon^{(-)}_{k+q}-\Delta^2\over E^{(-)}_kE^{(-)}_{k+q}}\bigr]\times
\nonumber\\
\times\bigl[
{1\over E_k+E_{k+q}-\omega }+{1\over E_k+E_{k+q}+\omega}\bigr],
\label{eq:54} 
\end{eqnarray}  
and
\begin{eqnarray}
\chi^{+-}_Q({\mathbf q},\omega)=
{1\over 2N}\sum'_k{\Delta (E^{(-)}_k+
E^{(-)}_{k+q})\over E^{(-)}_kE^{(-)}_{k+q}}\times
\nonumber\\  
\times\bigl[  
{1\over E_k+E_{k+q}-\omega }-{1\over E_k+E_{k+q}+\omega}\bigr].
\label{eq:55}   
\end{eqnarray}
In the large $U$ limit, Eq.~\ref{eq:52} becomes
\begin{equation}
\bar\chi^{+-}_0({\mathbf q},{\mathbf q},\omega)=\eta_q^2[{1\over\omega
+\omega_q}-{1\over\omega -\omega_q}],
\label{eq:56}   
\end{equation}
with
\begin{equation}
\eta_q^2={1\over 2}\sqrt{a_q-\gamma_q\over a_q+\gamma_q},
\label{eq:57} 
\end{equation}
\begin{equation}
\omega_q=2J\sqrt{a_q^2-\gamma_q^2},
\label{eq:58}
\end{equation}
\begin{equation}
\gamma_q={cos(q_xa)+cos(q_ya)\over 2},
\label{eq:59}
\end{equation}  
and
\begin{equation}
a_q=1+{J'\over J}(1-cos(q_xa)cos(q_ya))+{J''\over J}(1-\gamma_{2q}),
\label{eq:60} 
\end{equation}
with $J=4t^2/U$, $J'/J=(t'/t)^2$, $J''/J=(t''/t)^2$.  As befits a
Goldstone mode, $\omega_q=0$ at $q$ = $(0,0)$ and $(\pi ,\pi )$.  In this
case, $\Psi_{c,v}=\Phi_{c,v}^2$, with
\begin{equation}
\Phi_{c,v}=\bar\eta_q(\epsilon^{(-)}_k-\epsilon^{(-)}_{k+q})\pm\eta_q(\epsilon
^{(-)}_k+\epsilon^{(-)}_{k+q}),
\label{eq:61}
\end{equation}
with $\bar\eta_q=1/(2\eta_q)$, or
\begin{equation}
\Psi_{c,v}=16t^2\bigl[{a_q(\gamma^2_k+\gamma^2_{k+q})-2\gamma_k\gamma_{k+q}
\gamma_q\over\sqrt{a_q^2-\gamma_q^2}}\pm
(\gamma^2_k-\gamma^2_{k+q})\bigr].
\label{eq:62}
\end{equation}  

\begin{figure}
\epsfxsize=0.33\textwidth\epsfbox{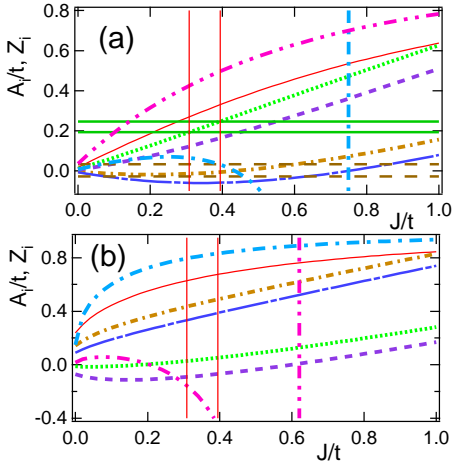}
\vskip0.5cm
\caption{Renormalized parameters for lower (a) and upper (b)
Hubbard bands: $Z$ (solid lines), $A_1$ (dot-dashed lines), and $A_2$   
(dotted lines), in comparison with $ZA_{01}$ (long-dashed-short-dashed 
lines), and $ZA_{02}$ (short-dashed lines), assuming parameters
$t=0.326eV$, $t'=-0.375t$, and $t''=0.15t$.  Horizontal lines =
experimental range for $A_1$ (long dashed lines) and $A_2$ (solid lines),
after Refs.~[\protect\onlinecite{LaR,Dur}]. Also shown are the individual
renormalization factors $Z_1$ (long-dash-dotted line) and $Z_2$
(long-dash-dot-dotted line).  Vertical lines delimit parameter values
consistent with experiment.}
\label{fig:67}
\end{figure}

Given $\omega_q$, $E^{c,v}_k$, and $\Psi_{c,v}$, Eq.~\ref{eq:51} can be
solved numerically to find both the coherent and incoherent parts of the
ARPES spectral weight.  However, the incoherent part contributes to a weak
background, and the experimental spectra are generally compared to the   
coherent part.  Hence, for present purposes what is needed is the
dispersion of the coherent part of $G$.  Following Chubukov and
Morr\cite{ChuM1} this can be simplified.  The Green's function has the
form
\begin{equation}
G(k,\omega )={Z\over \omega -\omega_{max}+\bar E_k-i\gamma (\omega
-\omega_{max})^2\Theta(\omega_{max} -\omega)},
\label{eq:63}   
\end{equation}
with quasiparticle residue $Z$, band edge $\omega_{max}$, damping
$\gamma$, dispersion $E_k$, with step function $\Theta (x)$ = 1(0) for $x$
$>(<)$ 0.  The quasiparticle residue can be found as
\begin{equation}
{1-Z\over Z^2}=\int{d^2q\over 4\pi^2}{\Psi (k_0,q)\over (\omega_q
+E_{k_0+q})^2},
\label{eq:64}
\end{equation}
where $k_0$ is the band-edge momentum: $\vec k_0$ = $(\pi /2,\pi /2)$
[$(\pi ,0)$] for hole [electron] doping.  (With the conventional signs
$t'<0$, $t''>0$; in the special case $t'=t''=0$, both energies are
degenerate.)  An equation for the dispersion can then be found by
substituting Eq.~\ref{eq:63} into Eq.~\ref{eq:51}, and setting $\omega
=\omega_{max}$:
\begin{equation}
\bar E_k^{c,v}=ZE_k^{c,v}-Z^2e_k^{c,v}
\label{eq:65a}
\end{equation}
\begin{equation}
e_k^{c,v}=\int{d^2q\over 4\pi^2}\bigl[
{\Psi (k,q)\over\omega_q +E_{k+q}}-
{\Psi (k_0,q)\over \omega_q +E_{k_0+q}}\bigr]],
\label{eq:6b}  
\end{equation}
(The damping adds a small correction to the dispersion, which we ignore.)
It is convenient to rewrite Eq.~\ref{eq:6a} as 
\begin{eqnarray}
E_k^{c,v}=4A_{01}(c_{01}+cos(k_xa)cos(k_ya))
\nonumber\\
+A_{02}(c_{02} +cos(2k_xa)+cos(2k_ya)),   
\label{eq:66}   
\end{eqnarray}
with $c_{01}$ = 0 (1), $c_{02}$
= 2 (-2) for the lower (upper) Hubbard band.  It is found that $\bar E_k$
satisfies a similar equation, with renormalized $A_{0i}\rightarrow A_i$.
In this case, the self-consistent equation Eq.~\ref{eq:65a} can be reduced
to a pair of equations at fixed $k$-values.  For example, at $k=(0,0)$ 
\begin{eqnarray}
4(1+c_{01})A_1+(2+c_{02})A_2
\nonumber\\     
=Z[4(1+c_{01})A_{01}+(2+c_{02})A_{02}]
-Z^2e_{(0,0)}^{c,v},
\label{eq:67a}  
\end{eqnarray}   
with a similar equation at $\vec k$ = $(\pi ,0)$ [or ($\pi /2,\pi /2)$].
Figure~\ref{fig:67} illustrates the self-consistent values of $Z$, $A_1$,
and $A_2$ as a function of $J$ for fixed $t'$, $t''$.
Note that any attempt to extract the bare parameters from the measured
dispersion is highly underdetermined.  Thus, while the band dispersion
$\bar E_k$ and spin wave dispersion $\omega_q$ depend explicitly on $J$,
$t'$, and $t''$, the vertex function depends on $t$, so there are four
parameters to determine, but only two parameters $A_1$ and $A_2$ can be
found from the ARPES dispersion.  Moreover, from Fig.~\ref{fig:67}a, the
value $A_1$ is insensitive to $J$ in the range of interest.  In principle,
the parameters can be determined from additional measurements, including
the Mott gap $\Delta$, the spin wave velocity $c_s$ (as $q\rightarrow 0$,
$\omega_q\rightarrow c_sq$),
\begin{equation}
c_s=2a\sqrt{J({J\over 2}+J'+2J'')},
\label{eq:68}
\end{equation}
or the maxima in the spin wave spectra, $\omega_{(\pi /2,\pi /2)}=2(J+J'
+2J'')$, $\omega_{(\pi ,0)}=2(J+2J')$.  

Given this indeterminancy, a simplified picture is assumed here to estimate 
parameter changes: the renormalized value of $t=0.326eV$ is assumed fixed,
to keep $\Psi$ and the experimental ratios $A_i/t$ constant, and further,
the ratio $t'/t''=-2.5$ was assumed constant.  Then the pairs of solid and
long-dashed horizontal lines in Fig.~\ref{fig:67}a give the experimental
ranges\cite{LaR,Dur} for $A_1$ and $A_2$ respectively.  A reasonable   
match can be found for a bare $t'=-0.375t$.  In this case, the value of
$A_2$ suggests a bare $J$ in the range $0.33-0.41t$, or $108-135meV$.  For
the same parameter range, the individual parameters are renormalized by
$Z_i=A_i/A_{0i}$, with $Z_1$ $\simeq$ 0.059-0.020 [0.8-0.84], $Z_2$
$\simeq$ 0.44-0.51 [-0.21 -- -0.54] for the lower [upper] Hubbard band.  
The ratio $j/t''$ must be renormalized by the SCBA, since $A_2$ and $A_{02}$ 
cross zero at different values of $J$, causing $Z_2$ to be negative for
the upper Hubbard band (it diverges when $A_{02}\rightarrow 0$).  
\begin{figure}
\epsfxsize=0.33\textwidth\epsfbox{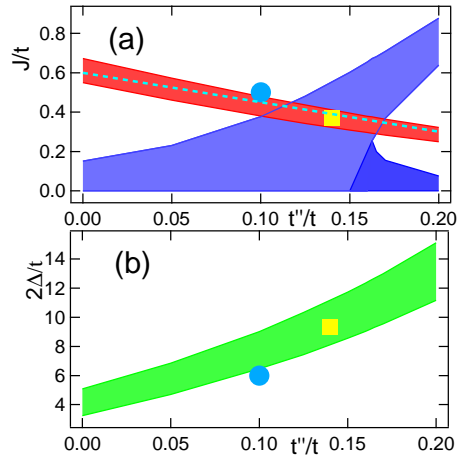}
\vskip0.5cm
\caption{Parameter values consistent with ARPES data for $J$ (a) and
$2\Delta$ (b).  Broad range determined by $A_1$; narrow range (in both a
and b) by $A_2$.  Circle = SCR result, square = best SCBA approximation.  
Dashed line in a: $J=0.6t-1.5t''$.}
\label{fig:68a}
\end{figure}

\begin{figure}
\epsfxsize=0.33\textwidth\epsfbox{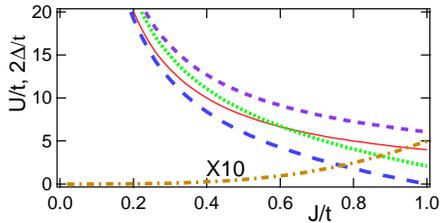}
\vskip0.5cm
\caption{Renormalized $U$ parameter as a function of $J$: solid line =
bare $U=4t^2/J$ in large gap limit; short-dashed line = renormalized $U$
from Eq.~\protect\ref{eq:68c}; long-dashed line = bare $\Delta$ corrected
for the small gap limit, Eq.~\protect\ref{eq:68c}; dotted line =
renormalized $U$ in the small gap limit from Eq.~\protect\ref{eq:68c};
dot-dashed line = $A_{03}$. }
\label{fig:66}
\end{figure}

\begin{figure}
\epsfxsize=0.33\textwidth\epsfbox{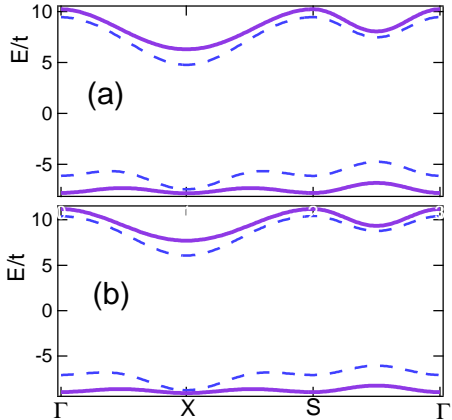}
\vskip0.5cm
\caption{Comparison of mean-field (dashed lines) and SCBA (solid lines) 
dispersions for $t'=-0.375t$, $t''=0.15t$, and two choices of $J$, 
$J/t$ = 0.42 (a) or 0.33 (b).}
\label{fig:66a}
\end{figure}

The above calculation can be repeated for different values of $t''$, and
the allowed parameter values for $J$ and $t''$ are shown in
Fig.~\ref{fig:68a}a.  The `best' SCBA value (square) differs from the SCR
value (circle) by less than a factor of two.  Since $J$ is reduced by
polaron coupling, $U=4t^2/J$ must increase, Fig.~\ref{fig:68a}b.  This can
be seen directly from the self-consistent equation for $G$.  The leading
edge of the band is found from $Re(G^{-1}(k_0,\omega_{max}))=0$, or
\begin{equation}
\omega_{max}=\omega_{max0}+
\int{d^2q\over 4\pi^2}{\Psi (k_0,q)\over\omega_q+E_{k_0+q}}.
\label{eq:68c}
\end{equation}
The gap $2\Delta$ is equal to the splitting between the upper and lower
Hubbard bands at $(\pi /2,\pi /2)$ -- it is {\it not} the sum of the
$\omega_{max}$'s for these two bands, since the bottom of the upper
Hubbard band lies at $(\pi ,0)$.  Correcting for the renormalization of
the dispersion at $(\pi /2,\pi /2)$ reduces the gap, but even so the
renormalized $\Delta$ (Fig.~\ref{fig:66}, short-dashed line) is larger
than the bare value (solid line).  Figure~\ref{fig:66a} shows that the
SCBA increases the Mott gap near half filling, which corrects a
shortcoming of the SCR model, noted above.

\subsection{Extension to Small $U$}

\begin{figure}
\epsfxsize=0.33\textwidth\epsfbox{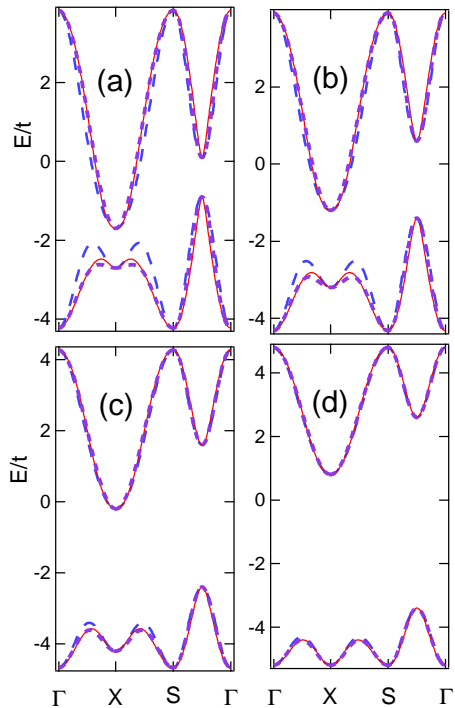}
\vskip0.5cm
\caption{Mean field band structure (solid lines) plus approximations
involving Eqs.~\protect\ref{eq:69},~\protect\ref{eq:70}, with
(short-dashed lines) or without (long-dashed lines) a finite $A_{03}$,
for $\Delta /t$ = 0.5 (a), 1.0 (b), 2.0 (c), and 3.0 (d).}
\label{fig:68}
\end{figure}

The above results were valid for the large-$U$ limit, where the gap
parameter $\Delta=U<S_{zi}>\rightarrow U/2>>t$, in which case
$J=2t^2/\Delta$.  As $\Delta$ decreases, certain modifications are
necessary.  The most important is a modification of $J$.  From the above
analysis, the susceptibility, spin wave dispersion, and renormalized band
parameters all depended on {\it the bare electronic dispersion}.  Hence,
the value of $J$ should be chosen to best approximate the bare $A_{0i}$,
Eqs.~\ref{eq:66a},~\ref{eq:66b}.  This can be accomplished by matching the
exact dispersion to the approximate form at $\vec k =(0,0)$, or
\begin{equation}
{J\over t}={\Delta\over 4t}[\sqrt{1+({4t\over\Delta})^2}-1],
\label{eq:69}
\end{equation}
Fig.~\ref{fig:68}.  Note that $J\rightarrow 1$ as $\Delta\rightarrow 0$,
Fig.~\ref{fig:66}.  It is interesting to note that when the
renormalization correction, Eq.~\ref{eq:68c}, is added in, the
renormalized $\Delta$ (dotted line in Fig.~\ref{fig:66}) lies close to the
perturbative result $\Delta =2t^2/J$ (solid line).

For small $\Delta$ an additional correction is required, to account for
quartic corrections in $t/\Delta$.  This can be done, as above, by
adding a term $A_{03}(1-c_{2x}c_{2y})$ to the model bare dispersion, which
allows a fit to the exact bare dispersion at $(\pi /2,0)$, if
\begin{equation}
{A_{03}\over t}={\Delta\over 2t}[\sqrt{1+({2t\over\Delta})^2}-1]
-{J\over 2t}.
\label{eq:70}   
\end{equation}
This yields a very good approximation to the dispersion down to $\Delta
=t/2$, Fig.~\ref{fig:68} (short dashed line).  The parameter $A_{03}$ is
plotted in Fig.~\ref{fig:66}.

\subsection{Summary of SCBA Results}

(1) Thus at half filling polaronic effects renormalize the bandwidth by
only a factor of $\sim$2, with some change in lineshape.  Polaronic
effects {\it reduce} the values of $J$, $t'$, and $t''$, and hence {\it
increase} the value of $U$.  Thus the gap is enhanced at half filling,
correcting a shortcoming of the SCR calculation
(Figs.~\ref{fig:7e},~\ref{fig:7em}).

While the present calculations are restricted to half filling, some
additional features can be extracted from the calculations of Kusunose and
Rice (KR)\cite{KuR}.  (2) While considerable weight is transferred to an
incoherent spectrum, the coherent spectrum is quite similar to that found
in RPA and SCR calculations, and it is this component which is mainly seen
in the ARPES spectra.  Possible evidence for the incoherent states is a
second peak seen in ARPES spectra of half-filled cuprates\cite{Ron}, about
0.6eV below the main peak of the LHB near the nodal point.  While KR find an
incoherent peak at half filling about 2.5$t$ below the first peak, its
intensity actually maximizes away from the nodal direction toward
$\Gamma$, while the experimental peak is stronger in the opposite
direction, towards $(\pi ,\pi )$.

(3) An important result of the RPA and SCR calculations is that $U$ must
decrease with doping to reproduce the experimentally observed crossover to
a large Fermi surface.  The same result has been found by S\'en\'echal and
Tremblay\cite{SeMSTr}.  The results of KR are consistent, in that KR kept
$U$ doping independent, and did not find this crossover.

(4) Whereas in the mean-field and SCR calculations, electron
doping shifts the Fermi level into the UHB without affecting the relative
weights of the two subbands, in the SCBA the UHB states below the Fermi
level are formed by spectral weight transfer from the LHB.  This spectral
weight transfer had been seen experimentally, and its absence was known to
be a shortcoming of mean field theory, which is thus seen to be corrected
in the SCBA.

(4) In lightly doped NCCO, Armitage, et al.\cite{nparm} found an
additional weak pseudogap -- actually a leading edge gap at the Fermi
level of the UHB -- which was not reproduced by the SCR calculation.  Such
a pseudogap is found by KR, and in an earlier calculation by Stanescu and
Phillips\cite{MNess}.  KR interpreted this as evidence that the filled
states were not actually part of the UHB but were {\it in-gap states}
close to the bottom of the UHB.  Similar in-gap states had been proposed
for hole-doped cuprates\cite{ingap}, and have been considered as evidence
for stripes.  [In LSCO, where stripes are most clearly observed, the added
states are close to mid-gap\cite{Ino}; in other hole-doped cuprates, the
evidence is less clear, but if in-gap states exist, they must lie close to
the top of the LHB.]  

The connection between polarons and stripes is a delicate issue: for very
light doping one would expect {\it magnetic polarons} to form for both
hole and electron doping.  These polarons are strongly dressed electrons,
with many features of second-phase inclusions, and have been suggested to 
act as precursors for nanoscale phase separation\cite{Nag,EmK}.  In
hole-doped cuprates, there is considerable evidence that these polarons
tend to cluster and form stripes.  In electron-doped cuprates there is
considerably less evidence for stripes, and it may be that polarons do not
form clusters.  Hence, the differences between a polaronic phase and a
stripe phase might be rather subtle.  

\section{Extension to Hole Doped Cuprates}

Thus, for electron-doped cuprates, a three-fold coincidence of Mott gap
collapse, Fermi surface crossover, and zero-T QCP is found.  SCR
theory predicts a similar triad for the hole doped cuprates, and the
present section explores the extent to which this is found experimentally.

\subsection{Pseudogap}

In hole doped cuprates, ARPES finds two features which are commonly
referred to as pseudogaps -- a `hump' feature found near $(\pi ,0)$ at
higher binding energy than the main, superconducting `peak', and the
`leading edge gap', a loss of spectral weight in the immediate vicinity of
the Fermi level.  This latter feature is not explained by the present
calculation; it may be the magnetic feature discussed in Section 
V.B\cite{MNess,KuR,SeMSTr}, or it may be associated with the onset of strong
superconducting fluctuations\cite{SPS,pg2}.  

\begin{figure}
\leavevmode
   \epsfxsize=0.33\textwidth\epsfbox{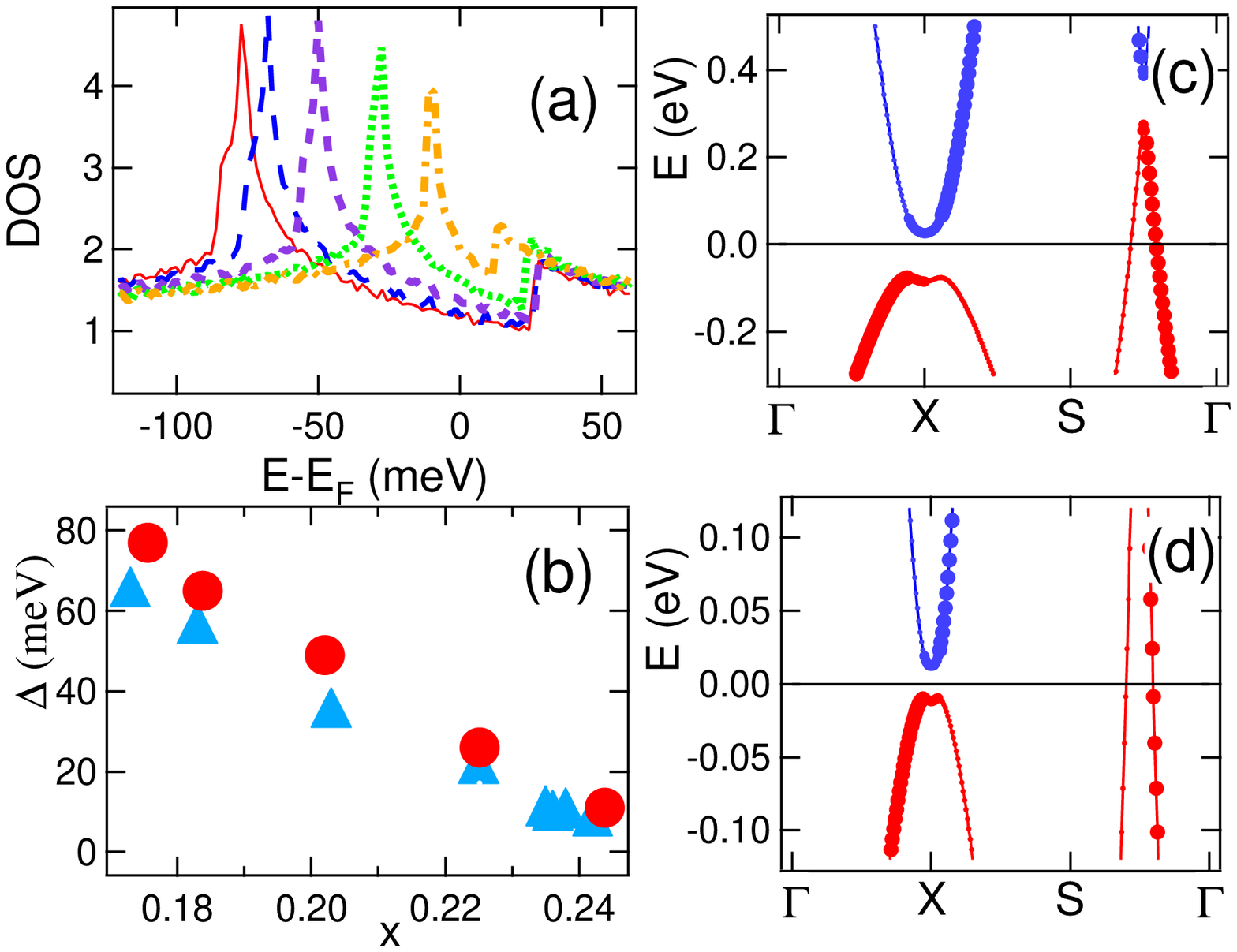}
\vskip0.5cm
\caption{(a) Calculated dos for a series of hole doped cuprates, assuming
$U_{eff}=2.3t$, with $x$ = 0.176 (solid line), 0.184 (long dashed line),
0.202 (short dashed line), 0.225 (dotted line), and 0.244 (dot-dashed
line). (b) Comparison of shift of lower dos peak (circles) from (a) with
representative tunneling data [\protect\onlinecite{Kras}] (triangles).
(c,d)  Band dispersion near the pseudogap for $x$ = 0.176 (c) and 0.244
(d).}
\label{fig:n11}
\end{figure}

On the other hand, the `hump' feature can be consistently interpreted as
the collapse of the Mott pseudogap\cite{SPS}.  Bilayer splitting cannot 
explain SIN tunneling measurements\cite{pg2,Miya,Kras} which find two
hump-like features, roughly symmetric about the Fermi level.  Correlation
with ARPES suggests that the tunneling peaks reflect structure near $(\pi
,0)$, and Figure~\ref{fig:n11} shows that semi-quantitative agreement with
experiment can be attained in terms of weakly split Hubbard bands, for a
screened $U=2.3t$ (see also Fig.~\ref{fig:0d}b). For simplicity, the
calculation is carried out at the mean-field level. Figure~\ref{fig:n11}c,d 
shows how the bottom of the UHB near $(\pi ,0)$ gradually merges into the
VHS of the LHB.  The intensities and positions of the two dos peaks reveal
a clear asymmetry.  As the Mott gap vanishes, the two peaks merge into the
VHS of the bare band.  (There may be complications due to nanoscale phase
segregation, since STM studies suggest that the peak and hump features are
spatially segregated\cite{patch4}.)

The above interpretation requires that for hole doping also the Mott gap
must collapse slightly above optimal doping.  This is consistent with
recent experimental observations of a QCP\cite{Tal1}.  
Moreover, the model predicts that at the QCP, where the pseudogap just
closes, the Fermi level is {\it exactly at the VHS} (H-point). This result
had been found experimentally in some lightly overdoped
cuprates\cite{Surv,Tal2}. 

In a recent confirmation of the QCP just beyond optimal hole
doping\cite{MZG}, it is suggested that this `has to do with the
restoration of the Fermi-liquid state in the overdoped regime {\it
characterized by a large Fermi surface}' (emphasis added); a similar
conclusion was made by Balakirev, et al.\cite{MOAB3}.  Thus two elements
of the QCP triad are present.  The third is more elusive.

\subsection{T=0 QCP}

In electron-doped cuprates, a finite N\'eel temperature persists all the
way to the QCP; by contrast, for hole doping $T_N\rightarrow 0$ at a
doping $x~ 0.02-0.03$, considerably below the proposed QCP.  Here it is
suggested that a T=0 magnetic transition persists out to the QCP, but the
correlation length grows so slowly that three-dimensional N\'eel order
is superceded by the superconducting transition.  Details are presented in
a related publication\cite{MKIV}, and only briefly summarized here.

The key insight is that the susceptibility must satisfy the
fluctuation-dissipation theorem\cite{Mor,Ful}
\begin{equation}
<M^2>=-\int{d\omega\over\pi}n(\omega )\int{d^2q\over (2\pi
)^2}({c_x+c_y\over 2})Im\chi (\vec q,\omega )
\label{eq:21x}
\end{equation}
where $<M^2>$ is the mean square local amplitude of nearest neighbor spin
fluctuations and $n$ is the Bose function.
For hole-doped cuprates the $q$-plateaus constitute a problem.  For
electron-doped cuprates, the plateau width is quite small, and the
susceptibility is large only over an area $\xi^{-2}$, so the sum rule
Eq.~\ref{eq:21x} is never saturated, and $\chi_Q$ and
$\xi$ both diverge exponentially with decreasing T.  For hole doping the
plateau width is large, Fig.~\ref{fig:41a}, and the curvature on the
plateau $Aq^2$ is relatively small, so as T decreases intensity grows all
across the plateau.  This tends to saturate the sum rule, leading to a
greatly weakened divergence of the correlation length, 
\begin{equation}
\xi^2={a\over T}-b,
\label{eq:22x}
\end{equation}
with $a$ and $b$ constants.  From Eqs.~\ref{eq:B30},~\ref{eq:B32}, it can
be shown that
\begin{eqnarray}
a\simeq {A\pi |\delta_0|\over 3ua^2q_c^2}
\nonumber \\
\simeq {8\pi^2A<M^2>\over U\chi_{0Q}q_c^2},
\label{eq:23x}
\end{eqnarray}
(where the latter form follows from Eq.~\ref{eq:21x}, and
$\chi_{0Q}=\chi_0(\vec Q,0)$) so $a\rightarrow 0$ at the QCP.

This result has a number of consequences: (1) neutron
diffraction\cite{BoRe,BalBo} measures the {\it plateau width},
Fig.~\ref{fig:41a} and not the correlation length.  (2) NMR\cite{OBAM}
measures the correlation length, and in YBCO
finds a weak $T\rightarrow 0$ divergence of $\xi$, as predicted.  Thus the
present results resolve a long-standing\cite{MMP,OBAM} controversy about the 
correlation length in hole-doped cuprates.  (3) In the cuprates, N\'eel order 
appears at $T>0$ only if the correlation length exceeds $100a$, where $a$
is the lattice constant\cite{Gre} (the connection between $T_N$ and $\xi$
is discussed in the next section).  This explains the broad range of hole
dopings where there is only T=0 AFM order: Fig.~\ref{fig:41a} shows that
the measured $\xi\rightarrow 100a$ only at $x=0.02$ in LSCO (see also
Fig.~\ref{fig:7de}, below).  (4) Moreover, the slope of the $T^{-1/2}$-term 
in $\xi$ decreases rapidly with doping, signalling a QCP just above
optimal doping.  Hence, the triad of features of the AFM QCP are also
present in the hole-doped cuprates, with the broad susceptibility plateaus
responsible for the striking differences from electron doping.

\subsection{Incommensurate Magnetism and Competing Phases}

The above analysis strongly suggests that at high energy scales the
physics of the cuprates is dominated by magnetic ordering.  This includes
the large pseudogap regime and the attendant QCPs.  None of this analysis
precludes interesting new physics on lower energy scales, including of
course superconductivity near the QCPs.  Another possibility is the
admixture of a second phase generating an enhanced gap -- a popular
choice being the flux phase\cite{AFflux}.

The physics associated with nanoscale phase separation, or `stripe'
physics, seems to also fall in this category.  Incommensurate magnetic
modulations are seen in several cuprates -- particularly the LSCO family
-- and while the SCR model does find an incommensurate susceptibility
particularly for hole doping (Fig.~\ref{fig:10}b) it probably cannot
reproduce the observed doping dependence of the incommensuration.  
Indeed, it has been noted\cite{Schulz} that the incommensurability
generally signals an instability toward phase separation.  Experiments
suggest that phase separation and/or stripe physics is present in the hole
doped cuprates\cite{Tran,patch4} down to arbitrarily small
dopings\cite{Birg}.  However, the temperature at which stripes are
stabilized seems too low\cite{Tran,Imai} for them to be directly
responsible for the pseudogap phenomena.  
 
A detailed discussion of this issue here is clearly out of the question,
but the following suggests a possible explanation.  Doped carriers in an
AFM are strongly dressed by their environment, forming magnetic
polarons\cite{Nag} in a pure Hubbard model, but in a more general
situation being sensitive to nearby competing phases\cite{Yone}.  Thus, it
is suggested that the physics of competing phases enters the problem at
the level of the properties of polarons, and different degrees of phase
separation and/or stripe formation in different cuprates have to do with
the tendency of polarons to cluster.  That is, the stripe physics should
enter the problem on a lower energy/temperature scale than the fundamental
pseudogap phenomena discussed in the present paper.  The $q$-plateau in
hole-doped cuprates greatly enhances the sensitivity to stripe physics,
since the system is close to instability over a wide range of
incommensurate modulations.  

\section{Magnetic Properties}

\subsection{Electron Doping}

While the present model was developed on the basis of ARPES data, the
collapse of the Mott gap should be clearly reflected in other properties
as well, in particular in the magnetic response.  

\begin{figure}
\leavevmode
   \epsfxsize=0.33\textwidth\epsfbox{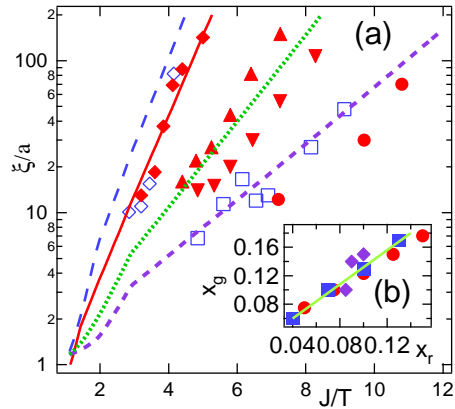}
\vskip0.5cm
\caption{Temperature dependence of correlation length $\xi$ in NCCO for
$x=0$ (solid line), -0.04 (long-dashed line), -0.085 (dotted line), and
-0.10 (short-dashed line).  Data are from Ref.~[\protect\onlinecite{Mat}]:
$x$ = 0 (open diamonds) and -0.15 (open squares); from
Ref.~[\protect\onlinecite{Gre}]: $x$ = 0 (solid diamonds), -0.10 (solid up
triangles), -0.14 (solid down triangles), and -0.18 (solid circles).
Fits are to Eq.~\protect\ref{eq:101}, with parameters appropriate to $x$
= 0 (solid line), -0.04 (long-dashed line), -0.085 (dotted line), and
-0.10 (short-dashed line). Temperatures are measured in units of $J$ =
125meV.  Inset: Plot of as-grown nominal doping $x_g$ vs reduced nominal
doping $x_r$ for fixed values of $T_N$ (circles), $M$ (squares), and $\xi$
(diamonds).  Solid line is $x_g=1.2x_r+0.012$.}
\label{fig:7dd}
\end{figure}

Indeed, Mang, et al.\cite{Gre} have recently measured the ordered moment
$M$ in reduced NCCO samples, and find good agreement with the present
model\cite{KLBM} (see Fig.~\ref{fig:7de}b below).  The correlation length
has not yet been measured in reduced (superconducting) NCCO, but there are
data for the as-grown material, which is insulating\cite{Mat,Gre},
Fig.~\ref{fig:7dd}.  The reasons for the striking differences between the
two types of sample are not fully understood, but there seems to be some
interstitial oxygen which localizes a fraction of the doped electrons, so
one must dope the as-grown samples more to produce a given reduction of
the magnetic properties (e.g., to get a certain value of $T_N$, the doping
of the as-grown sample $x_g$ must be about 0.02-0.03 larger than for the
reduced sample $x_r$, inset in Fig.~\ref{fig:7dd}).  The data for the
undoped sample were used to estimate $\rho_s(x=0,T=0)$, and thereby
$u^{-1}=0.384eV$.  Comparing this to the $\sigma$-model
calculations\cite{CHN,KoCha}, $\rho_s=JS^2$, gives $J=113meV$, in good
agreement with other estimates.  

However, a fit to Eq.~\ref{eq:16} could only be made by reducing the
($T$-dependent) prefactor $\xi_0$ by a factor of 16.  A similar problem
was encountered in the $\sigma$-model calculations: one-loop
renormalization\cite{KoCha} found $\xi_0\sim 1/\sqrt{T}$, as here (below
Eq.~\ref{eq:16}), while a two-loop calculation\cite{Tkhsh} found a
$T$-independent $\xi_0$.  Introducing a Castro Neto-Hone-like
interpolation formula\cite{HCN}, 
\begin{equation}
\xi_0={e\over 4}\sqrt{eA\over 2C(T+2\pi\rho_s)},
\label{eq:101}
\end{equation}
yields the solid-line fit in Fig.~\ref{fig:7dd}, with no adjustment of the
prefactor.  Moreover, the curves for the doped samples apply the {\it
same} correction factors.  The agreement in T-dependence is quite good;
while the theoretical $x_r$ is smaller than the experimental $x_g$, the
ratio is consistent with both those derived from $T_N$ and from the
magnetization $M$, inset in Fig.~\ref{fig:7dd}).  This strongly suggests
that {\it as far as magnetic properties are concerned} as-grown NCCO
behaves like reduced NCCO, with a few percent of the electrons localized
(however, as-grown NCCO never becomes superconducting).

In discussing the as-grown NCCO samples, mention should be made of the
`anomalous pseudogap'\cite{Ono} found in an as-grown sample near
$x$=-0.15 -- which should correspond to $x\sim -0.12$ in the reduced
samples.  From Fig.~\ref{fig:11a1}c, $\rho_s$ falls off in the range 200 -
1000K as $x$ varies from -0.15 to -0.10, signalling the opening of the
Mott (pseudo)gap.  In the as-grown sample, a pseudogap was found to open
below 240K, centered at 300meV.  From Figs.~\ref{fig:7e},~\ref{fig:7em},
the gap near $(\pi ,0)$ would be in this range.  Additional infrared and
Raman phonons were observed, beyond those allowed by tetragonal symmetry.
This could be associated with the orthorhombic symmetry of the magnetic
Brillouin zone.  Clearly more work needs to be done, but if this is the
correct interpretation, the present model predicts what the doping
dependence should be, and that similar features should be seen in the
reduced samples as well.

\subsection{Hole Doping}

\begin{figure}
\leavevmode
   \epsfxsize=0.33\textwidth\epsfbox{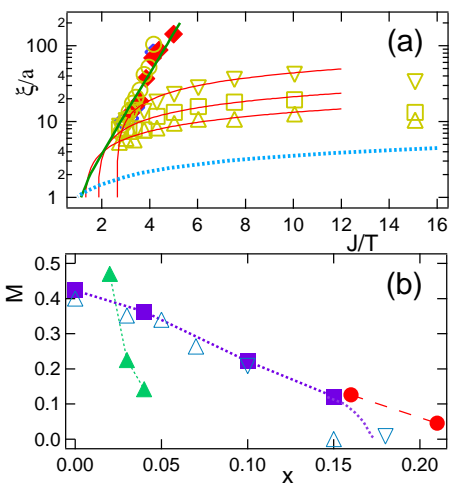}
\vskip0.5cm
\caption{(a) Temperature dependence of correlation length $\xi$ in LSCO for
$x=0$ (open circles), 0.02 (inverted triangles), 0.03 (squares), and
0.04 (triangles) [from Ref.~[\protect\onlinecite{Keim}]], compared with
Nd$_2$CuO$_4$ (open and filled diamonds, as in Fig.~\protect\ref{fig:7dd}).  
Thick solid curve = fit for undoped material from Fig.~\protect\ref{fig:7dd}).
Thin solid lines = fits to Eq.~\protect\ref{eq:22x}.
Dotted line = calculated value for $x=0.10$, as described in the text.
(b) Comparison of magnetization extracted from $\xi (T)$,
Eq.~\protect\ref{eq:23x} (triangles for LSCO\protect\cite{Keim}, circles
for YBCO\protect\cite{OBAM,MKIV}), with that for NCCO, taken from ARPES
fit, Ref.~[\protect\onlinecite{KLBM}] (squares), and from magnetization
(scaled to $M=0.4$ at x=0; triangles: Ref.~[\protect\onlinecite{RPD}], 
inverted triangles: Ref.~[\protect\onlinecite{Gre}].  All lines are simply
drawn to connect the data points, except for that part of the dotted 
line connected with the ARPES data (squares) extrapolated beyond $x=0.15$.
This represents a mean-field calculation, assuming that $U$ does not
change with doping over this range, and using the band parameters of
Ref.~[\protect\onlinecite{KLBM}].} 
\label{fig:7de} 
\end{figure}

The results on NCCO should be contrasted to those for LSCO\cite{Keim},
Fig.~\ref{fig:7de}a, where a saturation of the effective $\xi$ is observed
in all doped samples.  For undoped La$_2$CuO$_4$, the data (open circles)
largely overlap those of Nd$_2$CuO$_4$ (open and filled diamonds), but a
small change of slope may be present in the best fits.  
For lightly-doped LSCO\cite{Keim}, the data can be fit to Eq.~\ref{eq:22x}
down to $\sim$150K, Fig.~\ref{fig:7de}a, below which $\xi$ saturates or
decreases.  In principle, it should be possible to calculate this
saturation of $\xi$ directly from Eq.~\ref{eq:16}.  As noted in Section
III.C, the value of $A$ tends to be overestimated when the susceptibility
peak is incommensurate.  Thus, the dotted line in Fig.~\ref{fig:7de}a is
the calculated value of $\xi$, using Eq.~\ref{eq:101} with parameters
appropriate to $x=0.10$ hole doping, except that $A/a^2$ = 0.24, only 1/3
the value estimated from Fig.~\ref{fig:10}a.

From the $a$ coefficient of Eq.~\ref{eq:22x} it
should be possible to extract the magnetization, Eq.~\ref{eq:23x}.
However, as explained in Section VI, neutron scattering data tend to
measure the susceptibility plateau width $q_c$, strongly underestimating
$\xi$.  This is illustrated in Fig.~\ref{fig:7de}b, where magnetization
$M=\sqrt{<M^2>}$ derived from $\xi$ via Eq.~\ref{eq:23x} is compared to
$M$ in NCCO estimated from the ARPES data\cite{KLBM} (squares) and from
magnetization (upright\cite{RPD} and inverted\cite{Gre} triangles). The
$\xi$-derived data include the NCCO neutron data of Fig.~\ref{fig:7de}a
(triangles) and NMR data from YBCO\cite{OBAM} (circles) expected\cite{MKIV} 
to give a better estimate of $\xi$.  For both sets of data, the parameter 
$A/\chi_{0Q}Ua^2$ was taken as a constant, 2.8.  Except for the lowest
doping, the neutron data lead to an underestimate for $M$, confirming that
the measured $\xi$ is too small.  In contrast, the NMR data are consistent
with the electron-doped results, and strongly suggest the presence of a
QCP just above optimal doping.

More recent experiments on very lightly doped LSCO\cite{Waki} have
found that the magnetization at these dopings is actually incommensurate
-- consistent with diagonal stripes.  This points out an interesting
parallel with the present model: early experimental samples displayed
flat, diamond-shaped susceptibility plateaus near $(\pi ,\pi )$.  As
sample quality improves, incommensurate structure seems to become more
prominent: see, e.g., Fig. 1 of Ref.~\onlinecite{Mook}.  Related behavior
arises in the model: The susceptibility for hole-doped cuprates displays a
flat-topped plateau at high temperatures, $\delta >0$.  As the temperature
is lowered, $\delta\rightarrow 0$, incommensurate structures develop from
fine structure on top of the plateau, Fig. 3 of Ref.\onlinecite{MKIV},
gradually dominating the spectrum. However, in the calculations this
incommensurability is sensitive to sample `quality': it only shows up when
$\delta$ is very close to zero.  Hence, in real samples, the appearence of
such structure should be very sensitive to disorder or sample
inhomogeneity.  Finally, it should be noted that interpretation of the
incommensurability in terms of stripes remains controversial in cuprates
other than the LSCO family.  Reznik, et al.\cite{Rez} report an
approximately uniform ring of incommensurability in optimally doped YBCO,
which is dispersive and pushed up to finite frequencies by the spin gap in
the superconducting state.  A ring or diamond of incommensurability is
actually quite close to what is found here, and the extension of the
present calculations to the superconducting phase should be quite similar
to the results of Eschrig and Norman\cite{MEsN}.

\section{Three Dimensional N\'eel Order}

The (inverse) Stoner factor $\delta_q$, Eq.~\ref{eq:B18}, can 
be generalized to include interlayer coupling:
\begin{equation}
\delta_q(\omega )=\delta +Aq^2+A_zq_z^2-B\omega^2-iC\omega,
\label{eq:0B18}
\end{equation}
leading to a susceptibility 
\begin{equation}
\chi (\vec q,\omega )={\chi_Q\over 1+\xi^2[(\vec q-\vec Q)^2+a_z
(q_z-Q_z)^2]-\omega^2/\Delta^2-i\omega /\omega_{sf}},
\label{eq:15z}
\end{equation}
with $a_z=A_z/A$.  

In the physical cuprates, the interlayer hopping has an anomalous dispersion,
generally written as $t_z=t_{z0}(c_x-c_y)^2$.  This formula holds for
bilayer splitting, and in general when the CuO$_2$ planes are stacked {\it
uniformly}.  However, as explained in Appendix E, many of the cuprates,
including NCCO, have a {\it staggered layering}, with the Cu in one
CuO$_2$ plane laying above a vacancy in the neighboring CuO$_2$ sheet.
This leads to a {\it magnetic frustration}: the Cu in one sheet has four
nearest neighbors in the adjacent sheet, two with spin up, two with spin
down.  This frustration is reflected in a more complicated dispersion of $t_z$:
\begin{equation}
t_z=t_{z0}(c_x-c_y)^2\cos{k_xa\over 2}\cos{k_ya\over 2},
\label{eq:n21}
\end{equation}
which vanishes at $(\pi ,0)$ and $(0,\pi )$, and leads to a greatly
reduced interlayer coupling. (Effects of AFM frustration associated 
with layering have been discussed in Ref.~\onlinecite{PiL}.)

The consequences of both uniform and staggered stacking are explored in
Appendix E.  If the c-axis resistivity is coherent, it can be used to
estimate the interlayer hopping $t_{z0}$.  It is found that the value of
$t_{z0}$ needed to produce a given resistivity anisotropy is approximately
5 times smaller for uniform stacking, to account for the frustration in
the staggered stacking.  With the corresponding $t_{z0}$'s determined from
resistivity, both forms of interlayer coupling give rise to comparable
interlayer coupling, and hence a finite N\'eel temperature.  While the
optimal $Q$-vector depends on doping, at half filling both forms predict
$\vec Q=(\pi ,\pi ,0)$, consistent with experiment in La$_2$CuO$_4$.
Even for quite strong anisotropy, this mechanism can account for the
observed $T_N$s (in fact, tends to overestimate $T_N$), without the
necessity of invoking additional mechanisms, such as a
Kosterlitz-Thouless transition, with the reduced spin dimensionality
caused by spin-orbit coupling effects\cite{Ding,SinT,KAASh,KK}.

Within mode coupling theory\cite{STeW} (Appendix E), the N\'eel
temperature is found from the gap equation (Eqs.~\ref{eq:D1},~\ref{eq:B34e})
\begin{equation}
\chi_0(T)U=\eta +{3uTa^2\ln{({T\over T_{3D}})}\over\pi A},
\label{eq:21}
\end{equation}
where $T_{3D}\sim t_z^2$ is defined below Eq.~\ref{eq:B34i}.  It is found
that $T_{3D}$ is approximately constant, independent of doping in the
electron-doped regime.  Apart from a small numerical factor, 
Eq.~\ref{eq:21} differs from the isotropic three-dimensional result by
the logarithmic factor, which diverges ($T_N\rightarrow 0$) as
$t_z\rightarrow 0$.  

Equation~\ref{eq:21} can be rewritten in a suggestive form.  Approximating 
$\rho_s$ by $\rho_s^a=A(\chi_0U-\eta )/12ua^2$ (Eq.~\ref{eq:B40}), then, 
using Eq.~\ref{eq:16}, the N\'eel transition occurs when
\begin{equation}
J_z[{\xi(T_N)\over\xi_0(T_N)}]^2=\Gamma T_N
\label{eq:22a}
\end{equation}
where $J_z=J(t_{z0}/t)^2$, $J=4t^2/U$, and $\Gamma=4t_{z0}^2/UT_{3D}$.  A
very similar form was proposed earlier\cite{BiGGS}, and
experimentally\cite{Gre} N\'eel order seems to appear when $\xi\simeq
100a$.

\begin{figure}
\leavevmode   
   \epsfxsize=0.33\textwidth\epsfbox{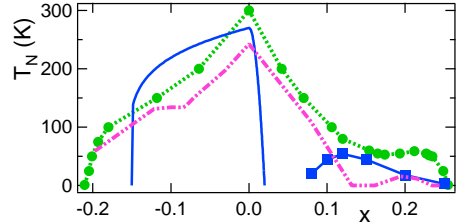}
\vskip0.5cm
\caption{Comparison of experimental N\'eel temperatures for NCCO and LSCO,
(solid line), and for the stripe (magnetic) ordering transitions observed
in Nd-substituted LSCO [\protect\onlinecite{Ich}] (solid line with
squares) with the model of interlayer coupling with staggered stacking and
$t_{z0}=t/10\sim 30meV$, plotted as $T_N/10$ (dot-dot-dash line). Also
included is the approximate expression, Eq.~\protect\ref{eq:22b} (dotted
line with circles).  (Note that there is a range of hole doping for which
$A$ is found to be negative; in this range $T_N$ was arbitrarily assumed
to vanish in the staggered model, $T_N=0$.) } 
\label{fig:nD7} 
\end{figure}

Figure~\ref{fig:nD7} compares the calculated value of $T_N$ with the
experimental values.  
While
the overall doping dependence is comparable, the calculated $T_N$ is about
an order of magnitude higher.  The calculation is for staggered stacking,
with $t_z$ adjusted to reproduce the observed resistivity anisotropy, but
Appendix E shows that the overestimate is generic: the coefficient of the
logarithm needs to be larger to reduce $T_N$.  Also shown in
Fig.~\ref{fig:nD7} (dotted line) is a simplified model, which assumes that
\begin{equation}
T_0^*={\pi A\over 3ua^2\ln ({T\over T_{3D}})}
\label{eq:22b}
\end{equation}
is doping independent, $T_0^*=1200K$.  This model reproduces qualitatively
the shape of the numerical calculation, but with a magnitude comparable to
experiment.  The magnitude of $T_N$ could be matched almost quantitatively
if $U_{eff}$ also has a significant temperature dependence, as discussed
in Appendix E . The overall doping dependence is also comparable
to experiment.  The agreement could be further improved by using a smaller
value of $t'$, which would shrink the doping range over which N\'eel order
occurs.  

Finally, it should be noted that a finite $T_N$ can change a continuous
QCP into a {\it first order}.  This follows beacuse the plateau width
increases with increasing temperature.  Hence, near the plateau edge, the
system can satisfy the Stoner criterion at some finite temperature, but
fail to satisfy it at a lower temperature, having fallen off of the
plateau edge.  Such a first order termination of the AFM state seems to be
found in the electron doped cuprates, most notably in
Pr$_{2-x}$Ce$_x$CuO$_4$ (PCCO)\cite{PCCO}, and in a related organic
material\cite{KIMK}.

\section{Discussion}

\subsection{Slater vs Mott Physics}

Theories of magnetism fall into two diametrically opposed
classes\cite{PWA}: band vs atomic models, or Slater vs Mott physics.  
In both approaches, the copper band is split by a Mott gap into two parts.
In Slater theory, long range magnetic order leads to a unit cell doubling,
so each subband remains conventional, with two electrons per unit cell.
In Mott theory, the bands are highly unconventional: the gap opening is
purely a local effect -- there is an energy penalty of $U$ for two
electrons to sit on the same copper site.  Since there is no change in 
lattice symmetry, the unit cell remains the same, and the bands hold only
half as many electrons as conventional bands.  Distinguishing between the 
two models is complicated, since in the Hubbard model, residual hopping   
proportional to $4t^2/U\sim J$ leads to AFM coupling of the electrons, and
can lead to parasitic N\'eel order, at a temperature $T_N$ much lower than
that at which the Mott gap opens.  On the other hand, strong  fluctuations
in the Slater model can greatly reduce $T_N$, leaving a pseudogap near 
the mean-field instability temperature.

While the mean field results might give a good qualitative picture of the
Slater regime, they are unlikely to be able to describe Mott physics.
However, it would still be hoped that the mean field results can give an
indication of when the crossover is likely to occur.  Here, two separate
indicators are presented.

\begin{figure}
\epsfxsize=0.33\textwidth\epsfbox{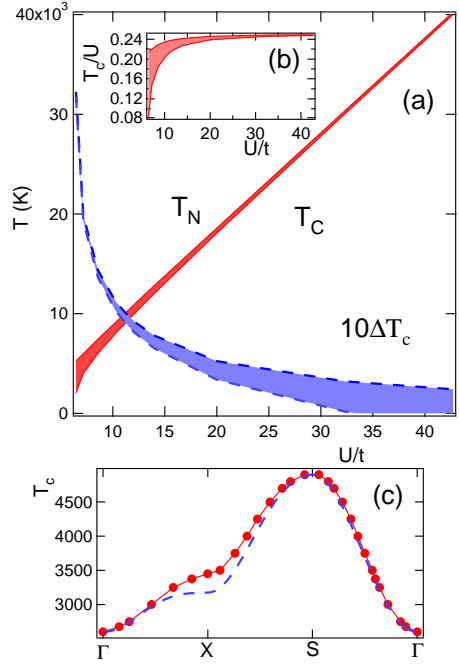}
\vskip0.5cm
\caption{(a) Mean field transition temperatures for N\'eel ($T_N$) and
ferromagnetic ($T_C$) orders and their difference $\Delta T_c=T_N-T_C$
(upper dashed line).  At any $U/t$ the ratio of the shaded area to the
total area below this line gives the fraction of the Brillouin zone which
is significantly excited ($T_c(\vec Q)\ge (1-\alpha_0)T_N$, for
$\alpha_0=0.01$).
(b) Replot of transition temperatures scaled to $U$.  (c) Plot of $T_c$  
vs $\vec q$, for $U=6t$.  The curve bears an uncanny resemblance to the
(scaled) electronic dispersion of the LHB, long-dashed line.}
\label{fig:2a}
\end{figure}

A first indication comes from looking at competing orders.  A Stoner
criterion $U\chi_q=1$ gives the onset temperature for {\it magnetic}
order at $\vec q$, ranging from AFM, $\vec q=\vec Q$ to ferromagnetic, $\vec
q=0$, Fig.~\ref{fig:2a}.  While at half filling for any value of $U$, AFM 
order dominates, the splitting decreases with increasing $U$.  The local,
or Mott physics should arise when fluctuations to all magnetic orders are
comparably likely, or the spread in transition temperatures, $\Delta T_c$,
is $<<T$.  Since the probability of a fluctuation of $N$ particles into a
phase with excess free energy $\Delta f$ is $\sim e^{-N\Delta f/k_BT}$,
one can crudely state that a phase will be significantly excited if
$T_N-T_c(\vec Q)\le\alpha_0T_N$, where $\alpha_0$ is a small numerical
constant.  The width of the $\Delta T_c$ curve in Fig.~\ref{fig:2a}a shows
the fraction of the Brillouin zone that is significantly excited 
for $\alpha_0=0.01$ (e.g., for $U/t\ge 32.5$, all modes are excited). This
suggests that for $U\ge 15t$, these fluctuations spread over a significant
fraction of the Brillouin zone, while for $U>30t$ virtually all magnetic
states are equally excited and the Slater picture is badly broken down.
However, the cuprates are generally found to be in the regime $U\le 12t$,
where a Slater picture should be reasonably accurate even close to the   
$T_N^*$ crossover.

\begin{figure}
\leavevmode
   \epsfxsize=0.33\textwidth\epsfbox{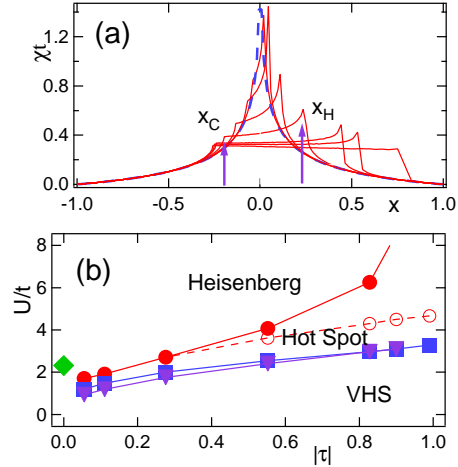}
\vskip0.5cm
\caption{(a) Bare susceptibility $\chi_0(\vec Q,0)$ at $T=1K$, for several
values of $\tau$ = -0.0552, -0.110, -0.276, -0.552, -0.828, -.9, -.99
(solid lines), and 0 (dashed line).
(b) Crossover couplings as a function of $\tau$: $U_V$ (triangles), $U_2$
(squares), and $U_1$ (circles). (Dashed line = $U_1$ for electron
doping.)}  
\label{fig:1x4}
\end{figure}

Alternatively, when the mean-field solution becomes insensitive to the
band structure, it is likely that a local picture is becoming dominant.
In the present instance, the band structure is determined by the ratio $t'/t$.
For any non-zero value of $t'$, the susceptibility has a generic doping
dependence\cite{AAA,BCT}, Fig.~\ref{fig:1x4} -- changing the sign of $t'$
merely interchanges electron and hole doping.  The role of the
susceptibility plateaus can be quantified, by defining ranges of $U$ where 
the nature of the transition changes, Fig.~\ref{fig:1x4}b.  Thus, for
$U<U_V$, there is no Mott transition at $x=0$, and the physics is    
dominated by an AFM transition at the VHS; for $U_V<U<U_2$, there is a
Mott transition at half filling, which terminates (on the electron-doped
side) before the plateau ends, and hence is controlled by dynamic critical
exponent $z=2$; for $U_2<U<U_1$, the Mott gap collapses in the enhanced
regime near the edge of the plateau; and for $U>U_1$ the Mott gap
terminates well off of the plateau, in a region of $z=1
$ physics.  For the present $\tau =-0.552$, 
the approximate values are $U_V/t$ = 2.4, $U_2/t$ = 2.6, and $U_1/t$ = 3.6
($x<0$) or 4.1 ($x>0$).  Note that the $z=2$ regime is quite narrow, and
can probably be subsumed into the VHS regime.  These values depend on     
$t'$, and the VHS moves to half filling as $t'\rightarrow 0$,
Fig.~\ref{fig:1x4}b.  Even when $t'=0$, Sen and Singh\cite{SenS} find a
crossover from SDW-like to Heisenberg-like behavior as correlations increase 
beyond $U_{0V}=3.26t$ (diamond in Fig.~\ref{fig:1x4}b). It must be
kept in mind that $U$ depends on doping, and the above estimates refer to 
$U$ near the plateau edge.  The bare $U_0=U(x=0)$ can be estimated by
assuming the doping is high enough to reach the Kanamori limit\cite{Kana},
$U=U_0/(1+U_0/8t)$.  This results in  $U_{0V}/t$ = 3.4, $U_{02}/t$ = 3.8,
$U_{01}/t$ = 6.5 or 8.4 for $\tau =-0.552$.  These last values are
comparable to but somewhat smaller that those estimated by the first
criterion.

Note that the cuprates are in the range $U_2<U<U_1$, where the plateau
edges form {\it natural phase boundaries} for the Stoner criterion,
thereby providing a natural explanation for the approximate electron-hole
symmetry of the QCPs.  

From the above discussion, it might be concluded that the cuprates 
are in the {\it Slater regime}, where mean field results are qualitatively
accurate.  While this is probably true for the electron-doped cuprates, it
must be kept in mind that {\it the criterion for the breakdown of the
Slater regime is the flatness of the susceptibility in $\vec q$} -- the
inability of Slater theory to deal with many competing phases.  In this
light it is interesting to speculate whether the flatness of the 
$q$-plateaus might be a signal of {\it enhanced Mott physics in the
hole-doped cuprates}.

\subsection{Magnon Bose Condensation and Non-Fermi Liquid Physics}

\begin{figure}
\leavevmode   
   \epsfxsize=0.33\textwidth\epsfbox{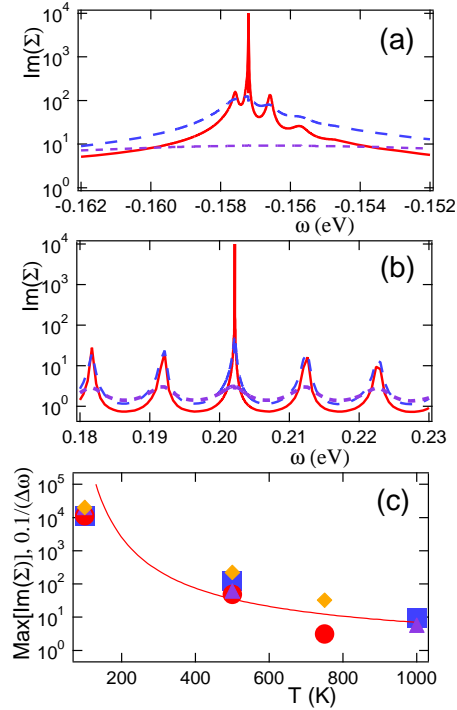}
\vskip0.5cm
\caption{(a,b): Blowups of $Im(\Sigma )$ for $x=0$ at $(\pi ,0)$ (a) and
$(\pi /2,\pi /2)$ (b) at $T$ = 100K (solid lines), 500K (long dashed
lines), and 1000K (a) or 750K (b) (short dashed lines). (c) Maximum of
$Im(\Sigma )$ vs $T$ for $(\pi ,0)$ (squares) and $(\pi /2,\pi /2)$
(circles); 0.1/(full width at half maximum) for $(\pi ,0)$ (triangles) and
$(\pi /2,\pi /2)$ (diamonds); solid line = corresponding $\xi (T)$,
Eq.~\protect\ref{eq:16}.} 
\label{fig:nD9} 
\end{figure}

Figure~\ref{fig:nD9} shows the sharp peak which arises in $Im \Sigma$ at
low T.  The growth is exponential, approximately matching that of the
coherence length, Eq.~\ref{eq:16}.  (Note that it requires a fine mesh in
the integral of Eq.~\ref{eq:18} to capture this growth.)  This peak arises
exactly at the incipient magnetic zone boundary, and turns into true Bragg
scattering at the transition to long range order: the increase in peak
height is almost exactly compensated by a decrease in the width of the
peak.  A simple physical explanation is that the SDW transition can be
interpreted as a {\it Bose condensation of the zone boundary magnons}.
Then the Mermin-Wagner theorem reduces to the fact that in a
two-dimensional system, Bose particles can only condense at $T=0$. 
A similar explanation for the transition has been presented
earlier\cite{VAT}.

In turn, the soft zone-boundary phonons explain one origin of non-Fermi
liquid physics in the model: Bragg scattering from a fluctuating
diffraction grating.  How does one define Luttinger's theorem when the
unit cell is strongly fluctuating?

\subsection{Comparison with Other Calculations}

As noted above, the present calculations predict that at the magnetic QCP
the (possibly $T=0$) Neel phase will terminate, the Mott gap will
collapse, and the Fermi pockets will merge into a large Fermi surface.
This result distinguishes the present calculations from many others in the
literature.  Here a number of Slater-like theories are discussed.

The present results are generally consistent with the $t-J$ model in the
low doping regime.  However, since the $t-J$ model cannot readily deal
with both Hubbard bands simultaneously, it is not appropriate in the
present analysis of electron doped cuprates, where (a) ARPES can detect
both bands (at least up to the Fermi level), and (b) the Mott gap is found
to collapse with doping, leading to an overlapping of both bands at the
Fermi level.  In the $t-J$ model double occupancy is forbidden in the LHB,
while in the UHB empty sites are forbidden.  Moreover, the Hubbard model
only allows $J$ values for $J\le 1$ (Figs.~\ref{fig:66},~\ref{fig:68}),
and near this upper limit significant modifications are needed.  In the
SCBA approach to the $t-J$ model, the parameter $A_1$ has a broad
peak\cite{MaHor} when $J\sim 0.8$, not found in the Hubbard model SCBA
calculations, suggesting that $t-J$ and Hubbard can be equivalent only for
$J<<0.8$ or for $U>>5$ -- that is, near half filling only.  It is
interesting to note that a recent $t-t'-t''-J$ model calculation seems
consistent with the first doped carriers forming weakly interacting
quasiparticles in pockets of the respective upper or lower Hubbard bands,
for either electron or hole doping\cite{LHN}.

The NAFL and spin fermion models are also based on Slater-type
physics, and should in principle make similar predictions to the present
SCR model. However, they tend to take their parameters from experiment,
which can lead to complications in the presence of stripe phases.  For
example, for hole doping, long range Neel order and diverging
susceptibilities terminate at a very low doping, $x\sim 0.02$.  While the
SCR model predicts $x_{QCP}\sim 0.25$, some empirical models take
$x_{QCP}\sim 0.02$.  In this case, the QCP is divorced from Mott gap,
since the Mott gap will clearly persist above $x=0.02$.  Even worse, 
Matsuda, et al.\cite{Birg} have shown that for doping between $x=0.02$ and
half filling the system is phase separated, so uniform AFM order exists
only at $x\le 0$.  

Three examples of spin fermion calculations will be given, to highlight
the differences and similarities.  (1) Abanov et al.\cite{AChS}
postulate a small-$x$ magnetic QCP.  They find that the magnetic resonance 
mode frequency goes to zero at this QCP, but also the superconducting gap 
vanishes at the same doping, which would have important consequences for
the mechanism of superconductivity.  This is in sharp contrast to the
present model, where the magnetic QCP is at much higher doping.  In this
model, the low-$x$ QCP is superconducting, presumably associated with
stripe effects.  (2) Chubukov and Morr\cite{ChuM1,ChuM2} studied the
crossover from small to large Fermi surfaces -- here driven at fixed
doping by reducing $U$.  They state\cite{ChuM2} that, ``as the system
moves away from half filling, the spectral weight transfers from the upper
band into the lower band and, near optimal doping, there exists just one
coherent band of quasiparticles.''  This suggests that the crossover is
due to a spectral weight shift, and not to the gap closing -- in
contrast to the present results and to experiment on NCCO.  However, it
should be noted that (a) their paper actually concentrates on changes at
the Fermi level, and did not explore how the UHB might have shifted with
$U$; and (b) it is possible that the {\it coherent} part of the UHB
collapses, while some weight remains in the incoherent part.
(3) On the other hand, Schmalian, et al.\cite{SPS} go beyond the SCBA,
summing both non-crossing and crossing diagrams via a generalization of a
technique of Sadovskii\cite{VSad}; their results for hole doping are
quite similar to the present results, with a magnetic QCP above
optimal doping -- but with $\xi$ adjusted at each doping to fit the
experiment. 

A number of groups have studied the Hubbard model using FLEX 
calculations, and have had considerable success in describing anomalous   
transport properties\cite{Yan}.
Here a pseudogap is found even though the FLEX model cannot describe the
splitting into UHB and LHB, and the pseudogap is derived from
superconducting fluctuations.  However, these models are consistent with
the present results, in that (1) the pseudogap they describe is clearly   
the lower, leading edge pseudogap which is not described by the present   
model, and (2) their calculation of the normal state properties require a
value of $U/t\sim 1.5-2.5$ much smaller than the values found at half
filling, and comparable to (or even smaller than) the doped values found
here.  [Spin fermion calculations also extract a small value of $U$ --
there called $g$ -- from experiments in near-optimally hole doped
cuprates\cite{ANorm}.]

The present calculations are in general consistent with the results of
Ref~\onlinecite{VAT}. These authors employ a (Two-Particle Self
Consistent) conserving approximation, and attempt to calculate $U(x)$
directly\cite{VCT}.  However, they incorporate the strong thermal
(Mermin-Wagner) fluctuations directly into their definition of $U$, so the
resulting doping dependence should not be compared to the form assumed
here.

A leading edge pseudogap can also arise in the Hubbard model in the
absence of superconductivity\cite{Hau,MNess,SeMSTr}, but only for large
$U>8t$\cite{SeMSTr}.

Some recent calculations have confirmed that $U$ must decrease with
electron doping to reproduce the ARPES data: in Kusonose and
Rice\cite{KuR} the demonstration is indirect -- the gap collapse does not
occur in a SCBA calculation if $U$ is kept large. S\'en\'echal and
Tremblay\cite{SeMSTr} give a more direct demonstration; their model can
also explain the hole-doped pseudogap near $(\pi ,0)$ in the absence of
stripe physics if $U$ does {\it not} decrease with hole doping. 

Finally, a proper study of the model incorporating QCP fluctuations is a
strong desideratum, but the problem of combining QCP and Mermin-Wagner
fluctuations has rarely\cite{Sach} been tackled in the literature.

\subsection{VHS}

Whether or not the VHS is responsible for the observed electron-hole
asymmetry, the present calculations reveal some novel features of Van Hove
physics.

\subsubsection{Temperature Dependent VHS}

As noted by Onufrieva and Pfeuty\cite{OPfeut}, the VHSs associated with
the susceptibilities (and hence with charge or spin nesting) are {\it
different} from those associated with the density of states (and
superconductivity).  Thus, whereas superconductivity will occur at the same
optimal doping for all temperatures, the doping of maximal nesting instability
{\it is a strong function of temperature}.

This contrasting behavior of nesting vs pairing {\it susceptibilities} is
related to a characteristic difference in the nature of the two {\it
instabilities}.  A superconducting instability has an intrinsic
electron-hole symmetry, which means that the gap is tied to the Fermi
level, and a full (s- or d- wave) gap can be opened at any doping level.
On the other hand, a nesting gap is dispersive, and only part of it lies
at the Fermi level (except in special cases).  Furthermore, a
(superlattice) Luttinger's theorem must be obeyed, requiring the presence
of residual Fermi surface pockets.  Stated differently, a full nesting gap 
can only open at integer filling, so {\it as the interaction strength
increases, any nesting instability must migrate to integral doping}
(e.g., half filling in the original band structure).  This same VHS
migration is mirrored in the T-dependence of the magnetic (or charge)
susceptibility.

\subsubsection{VHS Transitions}

We have seen that the doping-dependent $U_{eff}$ gives rise to a Mott gap
collapse near the edges of the susceptibility plateau in Fig.~\ref{fig:0a}.
If $U_{eff}$ is smaller (dot-dashed line: $U_{eff}$ reduced by 2/3), more
complicated behavior should arise.  Due to the peak in $\chi$ near the
H-point, there could be a reentrant transition, with one magnetic order
near half filling, and a second near the VHS.  For an even smaller
$U_{eff}$ (or replacing $U_{eff}\rightarrow J$)\cite{OPfeut}, the transition 
near $x=0$ can be eliminated, leaving a spin density wave transition near
the VHS.  In principle there could even be a phase separation {\it between
two AFM phases}: an insulating phase near half filling and a metallic
phase near the VHS.

\section{Conclusions}

The key conclusion to this work can be stated as follows: In doped cuprates 
there is a magnetic QCP where three factors coincide: the crossover from
small to large Fermi surface, Mott gap collapse, and Neel transition
termination.  In the SCR calculation there is no finite temperature Neel
transition, at least in the isotropic 2d limit, but the zero-temperature
Neel transition persists with doping up to a QCP controlled by a modified
Stoner criterion.  While the Mott gap opening is more of a crossover than
a sharp transition, nevertheless, the upper and lower Hubbard bands merge
at nearly the same point, and the Fermi surface pockets recombine to a
single large Fermi surface, consistent with band structure calculations.
Comparison with experiment suggests that this correctly describes the
situation in electron doped NCCO, both in ARPES (Section IV) and in
magnetization studies, Ref.~\onlinecite{Gre} and Section VII.

The hole doped case also appears to fit this model, but with complications 
associated with the $q$-plateau.  Thus, (a) the pseudogap collapses in a
QCP, as expected; (b) evidence for the Fermi surface crossover has recently 
been reported\cite{MOAB3,MZG}; (c) the correlation length appears
to diverge as $T\rightarrow 0$, but much more weakly than for
electron doping, due to a sum rule saturation\cite{MKIV}.

In more detail, the main results of this paper can be summarized:

\par\noindent $\bullet$
Fluctuation effects were added to the mean field Hubbard model via a
mode coupling calculation, which allowed satisfying of the Mermin-Wagner
theorem ($T_N=0$).  It was found that the mean-field gap $\Delta_{mf}$ and 
N\'eel temperature $T_N^{mf}$ evolved into a pseudogap $\Delta_{ps}\sim
\Delta_{mf}$ and an onset temperature $T^*\sim T_N^{mf}$ (as is familiar
from the related CDW results).  

\par\noindent $\bullet$
The resulting dispersions and Fermi surfaces are in excellent agreement
with photoemission experiments on electron-doped cuprates\cite{nparm},
while the pseudogap seems consistent with ARPES and tunneling results in
hole doped cuprates.  

\par\noindent $\bullet$
Magnetic properties -- saturation magnetization and coherence length --
are also well fit by the same model.  The good agreement between ARPES and
direct magnetic measurements leaves little doubt that the (large)
pseudogap is predominantly magnetic in origin.

\par\noindent $\bullet$
The zero-temperature N\'eel transition is controlled by a Stoner-like 
criterion, hence is sensitive to the bare susceptibility and in turn to
the Fermi surface geometry (hot spots).  This lead to an approximately
electron-hole symmetric QCP near optimal doping (termination of hot spot
regime), at which both zero temperature N\'eel transition and pseudogap
transition simultaneously terminate.

\par\noindent $\bullet$
The model leads to a NAFL-type susceptibility, and the calculation of the
NAFL parameters has been reduced to a calculation of the coupling
parameters $U$ and $u$, the former having a significant doping (and
possibly temperature) dependence. At present, $U(x)$ is estimated from
experiment, and the mode coupling $u$ via consistency with the $t-J$ model.
(A small portion of the renormalization of $U$ arises from quantum
corrections to the Stoner criterion.)

$\bullet$  The present theory differs from conventional NAFL theory
by the inclusion of two cutoff parameters, $q_c$ and $\omega_c^-$, which
shrink to zero at either the H- or C-points.  For example, $q_c$
is large near the H-point, but shrinks to zero at the C-point,
causing the $A$ parameter to have a strong temperature dependence in the
electron-doping regime.  

\par\noindent $\bullet$
Finally, a striking {\it temperature/frequency dependence} of the VHS 
susceptibility peak\cite{OPfeut}, causing it to shift to half filling at
high $T$, is interpreted in terms of Luttinger's theorem: if the coupling
is strong enough to open a full gap, the gap must fall at half filling.

{\bf Note:} After the present work was completed, I received a preprint
from A.-M.S. Tremblay reporting similar calculations for electron-doped
cuprates\cite{BKAT}.

{\bf Acknowledgments:}  
This work was supported by the Spanish Ministerio de Educaci\'on
through grant SAB2000-0034, and by the U.S.D.O.E. Contract W-31-109-ENG-38,
and benefited from the allocation of supercomputer time at the NERSC and
the Northeastern University Advanced Scientific Computation Center
(NU-ASCC).  Part of this work was done while I was on sabbatical at the 
Instituto de Ciencia de Materiales de Madrid, CSIC, Cantoblanco, E-28049 
Madrid, Spain.  I thank my hosts, Maria Vozmediano and Paco Guinea, for a
very stimulating visit, for numerous discussions, and for correcting
an error in the original calculation.

I thank Walter Harrison for stimulating conversations on calculating the
interlayer coupling, and Martin Greven and Antonio Castro-Neto for useful 
comments on the magnetic properties, and A.-M.S. Tremblay for a preprint
of his work.

\appendix
\section{Three Band Model}

A major simplification of the present calculation is to treat the cuprates in a
one-band model.  This is consistent with the Zhang-Rice picture\cite{ZR}, 
although the approximation is less drastic for electron doping, since the upper
Hubbard band is already predominantly copper-like.  Nevertheless, the model also
describes the doping dependence of the `lower Hubbard band', which is really a
charge transfer, predominantly oxygen-like band.  Here an explanation for why 
this simplification works is suggested.  

\begin{figure}
\leavevmode
   \epsfxsize=0.33\textwidth\epsfbox{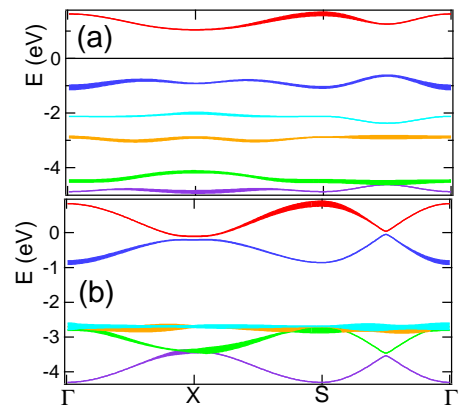}
\vskip0.5cm 
\caption{Dispersion of all six bands in three-band model, assuming
$m_Q$ = 0.3 (a) and 0.01 (b).} 
\label{fig:5}
\end{figure}

Even without carrying out self-consistent calculations, the nature of the
Mott transition can be understood by introducing a doping dependent gap.  
The energy bands can be calculated from the hamiltonian matrix
\begin{eqnarray}
H=\sum_j\Delta d^{\dagger}_jd_j
+\sum_{<i,j>}t_{CuO}[d^{\dagger}_jp_i+(c.c.)] \nonumber \\
+\sum_{<j,j'>}t_{OO}[p^{\dagger}_{j}p_{j'}+(c.c.)]
\nonumber \\
+Un_{j\uparrow}n_{j\downarrow}
+U_pn_{i\uparrow}n_{i\downarrow}\bigr),
\label{eq:100}
\end{eqnarray}
where $\Delta$ is the difference in on-site energy between copper and oxygen,
$t_{CuO}$ is the copper-oxygen hopping parameter, $t_{OO}$ the oxygen-oxygen 
hopping parameter and $U$ ($U_p$) the Hubbard interaction parameter on Cu 
(O). For good agreement with the doping dependence of the one band model,
it is necessary to properly incorporate the Hartree correction to the self
energy, $\Delta =\Delta_0+\Sigma_H$, $\Sigma_H=Un_{\downarrow}$ (for up 
spins), and $n_{\downarrow}=n/2-m_Q$, with $n$ the average electron
energy.  The resulting dispersions are shown in Fig.~\ref{fig:7en} for the
antibonding bands, and Fig.~\ref{fig:5} for the full dispersion.  In these
figures, the following parameters are assumed: $t_{CuO}=0.8eV$,
$t_{OO}=-0.4eV$, $\Delta_0=0$, $U=6eV$, and $U_p=3.75eV$. 

\begin{figure}
\leavevmode
   \epsfxsize=0.33\textwidth\epsfbox{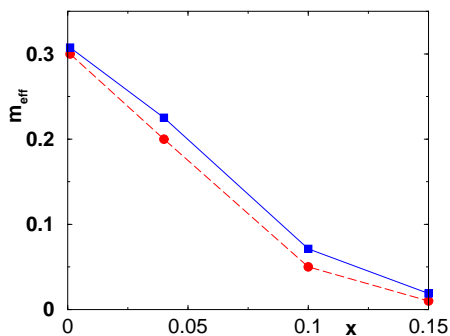}
\vskip0.5cm 
\caption{Effective magnetization $m_{eff}=mU/6t$ for the three-band
(circles) and one-band (squares) models.  The one-band result has been
multiplied by 3/4 to better agree with the three-band results.}
\label{fig:5d}
\end{figure}

The band dispersion is extremely similar to that found in the one band model, 
Fig.~\ref{fig:7em}, even though the lower band crosses over from 
the Zhang-Rice (hybridized copper-oxygen band) at half filling to a more copper 
like lower Hubbard band with increasing electron doping.  In addition, the 
effective magnetizations are proportional, Fig.~\ref{fig:5d}, although 
the one-band model overestimates the magnetization by $~1/3$.  This can be 
understood: in the three-band model, the shape of the Hubbard bands is fixed by
the combined effects of the magnetic instability and hybridization with the
oxygen band.  In the one band model, only the former effect is present, 
necessitating a larger value of $m$ to produce the same net splitting.

This remarkable agreement between one and three-band models goes well beyond the
Zhang-Rice model.  That model is restricted to the LHB in a small range of
doping near half filling; the present results compare both LHB and UHB over the
full range of electron doping. The result is nontrivial -- in the three band 
model, the bonding and non-bonding bands are also split into upper and lower 
Hubbard bands.  This degree of agreement comes about because the parameter 
$\Delta$ includes a large contribution from the magnetic Hartree term.  In
turn, this suggests that in the absence of magnetic effects 
the Cu and O energies are nearly degenerate -- as found in early LDA band 
structure calculations (see discussion in Ref.~\onlinecite{Hyb}).

\section{Charge Susceptibility and $U_{eff}$}

\begin{figure}
\leavevmode
   \epsfxsize=0.33\textwidth\epsfbox{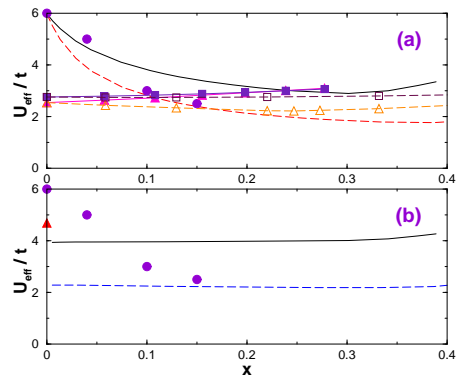}
\vskip0.5cm
\caption{Calculated $U_{eff}$ assuming (a) simple screening or (b) full vertex
correction of Chen, et al. [\protect\onlinecite{ChAT}].  In both cases, a bare 
$U=6.75t$ was assumed.  Solid lines = electron doping; long dashed lines = hole 
doping; triangles (squares) in a = paramagnetic screening of $U$, at $T$ =
1K (2000K); triangle in b = undoped; circles = data of 
Ref.~[\protect\onlinecite{nparm}].}
\label{fig:111}
\end{figure}

The proper choice of vertex corrections is an unresolved issue in the
analysis of the Hubbard model.  It is known to be of critical importance
for generating a pseudogap\cite{MVT}.  Here, by comparing simple
mean-field and SCR models to experiment, it is shown that the net effect
of vertex corrections is to make the coupling $U$ effectively doping (and
possibly temperature) dependent.  Kanamori\cite{Kana} 
showed that the effective Hubbard $U$ should decrease with doping, as an 
electron can hop around, and hence avoid, a second electron.  In the limit of a 
nearly empty (or full) band, this should lead to a correction of the form $U_
{eff}\sim U/(1+U/W)$, where $W=8t$ is the bandwidth.  It was 
found\cite{ChAT,BSW2} that Monte Carlo calculations of the susceptibility of a 
doped Mott insulator were approximately equal to the RPA susceptibility with 
suitable $U_{eff}$, and Chen, et al.\cite{ChAT} suggested the explicit form 
$U_{eff}=U/(1+<P>U)$, with P given by a vertex correction to the susceptibility 
and $<\cdot\cdot\cdot >$ an average over $\vec q$, at zero frequency.  
Figure~\ref{fig:111}b presents a calculation for $U_{eff}$ based on Chen, et al.
However, whereas Chen, et al. performed the average in the paramagnetic
phase, using bare Green's functions, here the dressed Green's functions
appropriate to the N\'eel phase are used, to approximately incorporate the
effect of this gap.  This makes little difference, since $P$ is dominated
by the intraband terms, and remains finite at half filling. Explicitly,
\begin{equation}
P=-{1\over N}\sum_{i,j,k}\hat U_{i,j}(k,k+q)\tilde F_{i,j}(k,k+q),
\label{eq:13a}
\end{equation}
\begin{equation}
\tilde F_{i,j}(k,k')={1-f_k^i-f_{k'}^j\over E_i(\vec k)+E_j(\vec k')-\omega-i
\delta},
\label{eq:13b}
\end{equation}
\begin{equation}
E_{\pm}(\vec k)={1\over 2}(\epsilon_k+\epsilon_{k+q}\pm E_0),
\label{eq:13c}
\end{equation}
\begin{equation}
E_0=\sqrt{(\epsilon_k-\epsilon_{k+q})^2+4\Delta^2},
\label{eq:13d}
\end{equation}
\begin{equation}
\hat U_{i,j}(k,k')={1\over 4}(1+iA_k)(1+jA_{k'})+ijB_kB_{k'},
\label{eq:13e}
\end{equation}
with $i,j$ summed over $+,-$, $\Delta$ the AFM gap,  
and $A_k=(\epsilon_k-\epsilon_{k+Q})/E_{0k}$, $B_k=\Delta /E_{0k}$.
In agreement with Chen, et al., the calculation finds $U$ to be
renormalized by a factor of $~2$ at finite doping, but does not recover a 
large $U$ near half filling, although different results are found
depending on whether $x=0$ from the start (triangle) or whether $x\rightarrow 0$
from the hole or electron doping sides.  

For modelling purposes, it is useful to have a $U_{eff}$ which evolves
smoothly from a large value at half filling to a reduced, Kanemori value
at finite doping.  A simple toy model consists of taking the RPA
screening of a charge response.  There should be a close connection
between the Kanemori mechanism and screening.  Screening involves creation
of a correlation hole about a given charge, while Kanemori's $U_{eff}$
involves the ability of a second charge to move around the first, while
avoiding double occupancy.  Near half filling, the second charge must move
in the correlation hole.  Approximating\cite{KLBM} the vertex correction
by the RPA screening of the charge susceptibility,
\begin{equation}
U_{eff}={U\over 1+<\chi >U},
\label{eq:12c} 
\end{equation}
it is possible to reproduce\cite{KLBM} the experimentally observed\cite{nparm} 
doping dependence, while matching the calculation of Chen, et al. away
from half filling, Fig.~\ref{fig:111}a.  

In this calculation, issues of self-consistency are also important.  To
minimize screening at half filling, it is necessary to reproduce the gap
in the susceptibility.  Hence, the susceptibility in Eq.~\ref{eq:12c} is
approximated by the charge susceptibility in the AFM state, $\bar
\chi^{00}_0$ from Eq. 2.24 of Ref.~\onlinecite{SWZ}, evaluated with the
bare $U=6.75t$.  (In principle, at finite doping there is a coupling to
the longitudinal magnetic susceptibility\cite{ChuF}, but this is neglected
for simplicity.)  The importance of using the AFM susceptibility is
illustrated in Fig.~\ref{fig:111}a: the solid and dashed lines show
$U_{eff}$ calculated using the charge susceptibility in the N\'eel state,
while the corresponding lines with triangles use the paramagnetic
susceptibility at low $T$.  The latter calculation finds a nearly doping
independent, but small $U_{eff}$; the former reproduces a large, weakly
screened $U$ near half filling.  Such a difference is expected in terms of
screening: when there is no gap at half filling, the enhanced susceptibility 
should be better able to screen $U$, resulting in a smaller $U_{eff}$.
This suggests that $U_{eff}$ should have an important {\it temperature
dependence} as the gap decreases -- which in turn will cause the gap to
close at a lower temperature.  Figure~\ref{fig:111}a also shows that there
is a weak temperature dependence of the screening.  The calculations
suggest that the large values of $U$ found in the cuprates are
characteristic mainly of the half filled regime and relatively low
temperatures.
A similar but larger screening effect was recently reported by Esirgen, et
al.\cite{Esi}.  

This procedure is still not fully self consistent.  If there is a
large difference between the bare $U$ and the screened $U_{eff}$, the gap
in $\chi$ should depend on the actual $U_{eff}$.  However, since
$U_{eff}\simeq U$ at half filling, any simple improvement will not
significantly change the overall doping dependence.  This is the same
kind of lack of self-consistency found for the SCR approach, and will 
be here neglected.



\section{Improved Solution of SCR Equation}

Approximating $coth(x)=max(1/x,1)$, and introducing the notation $\bar
Aq_c^2=Aq_c^2+\delta$, $\bar a_q=\bar Aq_c^2/\alpha_{\omega}$, and $t=2TC$, 
the solution to Eq.~\protect\ref{eq:B30} becomes
\begin{equation}
\delta -\delta_0={3ua^2\over\pi^2AC}\bigl[F_1+F_2\bigr],
\label{eq:B34b}
\end{equation}
with
\begin{eqnarray}
F_1=\int_{\delta}^{\delta+Aq_c^2}dy\int_{t}^{\alpha_{\omega}}dx{x\over
x^2+y^2}=
\nonumber \\
{\bar Aq_c^2\over 2}ln[{1+\bar a_q^2\over\bar a_q^2+(t/\alpha_{\omega})^2}]+
\nonumber \\
\alpha_{\omega}tan^{-1}(\bar a_q)
-{\delta\over 2}ln[{\delta^2+\alpha_{\omega}^2\over \delta^2+t^2}]-\alpha_
{\omega}tan^{-1}({\delta\over \alpha_{\omega}})
\label{eq:B34bi}
\end{eqnarray}
\begin{eqnarray}
F_2=t\int_{\delta}^{\delta+Aq_c^2}dy\int_0^t{dx\over x^2+y^2}=
\nonumber \\
=t\int_{\delta}^{\delta+Aq_c^2}{dy\over y}tan^{-1}({t\over y})=
\nonumber \\
=t[I_1({t\over\bar Aq_c^2})-I_1({t\over\delta})\bigr],
\label{eq:B34bii}
\end{eqnarray}
with 
\begin{equation}
I_1(x)=I_0(tan^{-1}(x))-tan^{-1}(x)ln(x),
\label{eq:B34c}
\end{equation}
\begin{equation}
I_0(x)=\int_0^xln(\tan{\theta})d\theta =L(x)+L({\pi\over 2}-x)-L({\pi\over 2}),
\label{eq:B34c1}
\end{equation}
and $L(x)=-\int_0^xln(\cos{t})dt$ is the Lobachevskiy function\cite{GRyd}.

For most purposes, it can be assumed that $\delta
<<t<<Aq_c^2,\alpha_{\omega}$, in which
case $I_0(tan^{-1}(x))=\theta (\ln{(\theta)}-1)$, with $\theta =
min\{x,1/x\}$, and then $F_2$, Eq.~\ref{eq:B34bii}, simplifies.  
\begin{eqnarray}
F_2=ln({t\over\delta})[\delta +ttan^{-1}({t\over\delta})]+\delta
-{t^2\over Aq_c^2}
\nonumber \\
\simeq {\pi\over 2}t \ln({t\over\delta}).
\label{eq:B34bx}
\end{eqnarray}
Defining 
$Z=1+(3ua^2/\pi^2AC)ln(\alpha_{\omega}/t)$, then 
\begin{equation}
Z\delta -\bar\delta_0={3ua^2T\over\pi A}ln({2CT\over \delta}),
\label{eq:B34e}
\end{equation}
which agrees with Eq.~\ref{eq:B34h} when $Z\rightarrow 1$.

\section{Parameter Evaluations}
                                         
At $T=0$, the imaginary part
of the susceptibility $\chi (\vec Q,\omega )$ can be calculated
analytically:
\begin{eqnarray}
Im(\chi (\vec Q,\omega ))=\sum_{\vec k}(f(\epsilon_{\vec k})-
f(\epsilon_{\vec k+\vec Q}))\delta (\epsilon_{\vec k+\vec Q}-\epsilon_{\vec k}
-\omega)
\nonumber \\
={F(\theta_1,\tilde k)-F(\theta_2,\tilde k)\over 4t},
\label{eq:C8aa}
\end{eqnarray}
where $F(\theta ,x)$ is an elliptic integral, $\tilde k=\sqrt{1-(\omega /8t)^2}
$, and $sin(\theta_i)=sin(\phi_i)/\tilde k$, with
\begin{equation}
\cos^2{(\phi_1)}=\cases{c^2_-&if $\omega\le\omega_c^-$\cr
                        \hat\omega /2&if $\omega >\omega_c^-$\cr},
\label{eq:C8bb}
\end{equation}
\begin{equation}
\cos^2{(\phi_2)}=\cases{c^2_+&if $\omega\le\omega_0$\cr
                        1&if $\omega >\omega_0$\cr},
\label{eq:C8cc}
\end{equation}
with $\hat\mu =\mu /2t$, $\hat\omega =\omega /4t$, $c^2_{\pm}=a_{\pm}+\sqrt{a_
{\pm}^2-\hat\omega^2}$, and $a_{\pm}=1-(\hat\mu\pm\hat\omega )/\tau$.
Similar results for $t'=0$ are discussed in Ref.~\onlinecite{BCT}.
The real part $Re\chi$ can be found from the Kramers-Kronig result,
\begin{equation}
Re\chi (\vec Q,\omega)={1\over\pi }\int_0^{\infty}{Im\chi (\vec Q,\omega')
\omega'd\omega'\over\omega^{'2}-\omega^2}.
\label{eq:C8g}
\end{equation}

\subsection{$A$ at the C-point}

\begin{figure}
\leavevmode
   \epsfxsize=0.33\textwidth\epsfbox{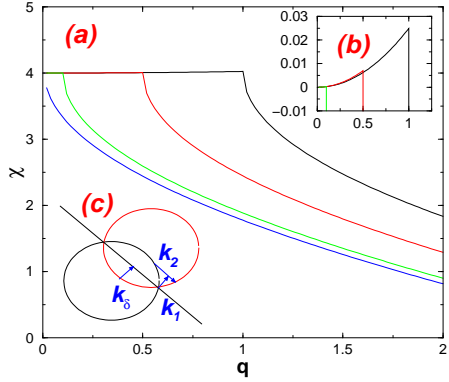}
\vskip0.5cm
\caption{(a) Calculated susceptibility $\chi (q)$ for several values of 
overlap $\delta$.  (b) Blowup of plateau region, for $\chi_q-\chi_{q=0}$. (c) 
Model of Fermi surfaces, defining $\delta$, $k_1$ ($k_{\perp}$) and $k_2$ ($k_
{\parallel}$).}
\label{fig:11b}
\end{figure}

To understand the $q$-plateau, and in particular the C-point, where the
plateau width shrinks to zero, it is convenient to introduce
a simplified model\cite{GMV}, for which the $q$-dependence of $\chi$ can
be calculated  {\it analytically}.  While the dashed lines in
Fig.~\ref{fig:44a} represented an $\omega$ shift, they can equally well
describe the $q$-shift of the energy denominator, Eq.~\ref{eq:0a}.  The
plateau edge corresponds to the point where the dashed line intersects the
q-shifted FS (horizontal arrows).  (Recall that $\vec q=\vec Q+\vec q'$.)
In the simplified model, the energy
denominator is linearized, so $\Delta\epsilon\propto k_{\perp}$, independent of 
$k_{\parallel}$.  Chosing $\vec q$ to point along the $(\pi ,\pi )$ direction, 
the FS can be approximated by two circles of radius $k_F$, centered at
$(\pi ,\pi )$ and $(-\pi ,-\pi )$ (for this choice of $\vec q$ the other
two circles at $(\pi ,- \pi )$ and $(-\pi ,\pi )$ can be ignored).  The
Q-shifted FS is then a circle centered at $\Gamma =(0,0)$.  The FS at
$(\pi ,\pi )$ and the Q-shifted FS are illustrated in Fig.~\ref{fig:11b}c.
To keep the picture symmetrical, both FSs are shifted (in opposite
directions) by $q'/2$ when $q'\ne 0$.

Adding the contributions of the overlap of the q-shifted FS with both the
FS
at $(\pi ,\pi )$ and the one at $(-\pi ,-\pi )$, $\chi_q\propto I_{k_\delta+
q'/2}+I_{k_\delta-q'/2}$, with 
\begin{equation}
I=\int_0^{k_c}{dk_{\perp}dk_{\parallel}\over k_{\perp}},
\label{eq:C6c}
\end{equation}
where the region of integration is over the part of the upper FS in 
Fig.~\ref{fig:11b}c not overlapped by the lower (q-shifted) FS, and $k_{\perp}$
ranges from zero at the apex of the wedge to the middle of the upper FS, $k_c=
k_F-k_\delta$, where $k_\delta$ is the overlap parameter defined in
Fig.~\ref{fig:11b}c.  To lowest order, for $k_\delta <<k_F$, 
\begin{eqnarray}
I=2k_F+\sqrt{k_Fk_\delta}\ln|\frac{1-\beta}{1+\beta}|
\label{eq:C6b}
\end{eqnarray}
with $\beta =\sqrt{k_\delta/k_F}$.
The expression for $I_{k_\delta-q'/2}$ must be modified when $q'>2k_\delta$ and 
the two FSs no longer overlap\cite{GMV}: $I_{k_\delta-q'/2}=2k_F[1-\gamma 
\tan^{-1}1/\gamma]$, with $\gamma =\sqrt{(q'-2k_\delta )/2k_F}$.  
The calculated susceptibilities, Fig.~\ref{fig:11b}a, display the flat topped 
plateaus with weak positive curvature ($A<0$, Fig.~\ref{fig:11b}b).  At the 
plateau edge the susceptibility falls sharply, $\chi\sim 1-\pi\gamma
/2\sim\sqrt{q'}$.  The C-point corresponds to $k_\delta =0$.

\subsection{$C$ and Plateaus in Frequency}

\begin{figure}
\leavevmode
   \epsfxsize=0.33\textwidth\epsfbox{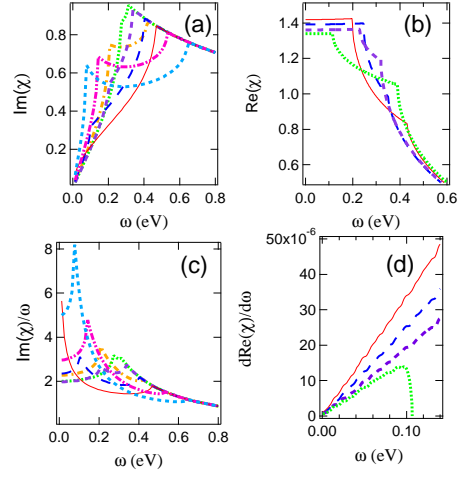}
\vskip0.5cm
\caption{(a) $Im\chi (\vec Q,\omega)$, (b) $Re\chi (\vec Q,\omega)$, (c) 
$Im\chi /\omega\equiv\hat C$, and (d) $dRe\chi (\vec Q,\omega)/d\omega$,
for (a,c): $\mu$ = 0 (solid line), -0.05 (long dashed line), -0.10
(dashed line), -0.15 (dotted line), -0.20 (dot-dashed line), -0.25
(dot-dot-dashed line), and -0.30eV (short dashed line); (b,d): $x$ = 0
(solid line), 0.04 (long dashed line), 0.10 (dashed line), and 0.15
(dotted line).}
\label{fig:44b}
\end{figure}

Figure~\ref{fig:44b} illustrates $Im\chi (\vec Q,\omega)$, $Re\chi (\vec Q,
\omega)$, and $Im\chi /\omega\equiv\hat C$.  While plateaus in $Re(\chi_0)$ 
have been noted above, Fig.~\ref{fig:0a}c, here the main interest
lies in $C=UIm\chi /\omega$.  This linear-in-frequency contribution to
$Im\chi$, generated by hot spots, is an important parameter in SCR and 
NAFL theories, and has been well studied.  The height of the plateau at
zero frequency $C=U\hat C(\omega =0)$ can be represented as a frequency
$\omega_1=1/C$, which can be found explicitly\cite{SCS}
\begin{equation}
C={1\over 2\pi Js_{x0}^2(1+\tau c_{x0})}={1\over\omega_1}
\label{eq:C8}
\end{equation}                                                          
(with $J=4t^2/U$, $s_{x0}^2=1-c_{x0}^2$). 

\begin{figure}
\leavevmode
   \epsfxsize=0.33\textwidth\epsfbox{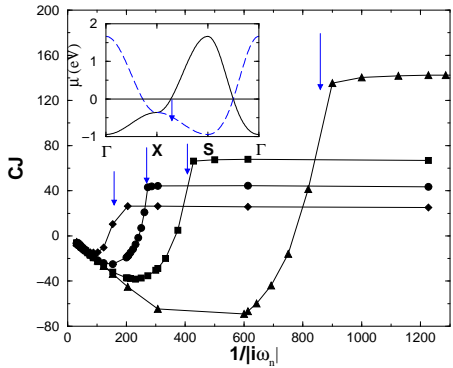}
\vskip0.5cm
\caption{$\hat C$ calculated for several values of $\mu$: $\mu$ =
-0.355 (diamonds), -0.357 (circles), -0.358 (squares), -0.359eV (triangles) 
[$\mu_v$ = -0.3599eV].  Inset: Band dispersion $\epsilon_{\vec k}$ (solid line) 
$\epsilon_{\vec k+\vec Q}$ (dashed line), for $\mu =0$.  Arrow = $\omega_c^-$.}
\label{fig:11}
\end{figure}

\begin{figure}
\leavevmode
   \epsfxsize=0.33\textwidth\epsfbox{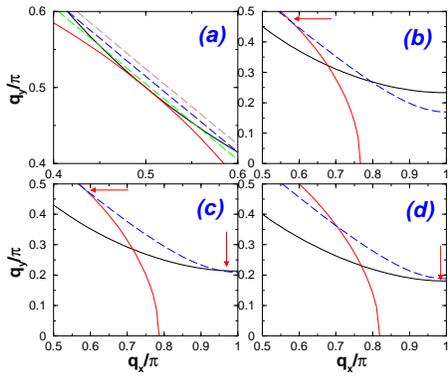}
\vskip0.5cm
\caption{Origins of critical cutoffs.  Thick solid line = FS; thin solid line =
Q-shifted FS; dashed lines = Eq.~\protect\ref{eq:C8d}, for several values of 
$\omega$.  Chemical potential $\mu$ = (a) 0, (b) -0.1, (c) -0.14, (d) -0.2eV.
horizontal arrows indicate $\omega_0$, vertical arrows $\omega_c^-$.}
\label{fig:44a}
\end{figure}

However, it is important to note that $C$ also approximates a plateau,
particularly near the H-point, Fig.~\ref{fig:11}, with a well-defined 
cutoff.  Moreover, the width of this plateau vanishes near {\it both} the
H- and C-points, controlled by two characteristic frequencies, $\omega_c^-$, 
Eq.~\ref{eq:00e}, and 
\begin{equation}
\omega_0={8t\over\tau}[\sqrt{1-\hat\mu\tau}-1],
\label{eq:C8e}
\end{equation}
respectively.  The origin of these critical frequencies 
can be understood from Fig.~\ref{fig:44a}.  The thick (thin) solid lines 
represent the original (Q-shifted) Fermi surfaces, while the dashed lines
represent
\begin{equation}
\omega =\epsilon_{\vec k+\vec Q}-\epsilon_{\vec k},
\label{eq:C8d}
\end{equation}
for various values of $\omega$.  Equation~\ref{eq:C8d} gives the points at which
the denominator of $\chi_0(\vec Q,\omega)$, Eq.~\ref{eq:0a}, vanishes. Thus at
$T=0$, $Im(\chi_0(\vec Q,\omega))$ is proportional to the length of the dashed
line lying between the original and Q-shifted FSs (i.e., where $f(\epsilon_{\vec
k})-f(\epsilon_{\vec k+\vec q})=\pm 1$).  Since the two FSs meet at an
angle, forming a wedge, $Im(\chi_0(\vec Q,\omega ))\sim\omega$. 

From Fig.~\ref{fig:44a}, the critical frequencies (denoted by arrows) are
points where the $\omega$ dependence of this length changes abruptly, leading 
to a sharp change in $Im\chi$.  Thus, near the H-point, the plateau width
is $\omega_c^-$ (inset, Fig.~\ref{fig:11}), while near the C-point it is
$\omega_0$. The vertical arrows in Fig.~\ref{fig:44a} indicate
$\omega_c^-$, where the dashed line 
(Eq.~\ref{eq:C8d}) intersects the FS at the zone boundary, while the horizontal 
arrows are\cite{BCT} $\omega_0$, where the dashed line ceases to intersect the 
Q-shifted FS.  There is a crossover at $\mu_c\simeq -0.14eV$: for $\mu >\mu_c$, 
$\omega_0=\omega_c^-$ while for $\mu <\mu_c$, $\omega_0=\omega_c^-$. Combining 
Eqs.~\ref{eq:00e},\ref{eq:C8e}, $\omega_0=\omega_c^-$ at $\mu_c=[1-z(2-\sqrt{z}
)^2]2t/\tau =-0.1384eV$, with $z=1-\tau$.  For $\omega >min\{\omega_0,\omega_c^
-\}$, $Im(\chi_0(\vec Q,\omega))\sim\omega^{1/2}$, so $C\sim 1/\omega^{1/2}$ 
-- i.e., the susceptibility is no longer on the plateau. 

Defining a width parameter $\alpha_{\omega}=min\{\alpha_{\omega}^-,\alpha_
{\omega}^0\}$, with $\alpha_{\omega}^0=\omega_0/\omega_1$, then
\begin{equation}
\omega_1/\omega_c^-={2\pi t(1-\tau )\over U}[{1+\tau c_{x0}\over -\tau}]
\equiv{1\over\alpha_{\omega}^-}.
\label{eq:C8b}
\end{equation}
This latter is in good agreement with the numerical results (arrows in 
Fig.~\ref{fig:44a}) and is similar to the result found by Onufrieva and 
Pfeuty\cite{OPfeut}, using a hyperbolic band approximation valid near a VHS, 
$\omega_1/\omega_c^-=2\pi t(1-\tau )/U$.  

Because of the dynamic scaling $\omega\sim q^z$, this crossover is also
reflected in the behavior on the plateau {\it in $\vec q$},
Fig.~\ref{fig:44}: for $\mu >-0.14eV$, the plateau has a negative
curvature, which can almost be scaled between different dopings, while for
$\mu <-0.14eV$, the plateau starts to fill in, ultimately developing a
peak at $\vec Q$.  (See also Fig. 3a in Ref.~\onlinecite{ICTP}.)
Note that the plateau width collapses in frequency at
both the $H$- and $C$-points, while the collapse in wave number
($q_c\rightarrow 0$) is only present near the $C$-point.

\subsection{$B$}

The parameter $B$ is small, and generally neglected.  However, it
enters into the evaluation of $u$, so will be discussed briefly.  The
expression for $B$ may be written exactly as the $\omega\rightarrow 0$
limit of
\begin{equation}
B=URe\sum_{\vec k}[{f(\epsilon_{\vec k})-f(\epsilon_{\vec k+\vec Q})\over
(\epsilon_{\vec k+\vec Q}-\epsilon_{\vec k})}]{1\over ((\epsilon_{\vec k+\vec 
Q}-\epsilon_{\vec k})^2-\omega^2}.
\label{eq:C6}
\end{equation}
It can be shown that $B$ has a logarithmic correction due to the hot
spots.  The integral can be approximately evaluated by (a) using symmetry
to reduce the integral to one over an octant of the Brillouin zone
containing one hot spot, (b) splitting the domain of integration into (i)
a circle of radius $k_c$ about the hot spot, and (ii) the remainder of the
domain, and (c) numerically evaluating the integral over domain (ii) while
providing an analytic approximation to that over (i).  
Then the $k$ integral over the hot spot circle can be written approximately as
\begin{eqnarray}
I=\int_0^{k_c}{(1-3\beta_{\theta}k)dk\over \alpha_{\theta}^2k^2-\omega^2}
\nonumber \\
\simeq {1\over\alpha_{\theta}^2}[{1\over k_c}-3\beta_{\theta}\log{\alpha_
{\theta}k_c\over\omega}].
\label{eq:C6a}
\end{eqnarray}
At $T=0$, the integral $I$ must then be integrated in $\theta$ over the wedge
where the difference in Fermi functions does not vanish.  
The integral from outside the hot spot circle will eliminate the 
$k_c$-dependence, but should not affect the $log(\omega )$ term.  

It is difficult to directly evaluate the two-dimensional principal value 
integral for $B$.  Instead, it is much simpler to evaluate $Re(\chi )$ via
Kramers-Kronig transformation of $Im(\chi )$ and find $B$ by numerical
differentiation.  When this is done, it is found that (a) $B$ is numerically
very small due to the plateau in $Re(\chi )$, Fig.~\ref{fig:44b}d, and (b)
the logarithmic correction is too small to determine accurately.

\subsection{$u$}

The quartic effective action is
\begin{eqnarray}
S={1\over 2}\sum_{\vec q,i\omega_n}\Pi_2(\vec q,i\omega_n)\phi(\vec q,i\omega_n)
\phi(-\vec q,-i\omega_n)
\nonumber \\
+{1\over 4(\beta N_0)^2}\sum{}^{'}\Pi_4(\vec q_i,i\omega_i)\phi(\vec q_1,i\omega
_1)\phi(\vec q_2,i\omega_2)\times
\nonumber \\
\times\phi(\vec q_3,i\omega_3)\phi(\vec q_4,i\omega_4),
\label{eq:B10}
\end{eqnarray}
where the prime in the second sum means summing over all $\vec q_i$, $\omega_i$,
such that $\sum_{i=1}^4\vec q_i=0$, $\sum_{i=1}^4\omega_i=0$, 
\begin{equation}
\Pi_2(\vec q,i\omega_n)={U\over 2}[1-U\chi_0(\vec q,i\omega_n)],
\label{eq:B11}
\end{equation}
\begin{eqnarray}
\Pi_4(\vec q_i,i\omega_n)={U^4\over 8}\sum_{\vec k,i\epsilon_n}
G_0(\vec k,i\epsilon_n)G_0(\vec k+\vec q_1,i\epsilon_n+i\omega_1)\times
\nonumber \\
\times G_0(\vec k+\vec q_1+\vec q_2,i\epsilon_n+i\omega_1+i\omega_2)
G_0(\vec k-\vec q_4,i\epsilon_n-i\omega_4),
\label{eq:B12}
\end{eqnarray}
with $u=\Pi_4/N_0\beta U^2$.

Since there is some controversy\cite{Mil,Chu} concerning $u$, it shall be
evaluated in detail.  Millis\cite{Mil} showed that for free electrons
(parabolic bands) this expression is in general well defined, but diverges
when $\vec Q$ is a `spanning' vector of the Fermi surface -- in the
present case, this would correspond to the H- and C-points.  Abanov, et
al.\cite{Chu} found a more severe divergence: $u$ diverges for all $\mu$
in the hot spot regime.  The problem lies in the limit of external frequencies 
$\rightarrow$ 0, momenta $\rightarrow$ 0 or $\vec Q$.  Taking this limit on the
momenta, the expression for $u$ can be written as 
\begin{eqnarray}
u={U^2\over N_0\beta}\sum_{\vec k,i\omega_n}{1
\over (\epsilon_{\vec k}-i\omega_n)(\epsilon_{\vec k}-i\omega_n+i\omega_4)}
\times
\nonumber \\
\times{1\over (\epsilon_{\vec k+\vec Q}-i\omega_n-i\omega_1)(\epsilon_{\vec k+
\vec Q}-i\omega_n-i\omega_1-i\omega_2)}.
\label{eq:C4}
\end{eqnarray}
The sum over Matsubara frequencies yields
\begin{eqnarray}
u=U^2\sum_{\vec k}[{f(\epsilon_{\vec k})\over
i\omega_4}\bigl({1\over (i\omega_3-\Delta\epsilon)(i\omega_3+i\omega_2-
\Delta\epsilon)}-
\nonumber \\
-{1\over (i\omega_1+\Delta\epsilon)(i\omega_1+i\omega_2+\Delta\epsilon)}\bigr)
\nonumber \\
+{f(\epsilon_{\vec k+\vec Q})\over i\omega_2}\bigl({1\over (i\omega_3-\Delta
\epsilon)(i\omega_1+i\omega_2+\Delta\epsilon)}-
\nonumber \\
-{1\over (i\omega_1+\Delta\epsilon )(i\omega_3+i\omega_2-\Delta\epsilon)}\bigr)
],
\label{eq:C4a}
\end{eqnarray}                                                                
where $\Delta\epsilon=\epsilon_{\vec k}-\epsilon_{\vec k+\vec Q}$.  Letting 
$\omega_{i,\pm}=(\omega_i\pm\omega_{i+2})/2$ ($i=1,2$), and noting that $\omega_
{1+}=-\omega_{2+}$, this simplifies to 
\begin{equation}
u=2U^2\sum_{\vec k}{(f(\epsilon_{\vec k+\vec Q})-f(\epsilon_{\vec k}))W_-
\over (W_-^2+\omega_{1+}^2)(W_-^2+\omega_{2-}^2)},
\label{eq:C4b}
\end{equation}
where
\begin{equation}
W_-=(i\omega_{1-}+\Delta\epsilon).
\label{eq:C4c}
\end{equation}
Thus in Matsubara frequency space, $u$ is largest for $\omega_{1+}=
\omega_{2-}=0$, so it should indeed be reasonable to estimate it in that limit:
\begin{equation}
u(i\omega_1,0,0)=U^2{\partial^2\over\partial (i\omega_1)^2} \sum_{\vec k}{f(
\epsilon_{\vec k+\vec Q})-f(\epsilon_{\vec k})\over i\omega_1+\Delta\epsilon}.
\label{eq:C4d}
\end{equation}
In turn, it should be possible to approximate $u$, Eq.~\ref{eq:C4d}, by
its $\omega_1\rightarrow 0$ limit, if this is nonsingular.  From
Eq.~\ref{eq:B18}, $U\chi_0(\vec Q,\omega )=B\omega^2
+iC\omega +1-\delta_0$. 
Thus, the analytic continuation $i\omega_1\rightarrow \omega+i\delta$ yields
\begin{equation}
u(0,0,0)=U^2\lim_{\omega\rightarrow 0}{\partial^2\chi_0(\vec Q,\omega )\over
\partial \omega^2}\simeq 
2BU.
\label{eq:C4e}
\end{equation}
Due to the plateau in $\chi (\vec Q,\omega )$, $B$ (Table I) and hence $u$ are
extremely small.  The smallness of $u$ is true only in the limit that all
external frequencies are small, which means that a more complicated expression
should be used to evaluate $u$.  Moreover, there is an additional problem: 
as found above, $B$ has a correction in $\ln{(\omega )}$, which would formally 
be divergent.  Hence, the model is not fully self-consistent, and $u$ will
be treated as an empirical parameter.  The weak logarithmic divergence
will be neglected, and $u$ approximated by a constant.

\section{Interlayer Coupling}

\subsection{Dispersion of $t_z$: Direct and Staggered Stacking}

Andersen, et al.\cite{ALJP} demonstrated that the anomalous form of
interlayer hopping in the cuprates, $t_z=t_{z0}(c_x-c_y)^2$, could be
understood by coupling the Cu$_{d_x^2-d_y^2}$ and O$_p$ orbitals to the
Cu$_{4s}$ orbitals, which have significant interlayer coupling.  Here, I
provide a simplified calculation including only these orbitals, and show
how the dispersion is modified by staggered stacking of the CuO$_2$
layers.  For uniform stacking (Cu above Cu), the hopping matrix becomes
\begin{equation}
H=\left(\matrix{\Delta&-2ts_x&2ts_y&0\cr
                -2ts_x&0&0&-2t_{ps}s_x\cr
                 2ts_y&0&0&-2t_{ps}s_y\cr 
                 0&-2t_{ps}s_x&-2t_{ps}s_y&\Delta_s+E_{sz}\cr},\right)
\label{eq:D1}
\end{equation}
with $s_i=\sin{k_ia/2}$.  
Here the first (last) row is for the Cu$_{d_x^2-d_y^2}$ (Cu$_{4s}$)
orbital, and the middle rows are for the O$_{px}$ and O$_{py}$ orbitals, 
with $E_{sz}=-4t_{sz}\cos{k_zc}$.  In the limit $\Delta_s+E_{sz}>>\Delta>>
t,t_{ps}$, the antibonding band has dispersion
\begin{equation}
E=\Delta -{2t^2\over\Delta}(c_x+c_y-2)-{4t^2t^2_{ps}\over\Delta^2(\Delta_s
+E_{sz})}(c_x-c_y)^2,
\label{eq:D2}
\end{equation}
so if $t_{sz}<<\Delta_s$, the interlayer hopping has the form
$t_{z0}\cos{k_zc}(c_x-c_y)^2$, with $t_{z0}=-16t^2t^2_{ps}t_{sz}/
\Delta^2\Delta_s^2$.  
While this form had been suggested earlier\cite{CAnd} and found
experimentally for the bilayer splitting in BSCCO\cite{Blay}, it should be
noted that it is only
approximate, and that, at least in YBCO, there is considerable splitting
of the bilayer bands along the zone diagonal\cite{ALJP}.  Nevertheless,
this form is adequate for the present purposes.

When successive layers are staggered, the only modification to the hopping
matrix is in the form of $E_s(k_z)$, which now acquires an in-plane
dispersion,
\begin{eqnarray}
E_s(k_z)=-4t_{sz}\cos{k_zc}[\cos{(k_x+k_y)a/2}+\cos{(k_x-k_y)a/2}]
\nonumber \\  
=-8t_{sz}\cos{k_zc}\cos{k_xa/2}\cos{k_ya/2},
\label{eq:D3}
\end{eqnarray}
which leads to Eq.~\ref{eq:n21}.

\subsection{Estimation of $t_z$ from Resistivity Anisotropy}

\begin{figure}
\leavevmode   
   \epsfxsize=0.33\textwidth\epsfbox{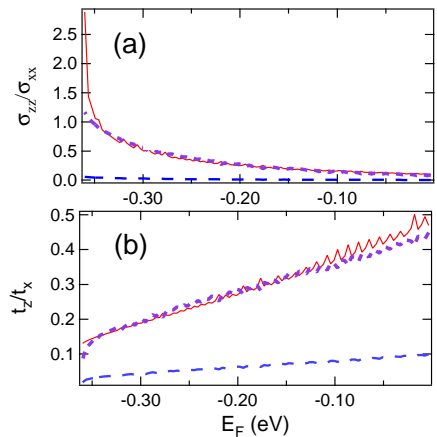}
\vskip0.5cm
\caption{(a) Normalized conductivity ratio,$\hat\sigma_{zz}/
\sigma_{xx}$ vs doping $E_F$, for uniform (solid line) and staggered
stacking (long dashed line and short dashed line, ($\times$20)); and (b) 
resulting normalized interlayer hopping $\hat t_{z0}$ for staggered (solid
line) and uniform stacking (long dashed line and short dashed line,
($\times$4.5).}
\label{fig:nD0}
\end{figure}

The dc conductivity can be estimated 
\begin{equation}
\sigma_{ii}={2e^2\over\Omega}\sum_{\vec k}v_i^2\delta (\epsilon_{\vec
k}-\mu )\tau_{\vec k},
\label{eq:D2a}
\end{equation}
$i=x,y,z$, with $\Omega$ the unit cell volume, $v_i=\hbar^{-1}d 
\epsilon_{\vec k}/dk_i$, and $\tau_{\vec k}$ the scattering rate.  Recent
ARPES data suggest that, when bilayer splitting is resolved, $\tau_{\vec
k}$ is relatively isotropic over the Fermi surface\cite{PashAli}.  Taking
$\tau_{\vec k}$ independent of $\vec k$, the conductivities are given by
integrals over the Fermi surface.  Figure~\ref{fig:nD0}a shows a
normalized conductivity ratio,
\begin{equation}
{\hat\sigma_{zz}\over\sigma_{xx}}={at^2\over ct_{z0}^2}{\sigma_{zz}
\over\sigma_{xx}},
\label{eq:D2b}
\end{equation}
while Fig.~\ref{fig:nD0}b shows the resulting normalized interlayer
hopping $\hat t_{z0}=t_{z0}\sqrt{c/a}$, which would be required to produce
a resistivity anisotropy $\rho_{zz}/\rho_{xx}=1000$.  For simplicity, it
is assumed that $t_{z0}$ is small, and $\hat\sigma_{zz}/\sigma_{xx}$ is
evaluated in the limit $t_{z0}\rightarrow 0$.  It can be seen that
(a) the staggered stacking reduces the conductivity by approximately a
factor of 20, independent of doping (except near the VHS), so (b) assuming
the resistivity anisotropy is 1000 for optimally doped LSCO, it is
estimated that $t_{z0}/t$ = 0.11 for staggered stacking; by contrast, if
the stacking had been uniform, a value of $t_{z0}/t$ = 0.025 would have
been required.  This calculation has recently been extended\cite{LSMB} 
to the bilayer compound Bi$_2$Sr$_2$CaCu$_2$O$_8$ (Bi2212), which has both
uniform and staggered stacking.  By comparison with a band structure
calculation, the approximate values $t_z/t$ = 0.38 for the
intracell hopping (corresponding to uniform stacking), and 0.14 for the
intercell (staggered) hopping were found.

\subsection{z-Component of Ordering Vector}

\begin{figure}
\leavevmode   
   \epsfxsize=0.33\textwidth\epsfbox{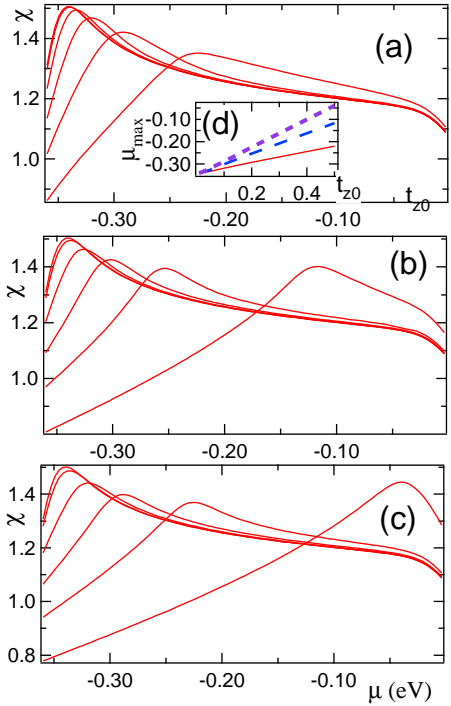}
\vskip0.5cm
\caption{$\chi_0(\vec Q,Q_z)$ at $T=100K$ vs chemical potential $\mu$, for
uniform stacking and $Q_z$ = $\pi$ (a), $\pi /2$ (b), and 0 (c).  The
various curves correspond to $t_{z0}/t$ = 0.01, 0.02, 0.05, 0.1, 0.2, and
0.5, with the peak in $\chi_0$ shifting to the right with increasing
$t_{z0}$.  Inset (d): position of peak, $\mu_{max}$, vs $t_{z0}$ for $Q_z$
= $\pi$ (solid line), $\pi /2$ (long dashed line), and 0 (short dashed
line).}
\label{fig:nD1}
\end{figure}

\begin{figure}
\leavevmode   
   \epsfxsize=0.33\textwidth\epsfbox{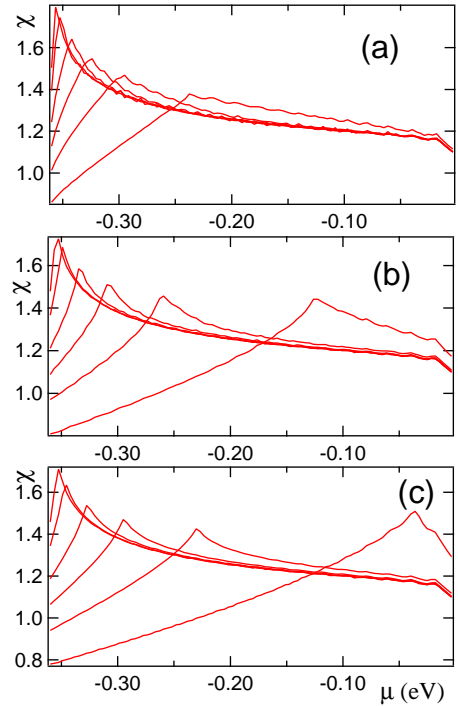}
\vskip0.5cm
\caption{$\chi_0(\vec Q,Q_z)$ vs chemical potential $\mu$, as in Fig.
\protect\ref{fig:nD1}, but at $T=10K$.}
\label{fig:nD2}
\end{figure}

Given a finite interlayer hopping $t_z$, the first issue is to identify
the three-dimensional ordering vector: what $Q_z$ minimizes the
free energy?  At mean field level, the initial magnetic instability will
be associated with the state for which the RPA denominator first diverges,
i.e., the state with the largest value of $Re\chi_0(\vec Q,Q_z)$.  (Note
that these calculations implicitly assume that the two-dimensional ground
state involves commensurate order at $\vec Q$.) For uniform stacking, a 
complicated dependence on doping, temperature, and $t_z$ is found. Figures
\ref{fig:nD1},\ref{fig:nD2} plot $\chi_0$ vs chemical potential for $T$ =
100K, 10K, respectively.  The shift of the susceptibility peak with doping
can readily be understood by comparison with Fig.~\ref{fig:0a}.  Both
temperature and interlayer coupling act to smear out the VHS, and in both
cases cause the susceptibility peak to shift to smaller chemical potential
(lower hole doping), Fig.~\ref{fig:nD1}d.  Note that the peak shifts at
different rates for different $Q_z$-values, showing that the band is
developing a considerable c-axis dispersion.  The fastest shift (short
dashed line in Fig.~\ref{fig:nD1}d, corresponding to $Q_z=0$) can thus be
considered as representing a crossover from quasi-two-dimensional to fully
three dimensional dispersion.


\begin{figure}
\leavevmode   
   \epsfxsize=0.33\textwidth\epsfbox{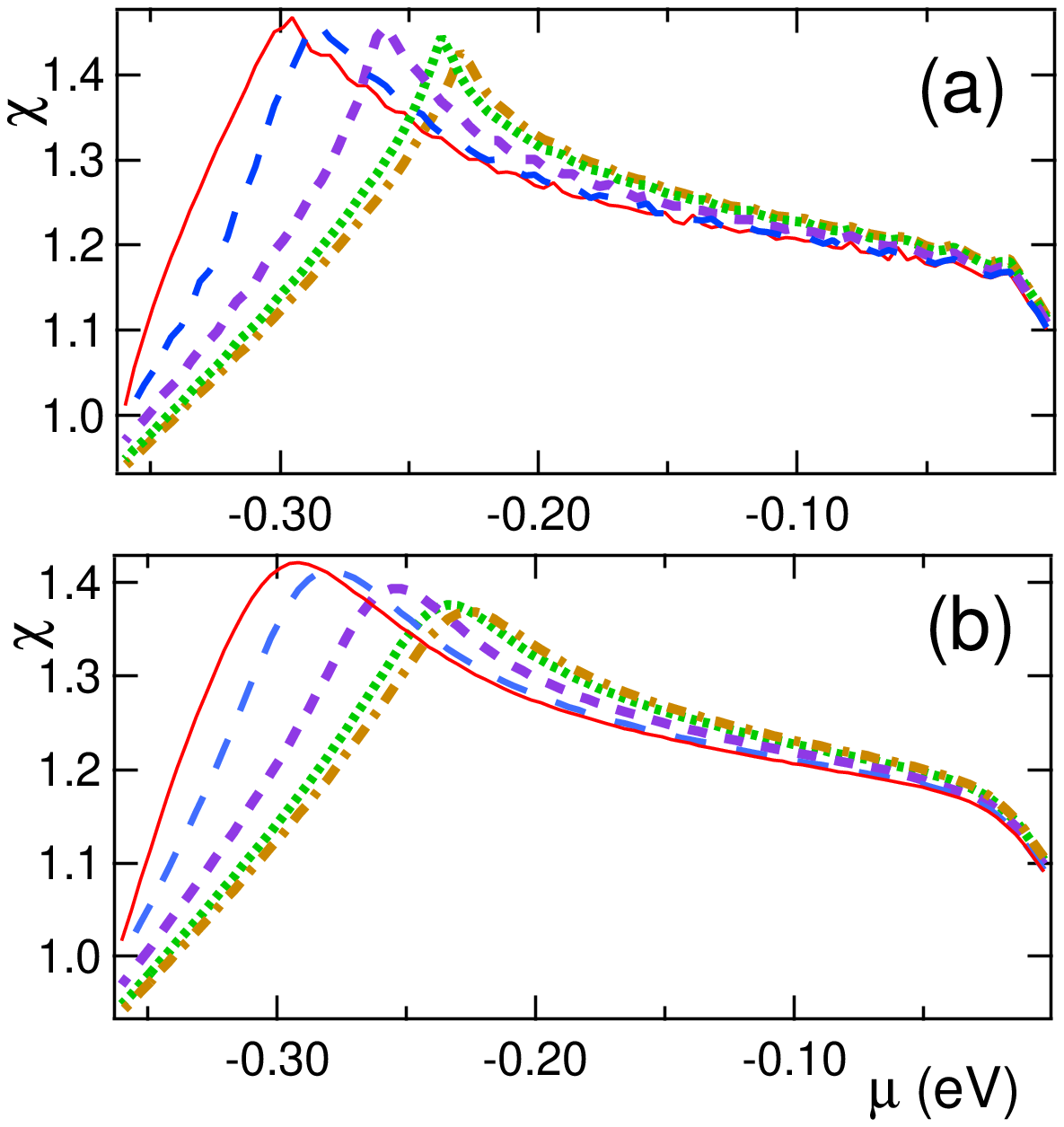}
\vskip0.5cm
\caption{$\chi_0(\vec Q,Q_z)$ vs chemical potential $\mu$, for uniform stacking 
and $t_{z0}=0.2t$, and $T=10K$ (a), or 100K (b), with $Q_z/\pi$ = 1 (solid
line), 0.75 (long dashed line), 0.5 (short dashed line), 0.25 (dotted
line), 0 (dot-dashed line).} 
\label{fig:nD3}
\end{figure}

This dispersive shift of the peak in $\chi_0$ leads to a doping dependence
of the optimal $Q_z$, as illustrated in Fig.~\ref{fig:nD3} for
$t_{z0}=0.2t$.  For large hole doping, near the $t_{z0}=0$ VHS, 
the susceptibility maximum corresponds to $Q_z=\pi /c$, while near the
susceptibility peak, the spin modulation becomes incommensurate
(intermediate values of $Q_z$ have the largest susceptibility).  There is
a rapid evolution of the optimal $Q_z$, and beyond the peak regime, over
essentially the entire electron-doped regime, the optimal $Q_z$ is $0$.
This same pattern is repeated for smaller $t_{z0}$, with only the region
of the susceptibility peak changing.  The results are essentially
independent of the sign of $t_z$.

\subsection{Calculation of $A_z$}

\subsubsection{Uniform Stacking}

Given $t_z$ and $Q_z$, the parameter $A_z$ of Eq.~\ref{eq:0B18} can be
evaluated: $U\chi (\vec Q+q_z\hat z,\omega =0)=U\chi (\vec Q+Q_z\hat z,0)+
A_z(q_z-Q_z)^2$.  The dominant ordering vectors, $Q_z=\pi /c$ and $Q_z=0$,
can be analyzed in more detail.  For the former choice, 
\begin{eqnarray}
A^{\pi}_z={Uc^2\over 4}\sum_{\vec k}\Bigl[{t_zc_z\over \epsilon_{\vec
k}-\epsilon_{\vec k+\vec Q}+i\delta}\bigr(2{f_{\vec k}-f_{\vec k+\vec
Q}\over \epsilon_{\vec k}-\epsilon_{\vec k+\vec Q}+i\delta}
\nonumber \\
-[f'_{\vec k}+f'_{\vec k+\vec Q}]\bigr)
-2t_z^2s_z^2\bigl({f''_{\vec k}-f''_{\vec
k+\vec Q}\over \epsilon_{\vec k}-\epsilon_{\vec k+\vec
Q}+i\delta}\bigr)\Bigr],
\label{eq:D4}
\end{eqnarray}
with $f'_{\vec k}=-f_{\vec k}(1-f_{\vec k})/k_BT$, $f''_{\vec k}=-f'_{\vec
k}(1-2f_{\vec k})/k_BT$, $c_z=\cos{k_zc}$, $s_z=\sin{k_zc}$.  For the
latter case
\begin{eqnarray}
A^0_z={-Uc^2\over 4}\sum_{\vec k}\Bigl[t_zc_z({f'_{\vec k}-f'_{\vec k+\vec
Q}\over \epsilon_{\vec k}-\epsilon_{\vec k+\vec Q}+i\delta})
\nonumber \\
+2t_z^2s_z^2\Bigl[{f''_{\vec k}-f''_{\vec
k+\vec Q}\over \epsilon_{\vec k}-\epsilon_{\vec k+\vec
Q}+i\delta}
\nonumber \\
+8{f_{\vec k}-f_{\vec k+\vec Q}\over (\epsilon_{\vec
k}-\epsilon_{\vec k+\vec Q}+i\delta )^3}-4{f'_{\vec k}+f'_{\vec k+\vec
Q}\over (\epsilon_{\vec k}-\epsilon_{\vec k+\vec Q}+i\delta )^2}\Bigr].
\label{eq:D5}
\end{eqnarray}

\begin{figure}
\leavevmode   
   \epsfxsize=0.33\textwidth\epsfbox{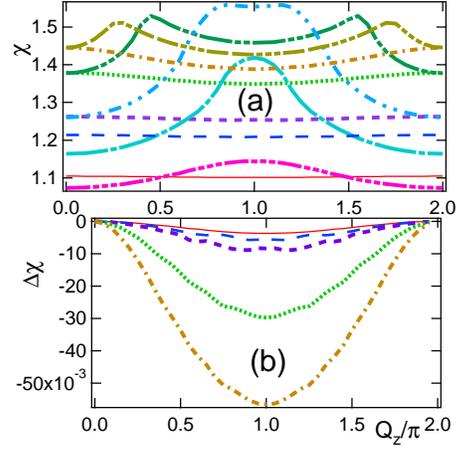}
\vskip0.5cm
\caption{(a)$\chi_0(\vec Q,Q_z)$ vs $Q_z$ for $t_{z0}=0.1t$, and $T=10K$, 
and a variety of chemical potentials $\mu$ = -0.003559 (solid line),
-0.08898 (long dashed line), -0.1779 (short dashed line), -0.2669 (dotted
line), -0.2847 (dot-dashed line), -0.2954 (long-long-short-short-short
dashed line), -0.3025 (long-short-short dashed line), -0.3203
(dash-dot-dot line), -0.3381 (long-short dashed line), and -0.3559 meV
(long-short-short-short dashed line). (b) $\Delta\chi = \chi_0(\vec
Q,Q_z)-\chi_0(\vec Q,Q_z=0)$, where the curves have the same meaning as in
frame (a).} 
\label{fig:nD4}
\end{figure}

\begin{figure}
\leavevmode   
   \epsfxsize=0.33\textwidth\epsfbox{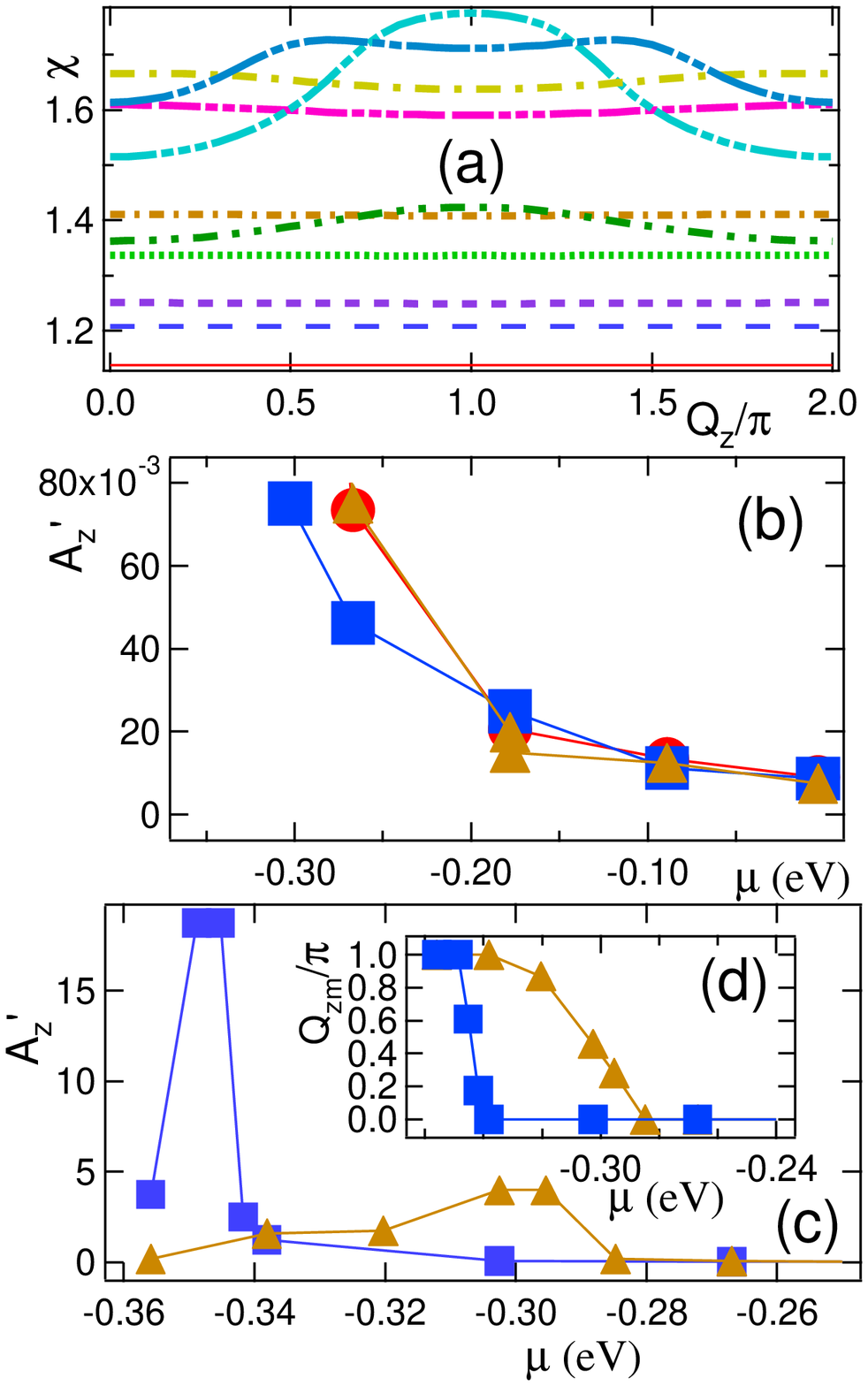}
\vskip0.5cm
\caption{(a)$\chi_0(\vec Q,Q_z)$ vs $Q_z$ for $t_{z0}=0.02t$, and $T=10K$, 
and a variety of chemical potentials $\mu$ = -0.003559 (solid line),
-0.08898 (long dashed line), -0.1779 (short dashed line), -0.2669 (dotted
line), -0.3025 (dot-dashed line), -0.3381 (long-long-short-short-short
dashed line), -0.3417 (long-dashed-dotted line), -0.3452
(long-short-short dashed line), -0.3488 (long-short-short-short dashed
line), and -0.3559 meV (long-dash-dot-dotted line). (b,c) $A_z'=A_z/Uc^2$
vs $\mu$ for $t_{z0}/t$ = 0.02 (squares, $A_z'\times 25$) and 0.1
(triangles,circles). (d) $Q_{zm}$ vs $\mu$ for $t_{z0}/t$ = 0.02 (squares)
and 0.1 (triangles).}
\label{fig:nD5}
\end{figure}

Figure~\ref{fig:nD4} (~\ref{fig:nD5}a) shows how $\chi_0(\vec Q,Q_z)$
varies with $Q_z$ for $t_{z0}=0.1t$ ($0.02t$), for a number of different
dopings. For the entire electron-doped regime, the peak is at $Q_{zm}=0$
(Fig.~\ref{fig:nD4}b,~\ref{fig:nD5}d), crossing over to $Q_{zm}=\pi /c$ in
the hole doped regime.  Away from the peak, the susceptibility varies as
$\hat A_zq_z^2$, with $q_z=Q_z-Q_{zm}$, and in the electron-doped regime
the full variation can be approximated by a cosine.  The amplitude of the
cosine falls to zero as the C-point is approached.  In the 
quasi-two-dimensional regime this amplitude scales with $t_{z0}^2$.
Figure~\ref{fig:nD5}b,c shows plots of the best parabolic fit to 
$A_z'=\hat A_z/c^2$ for $t_{z0}/t$ = 0.02 (squares) and 0.1 (triangles).
For $t_{z0}/t$ = 0.1, an alternative $A_z'$ is shown, found by fitting the
full susceptibility as a cosine in $q_z$ (circles).  The good agreement
between the two techniques shows that this is a reasonable approximation
in the electron-doped regime ($-0.2eV\le\mu\le 0$).  Near the
susceptibility peak, the variation is nonsinusoidal, and the parabolic fit
leads to a large value for $A_z'$.


\subsubsection{Staggered Stacking}

\begin{figure}
\leavevmode   
   \epsfxsize=0.33\textwidth\epsfbox{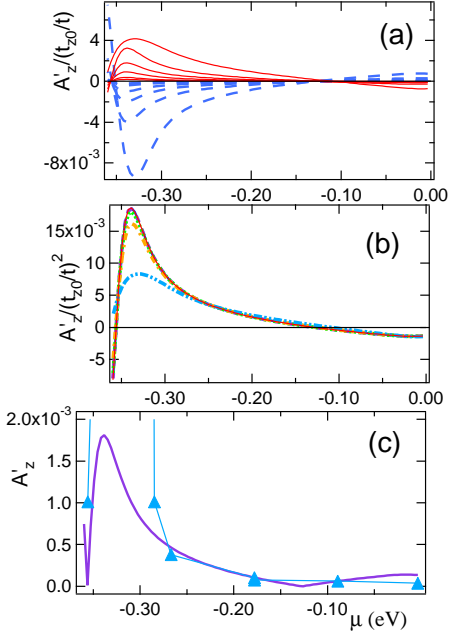}
\vskip0.5cm
\caption{(a)$A'_z=A_z/Uc^2$ vs chemical potential $\mu$ for $Q_z$ = 0
(solid lines) or $\pi$ (dashed lines), for a variety of
values of $t_{z0}$ and $T=100K$.  In order of increasing amplitude, the
values are $t_{z0}/t$ = 0.01, 0.02, 0.05, 0.1, 0.2, and 0.5.  (b) Scaling
of $A_z^{\prime (0)}$ with $(t_{z0}/t)^2$. Curves are $t_{z0}/t$ = 0.01
(solid line), 0.02 (long dashed line), 0.05 (short dashed line), 0.1
(dotted line), 0.2 (dot-dashed line), and 0.5 (dot-dot-dashed line). (c)
Comparison of $max(A_z)$ for staggered stacking (solid line) and uniform
stacking (triangles, $\times$1/20) at $t_{z0}/t$ = 0.1.} 
\label{fig:nD4b}
\end{figure}

The same calculations can be repeated for the $t_z$ of Eq.~\ref{eq:n21},
associated with staggered stacking; Fig.~\ref{fig:nD4b}a shows $A_z$
calculated from Eqs.~\ref{eq:D4},~\ref{eq:D5} at $Q_z$ = 0 (solid lines)
and $\pi$ (dashed lines). The frustration induced by staggering
of the CuO$_2$ layers is reflected in a strong suppression of the
$q_z$-dependence of $\chi$, which leaves a small residual contribution
{\it quadratic} in $t_{z0}$, Fig.~\ref{fig:nD4b}b.  Since $t_z$ vanishes
at $(\pi ,0)$, there is no shift of the susceptibility peak with doping.
Note the symmetry of the $A_z$ values between $0$ and $\pi$.  In fact,
$\chi (Q_z)$ is closely sinusoidal, particularly for small $t_{z0}$, with
maxima either at $\pi$ or $0$.  Thus, near either the H- or C-points, the
maximum of $\chi$ corresponds to $Q_z=\pi$.  For intermediate dopings,
$Q_z=0$ is favored.  At two distinct chemical potentials, the amplitude of
the cosine collapses and changes sign.  At the crossing points, $\chi$ is
independent of $Q_z$, leading formally to $T_N\rightarrow 0$.  Note from
Fig.~\ref{fig:nD4b}c that the suppression of $A_z$ is approximately in the 
same ratio as that of the resistivity, found above.

\subsection{Calculation of $T_N$}

\begin{figure}
\leavevmode   
   \epsfxsize=0.33\textwidth\epsfbox{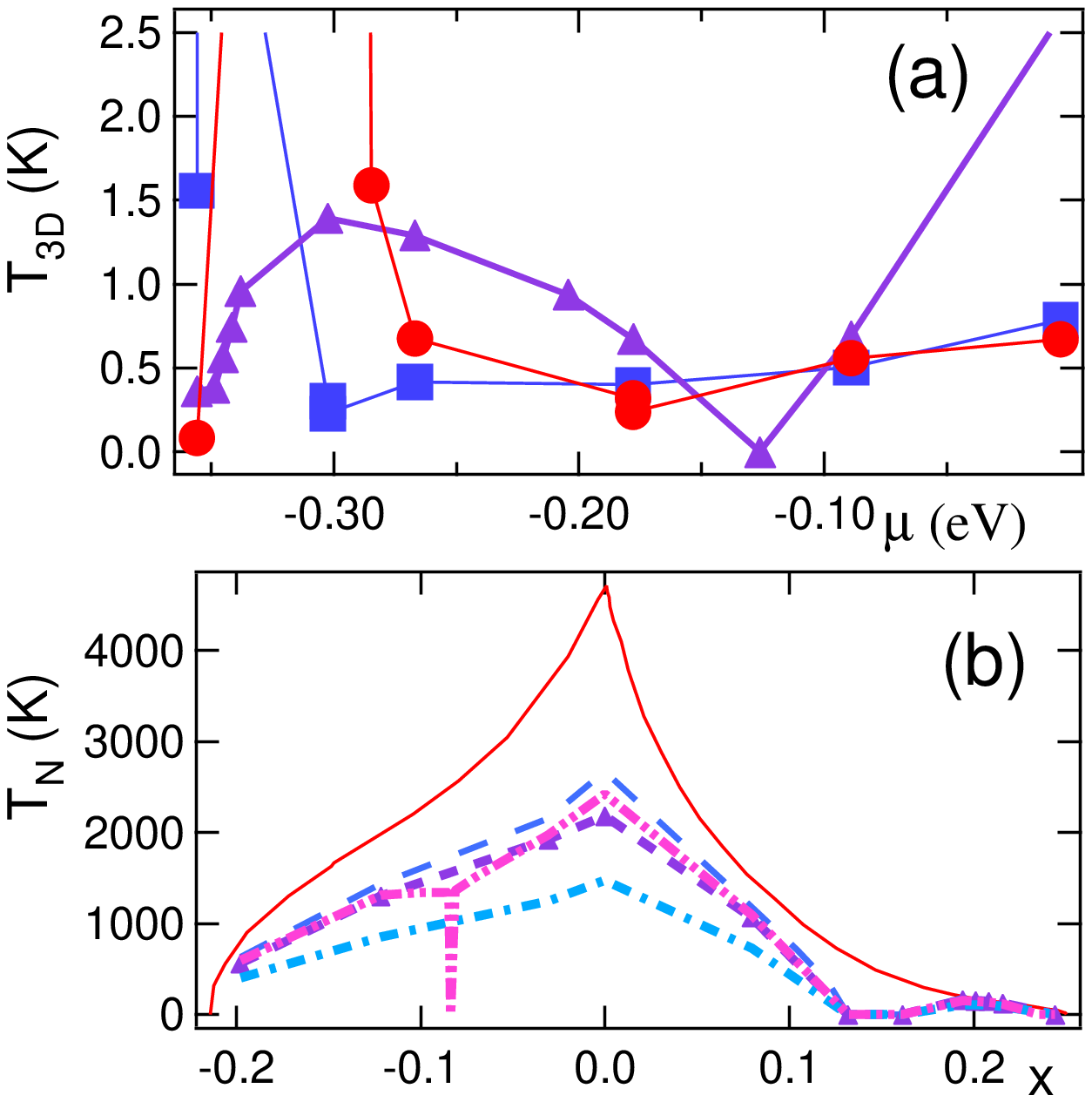}
\vskip0.5cm
\caption{(a)$T_{3D}$ vs $\mu$ for $T=10K$ and uniform stacking with
$t_{z0}=0.1t$ (circles, $\times$1/25) or 0.02 (squares), or staggered
stacking with $t_{z0}$=0.02 (triangles). 
(b) $T_N$ vs $x$, comparing mean field transition (solid line) with
interlayer coupling models (uniform stacking) assuming $t_{z0}/t$ = 0.1
(long dashed line), 0.02 (short dashed line), and 2$\times 10^{-6}$
(dot-dashed line), and the staggered stacking model assuming $t_{z0}/t$ = 
0.1 (dot-dot-dash line).} 
\label{fig:nD6} 
\end{figure}

When there is a finite interlayer hopping $t_z$, Eq.\ref{eq:B34h} becomes
\begin{eqnarray}
\delta-\bar\delta_0={6uTa^2c\over\pi^2A}\int_{0}^{\pi\over 
c}{dq_z\over\pi}\int_{y_0}^{y_0+Aq_c^2}{dy\over y}
tan^{-1}({2TC\over y})
\nonumber \\
\simeq{3uTa^2\over\pi A}\ln ({T\over T_{3D}})],
\label{eq:B34i}
\end{eqnarray}
where $y_0=\delta+A_zq_z'^2$ and $T_{3D}=\pi^2A_z/2Ce^2c^2$.  (A small
correction to $\bar\delta_0$ is neglected.  Treating the $q_z$ dependence
as a cosine rather than a cutoff quadratic
leads to qualitatively similar results.)  Thus a finite $A_z$ always cuts
off the divergence found in Eq.\ref{eq:B34h}, leading to a finite $T_N$
whenever there is a zero-temperature Neel state (e.g., up to a QCP).
It should be noted that the above calculation implicitly assumed that
$T>T_{3D}\sim A_z$: for $T<T_{3D}$ the logarithm is cut off and the
system behaves like an anisotropic three-dimensional magnet.  For
$t_{z0}/t<0.1$, the system is generally in the quasi-two-dimensional
limit, Fig.~\ref{fig:nD6}a.  Figure~\ref{fig:nD6}b compares the mean-field
Neel transition with the Neel transition found assuming uniform
stacking and finite interlayer couplings $t_{z0}/t$ = 0.1, 0.02, and
$2\times 10^{-6}$ [the last found by scaling the $T_{3D}$ for
$t_{z0}/t=0.02$ by the ratio of $t_{z0}^2$'s].  It is seen that
$T_N\rightarrow 0$ as $t_{z0}\rightarrow 0$, albeit exceedingly slowly.

The above calculations are for uniform stacking.  For staggered stacking
$A_z$ is reduced, in approximately the same ratio as the resistivities.
Hence, the staggered stacking with $t_{z0}/t$ = 0.1 should be comparable
to uniform stacking with $t_{z0}/t$ = 0.02, as observed,
Fig.~\ref{fig:nD6}.  While $T_N$ technically goes to zero for
staggered stacking near $x=-0.0838$, the decrease is logarithmic, and in
practice no more than a weak dip is expected to be observed (the point
with $T_N=0K$ is omitted from the plot in Fig.~\ref{fig:nD7}). Hence, if
$t_{z0}$ is estimated from the resistivity, it will be nearly impossible
to distinguish uniform from staggered stacking via measurements of $T_N$.

In the above calculations, a constant value of $A$ was assumed for each
doping, as given in Fig.~\ref{fig:10}.  In fact, for the electron-doped
cuprates, $A\sim 1/T^{1.5}$ for $T>T_A^*$, Fig.~\ref{fig:45}.  This would
cause an enhancement of the logarithmic correction, $\sim T^{2.5}$,
tending to pin $T_N$ close to $T_A^*$.  For the present parameter values,
this could reduce $T_N$ by roughly a factor of two, still larger than the
experimental values.

\begin{figure}
\leavevmode   
   \epsfxsize=0.33\textwidth\epsfbox{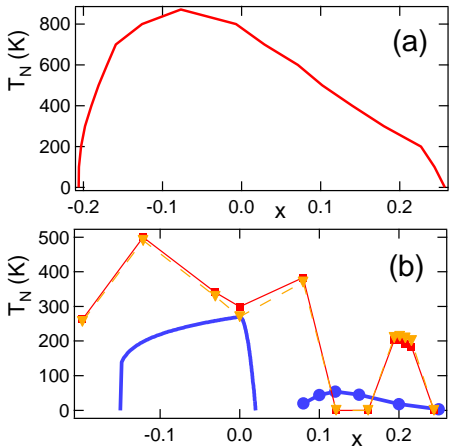}
\vskip0.5cm
\caption{(a) Mean field $T_{N}$ vs $x$ assuming paramagnetic $U_{eff}$
(Appendix B).  
(b) Corresponding $T_N$ vs $x$, calculated using Eq.~\protect\ref{eq:B34i}.
Squares = staggered stacking with $t_{z0}/t$ = 0.1; triangles = uniform
stacking with $t_{z0}/t$ = 0.02; solid line and circles = data, as in 
Fig.~\protect\ref{fig:nD7}.}
\label{fig:nD8} 
\end{figure}

A more likely source of the discrepancy is the possible temperature
dependence of $U_{eff}$, Appendix B.  The large $U_{eff}$ at half filling
arises from lack of screening, in the presence of a Mott gap -- and is
appropriate in analyzing the low-$T$ Fermi surfaces found in ARPES.  For
calculating the onset of the Mott gap, the mean field $T_N$, it is more
appropriate to use the paramagnetic susceptibility, as in
Fig~\ref{fig:111}a.  When this is done, considerably smaller transition
temperatures are found, both at the mean field level, Fig.~\ref{fig:nD8}a,
and when fluctuations and interlayer hopping are included,
Fig.~\ref{fig:nD8}b.  While the latter are closer to the experimental
values, no attempt has been made to correct $U_{eff}$ for the short range
gap.

\begin{references}
\bibitem{VTwarn}Vilk and Tremblay\cite{VAT} have cautioned that a naive
extension of their conserving approximation to include SCBA corrections
could lead to conservation law violations.
\end{references}
\end{document}